\documentclass[conference]{IEEEtran}

\usepackage{amsmath,amsfonts}

\let\labelindent\relax

\usepackage{algorithmic}
\usepackage{graphicx}
\usepackage{textcomp}
\usepackage{xcolor}
\usepackage{cite}
\usepackage{float}

\usepackage{subfigure}
\usepackage{booktabs}
\usepackage{xspace}

\newcommand{\etal}{{\textit{et al. }}}
\newcommand{\name}{HyperVision\xspace}
\newcommand{\Name}{H.V.\xspace}

\usepackage{threeparttable}
\usepackage{multicol}
\usepackage{multirow}
\usepackage{amssymb}
\usepackage{mathrsfs}
\usepackage{bm}
\usepackage{enumitem}
\usepackage{balance}

\usepackage[linesnumbered,ruled,vlined]{algorithm2e}
\newcommand{\tabincell}[2]{\begin{tabular}{@{}#1@{}}#2\end{tabular}}

\begin{document}

\title{
    Detecting Unknown Encrypted Malicious Traffic in Real Time via Flow Interaction Graph Analysis
    \vspace{-10mm}
}

\author{
    \IEEEauthorblockN{
        Chuanpu Fu\IEEEauthorrefmark{1}, Qi Li\IEEEauthorrefmark{2}\IEEEauthorrefmark{3}, Ke Xu\IEEEauthorrefmark{1}\IEEEauthorrefmark{3}
    }
    \IEEEauthorblockA{
        \IEEEauthorrefmark{1}Department of Computer Science and Technology, Tsinghua University
    }
    \IEEEauthorblockA{
        \IEEEauthorrefmark{2}Institute for Network Sciences and Cyberspace, Tsinghua University
        \IEEEauthorrefmark{3}Zhongguancun Lab
    }
}

\IEEEoverridecommandlockouts
\makeatletter\def\@IEEEpubidpullup{5.0\baselineskip}\makeatother
\IEEEpubid{\parbox{\columnwidth}{
    Network and Distributed System Security (NDSS) Symposium 2023\\
    27 February - 3 March 2023, San Diego, CA, USA\\
    ISBN 1-891562-83-5\\
    https://dx.doi.org/10.14722/ndss.2023.23080\\
    www.ndss-symposium.org
}
\hspace{\columnsep}\makebox[\columnwidth]{}}

\maketitle

\begin{abstract}

Nowadays traffic on the Internet has been widely encrypted to protect its confidentiality and privacy. However, traffic encryption is always abused by attackers to conceal their malicious behaviors. Since the encrypted malicious traffic has similar features to benign flows, it can easily evade traditional detection methods. Particularly, the existing encrypted malicious traffic detection methods are supervised and they rely on the prior knowledge of known attacks (e.g., labeled datasets). Detecting unknown encrypted malicious traffic in real time, which does not require prior domain knowledge, is still an open problem.

In this paper, we propose \name, a realtime unsupervised machine learning (ML) based malicious traffic detection system. Particularly, \name is able to detect unknown patterns of encrypted malicious traffic by utilizing a compact in-memory graph built upon the traffic patterns. The graph captures flow interaction patterns represented by the graph structural features, instead of the features of specific known attacks. We develop an unsupervised graph learning method to detect abnormal interaction patterns by analyzing the connectivity, sparsity, and statistical features of the graph, which allows \name to detect various encrypted attack traffic without requiring any labeled datasets of known attacks. Moreover, we establish an information theory model to demonstrate that the information preserved by the graph approaches the ideal theoretical bound. We show the performance of \name by real-world experiments with 92 datasets including 48 attacks with encrypted malicious traffic. The experimental results illustrate that \name achieves at least 0.92 AUC and 0.86 F1, which significantly outperform the state-of-the-art methods. In particular, more than 50\% attacks in our experiments can evade all these methods. Moreover, \name achieves at least 80.6 Gb/s detection throughput with the average detection latency of 0.83s.

\end{abstract}

\section{Introduction} \label{section:introduction}

Traffic encryption has been widely adopted to protect the information delivered on the Internet. Over 80\% websites adopted HTTPS to prevent data breach in 2019~\cite{CiscoETA,C&S21-EncryptedTraffic}. However, the cipher-suite for such protection is double-edged. In particular, the encrypted traffic also allows attackers to conceal their malicious behaviors, e.g., malware campaigns~\cite{AISEC16-Encrypted}, exploiting vulnerabilities~\cite{CCS17-Deemon}, and data exfiltration~\cite{IMC17-DataStolen}. The ratio of encrypted malicious traffic on the Internet is growing significantly~\cite{CONEXT12-BotFinder,AISEC16-Encrypted,KDD17-EncryptedTraffic} and exceeds 70\% of the entire malicious traffic~\cite{CiscoETA}.

However, encrypted malicious traffic detection is not well addressed due to the low-rate and diverse traffic patterns~\cite{AISEC16-Encrypted, CCS13-weakpwd,IMC17-DataStolen}. The traditional signature based methods that leverage deep packet inspection (DPI) are invalid under the attacks with the encrypted payloads~\cite{USEC08-Botsniffer}. Table~\ref{table:example} compares the existing malicious traffic detection methods. Different from plain-text malicious traffic, the encrypted traffic has similar features to benign flows and thus can evade existing machine learning (ML) based detection systems as well~\cite{C&S21-EncryptedTraffic,KDD17-EncryptedTraffic,AISEC16-Encrypted}. Particularly, the existing encrypted traffic detection methods are supervised, i.e., relying on the prior knowledge of known attacks, and can only detect attacks with known traffic patterns. They extract features of specific known attacks and use labeled datasets of known malicious traffic for model training~\cite{AISEC16-Encrypted,CONEXT12-BotFinder,KDD17-EncryptedTraffic}. Thus, they are unable to detect a broad spectrum of attacks with encrypted traffic~\cite{IMC17-DataStolen,CCS13-weakpwd,CCS17-Deemon,SP13-Crossfire}, which are constructed with unknown patterns~\cite{USEC14-ScanScan}. Besides, these methods are incapable of detecting both attacks constructed with and without encrypted traffic and unable to achieve generic detection because features of encrypted and non-encrypted attack traffic are significantly different~\cite{AISEC16-Encrypted,KDD17-EncryptedTraffic}.

In a nutshell, the existing methods cannot achieve unsupervised detection and they are unable to detect encrypted malicious traffic with unknown patterns. In particular, the encrypted malicious traffic has stealthy behaviors, which cannot be captured by these methods~\cite{AISEC16-Encrypted,CONEXT12-BotFinder} that detect attacks according to the patterns of a single flow. However, it is still feasible to detect such attack traffic because these attacks involve multiple attack steps with different flow interactions among attackers and victims, which are distinct from benign flow interactions patterns~\cite{SP17-lustrum,DSN17-DynaMiner,CCS13-weakpwd,CCS20-MySC,USEC20-Sunrise}. For example, the encrypted flow interactions between spam bots and SMTP servers are significantly different from the legitimate communications~\cite{USEC20-Sunrise} even if the single flow of the attack is similar to the benign one. Thus, this paper  explores utilizing interaction patterns among various flows for malicious traffic detection.

\newcommand{\bst}[1]{\color[rgb]{0.117, 0.447, 0.999}#1}
\newcommand{\wor}[1]{\color[rgb]{0.753,0,0}#1}

\newcommand{\cmark}{\bm{$\checkmark$}}
\newcommand{\xmark}{\bm{$\times$}}

\newcommand{\y}{\bst{\cmark}}
\newcommand{\n}{\wor{\xmark}}

\renewcommand{\arraystretch}{1.1}
\begin{table*}[!t]
    \footnotesize
    \centering
    \vspace{-6mm}
    \setlength\tabcolsep{3.8pt}
    \caption{The comparison with the existing methods of malicious traffic detection.}
    \vspace{-4.2mm}
    \begin{center}
    \begin{threeparttable}
    \begin{tabular}{c|c|c|cc|ccc|cc}
    \toprule
    \multirow{3}{*}{\tabincell{c}{Data Source\\Categories}} & \multirow{3}{*}{\tabincell{c}{Data Sources}} & \multirow{3}{*}{\tabincell{c}{Typical Methods}} & \multicolumn{2}{c|}{\tabincell{c}{Data for Detection}} & \multicolumn{3}{c|}{\tabincell{c}{Design Goals}} & \multicolumn{2}{c}{\tabincell{c}{Detection Performance}} \\
    \cline{4-10}
    & & &\multirow{2}{*}{\tabincell{c}{Unlabeled\\Datasets}} & \multirow{2}{*}{\tabincell{c}{Multi-Flow \\Features}} &  \multirow{2}{*}{\tabincell{c}{Generic\\Detection}} &  \multirow{2}{*}{\tabincell{c}{Realtime\\Detection}} & \multirow{2}{*}{\tabincell{c}{Unknown\\Attacks}} & \multirow{2}{*}{\tabincell{c}{Low\\Latency}} & \multirow{2}{*}{\tabincell{c}{High\\Throughput}}\\
    & & & & & & & & & \\
    \midrule
    \multirow{5}{*}{\tabincell{c}{Encrypted Traffic}} & \multirow{2}{*}{\tabincell{c}{Protocol Headers}}  & TLS Extensions~\cite{CiscoETA} & \n & \n & \n & \n & \n & \n & \y \\
    & & HTTPS Headers~\cite{KDD17-EncryptedTraffic} & \n & \n & \n & \n & \n & \n & \n \\
    \cline{2-3}
    & \multirow{3}{*}{\tabincell{c}{Related Flows}} & Time Series~\cite{CONEXT12-BotFinder} & \n & \n & \n & \n & \n & \n & \n \\
    & & TLS Handshakes~\cite{AISEC16-Encrypted} & \n & \n & \n & \n & \n & \n & \n \\
    & & Flow Statistics~\cite{TIFS18-Qi} & \y & \n & \n & \y & \n & \n & \y \\ 
    \cline{1-10}
    \multirow{5}{*}{\tabincell{c}{Plain-text and\\Encrypted Traffic}} & \multirow{2}{*}{\tabincell{c}{Network Logs}} & Intrusion Events~\cite{CCS17-Deeplog} & \y & \n & \n & \n & \y & \n & \n \\
    & & Sampled Connections~\cite{ACSAC12-Disclose} & \y & {\y}\tnote{1} & \n & \y & \n & \n & \y \\
    \cline{2-3}
    & \multirow{3}{*}{\tabincell{c}{Traffic Features}} & Per-Packet Features~\cite{NDSS18-Kitsune} & \y & \n & \n & \n & \y & \y & \n \\
    & & Per-Flow Features~\cite{NDSS21-Flowlens} & \n & \n & \n & \y & \n & \y & \n \\
    \cline{3-10}
    & & \tabincell{c}{\textbf{Flow Interaction Graph}} & \y & \y & \y & \y & \y & \y & \y \\
    \bottomrule
    \end{tabular}
    \begin{tablenotes}
        \scriptsize
        \item[1] Existing multi-flow features can only represent the features of specific flows, which cannot be used to represent complicated interaction patterns among various flows.
    \end{tablenotes}
    \end{threeparttable}
    \end{center}
    \label{table:example}
    \vspace{-6mm}
\end{table*}

To the end, we propose \name, a realtime detection system that aims to capture footprints of encrypted malicious traffic by analyzing interaction patterns among flows. In particular, it can detect encrypted malicious flows with unknown footprints by identifying abnormal flow interactions, i.e., the interaction patterns that are distinct from benign ones. To achieve this, we build a compact graph to capture various flow interaction patterns so that \name can perform detection on various encrypted traffic according to the graph. The graph allows us to detect attacks without accessing packet payloads, while retaining the ability of detecting traditional (known) attacks with plain-text traffic. Therefore, \name can detect the malicious traffic with unknown patterns by learning the graph structural features. Meanwhile, by learning the graph structural features, it realizes unsupervised detection, which does not require model training with labeled datasets.

However, it is challenging to build the graph for realtime detection. We cannot simply use IP addresses as vertices and traditional four-tuple of flows~\cite{Netflow,IPFIX} as edges to construct the graph because the resulting dense graph cannot maintain interaction patterns among various flows, e.g., incurring the dependence explosion problem~\cite{NDSS21-WATSON}. Inspired by the study of the flow size distribution~\cite{SIGCOMM18-ElasticSketch,TOCS03-FlowDistribution}, i.e., most flows on the Internet are short while most packets are associated with long flows, we utilize two strategies to record different sizes of flows, and process the interaction patterns of short and long flows separately in the graph. Specifically, it aggregates the short flows based on the similarity of massive short flows on the Internet, which reduces the density of the graph, and performs distribution fitting for the long flows, which can effectively preserve flow interaction information.

We design a four-step lightweight unsupervised graph learning approach to detect encrypted malicious traffic by utilizing the rich flow interaction information maintained on the graph. First, we analyze the connectivity of the graph by extracting the connected components and identify abnormal components by clustering the high-level statistical features. By excluding the benign components, we also significantly reduce the learning overhead. Second, we pre-cluster the edges according to the observed local adjacency in edge features. The pre-clustering operations significantly reduce the feature processing overhead and ensure realtime detection. Third, we extract critical vertices by solving a vertex cover problem using Z3 SMT solver~\cite{Z3} to minimize the number of clustering. Finally, we cluster each critical vertex according to its connected edges, which are in the centers of the clusters produced by the pre-clustering, and thus obtain the abnormal edges indicating encrypted malicious traffic.

Moreover, to quantify the benefits of the graph based flow recording of \name over the existing approaches, we develop a \textit{flow recording entropy model}, an information theory based framework that theoretically analyzes the amount of information retained by the existing data sources of malicious traffic detection systems. By using this framework, we show that the existing sampling based and event based traffic data sources (e.g., NetFlow~\cite{Netflow} and Zeek~\cite{Zeek}) cannot retain high-fidelity traffic information. Thus, they are unable to record flow interaction information for the detection. But the graph in \name captures near-optimal traffic information for the graph learning based detection and the amount of the information maintained in the graph approaches the theoretical up-bound of the idealized data source with infinite storage according to the data processing inequality~\cite{TIT98-DPI}. Also, the analysis results demonstrate that the graph in \name achieves higher information density (i.e., amount of traffic information per unit of storage) than all existing data sources, which is the foundation of the accurate and efficient detection. 

We prototype \name\footnote{Source code and datasets: https://github.com/fuchuanpu/HyperVision.} with Intel’s Data Plane Development Kit (DPDK)~\cite{DPDK}. To extensively evaluate the performance of the prototype, we replayed 92 attack datasets including 80 new datasets collected in our virtual private cloud (VPC) with more than 1,500 instances. In the VPC, we collected 48 typical encrypted malicious traffic, including (i) encrypted flooding traffic, e.g., flooding target links~\cite{SP13-Crossfire}; (ii) web attacks, e.g., exploiting web vulnerabilities~\cite{CCS17-Deemon}; (iii) malware campaigns, including connectivity testing, dependency update, and downloading. In the presence of the background traffic by replaying the backbone network traffic~\cite{WIDE}, \name achieves 13.9\% $\sim$ 36.1\% accuracy improvements over five state-of-the-art methods. It detects all encrypted malicious traffic in an unsupervised manner with more than 0.92 AUC, 0.86 F1, where 44 of the real-world stealthy traffic cannot be identified by all the baselines, e.g., an advanced side-channel attack exploiting the CVE-2020-36516~\cite{CCS20-MySC} and many newly discovered cryptojacking attacks~\cite{CCS19-cryptojacking}. Moreover, \name achieves on average more than 100 Gb/s detection throughput with the average detection latency of 0.83s.

In summary, the contributions of our paper are five-fold:
\begin{itemize}
\vspace{-1mm}
    \item We propose \name, the first realtime unsupervised detection for encrypted malicious traffic with unknown patterns by utilizing the flow interaction graph.
    \item We develop several algorithms to build the in-memory graph that allows us to accurately capture interaction patterns among various flows.
    \item We design a lightweight unsupervised graph learning method to detect encrypted traffic via graph features.
    \item We develop a theoretical analysis framework established by information theory to show that the graph captures near-optimal traffic interaction information.
    \item We prototype \name and use the extensive experiments with various real-world  encrypted malicious traffic to validate its accuracy and efficiency.
\end{itemize}

\begin{figure*}[t]
    \vspace{-6mm}
	\begin{center}
	\includegraphics[width=0.97\textwidth]{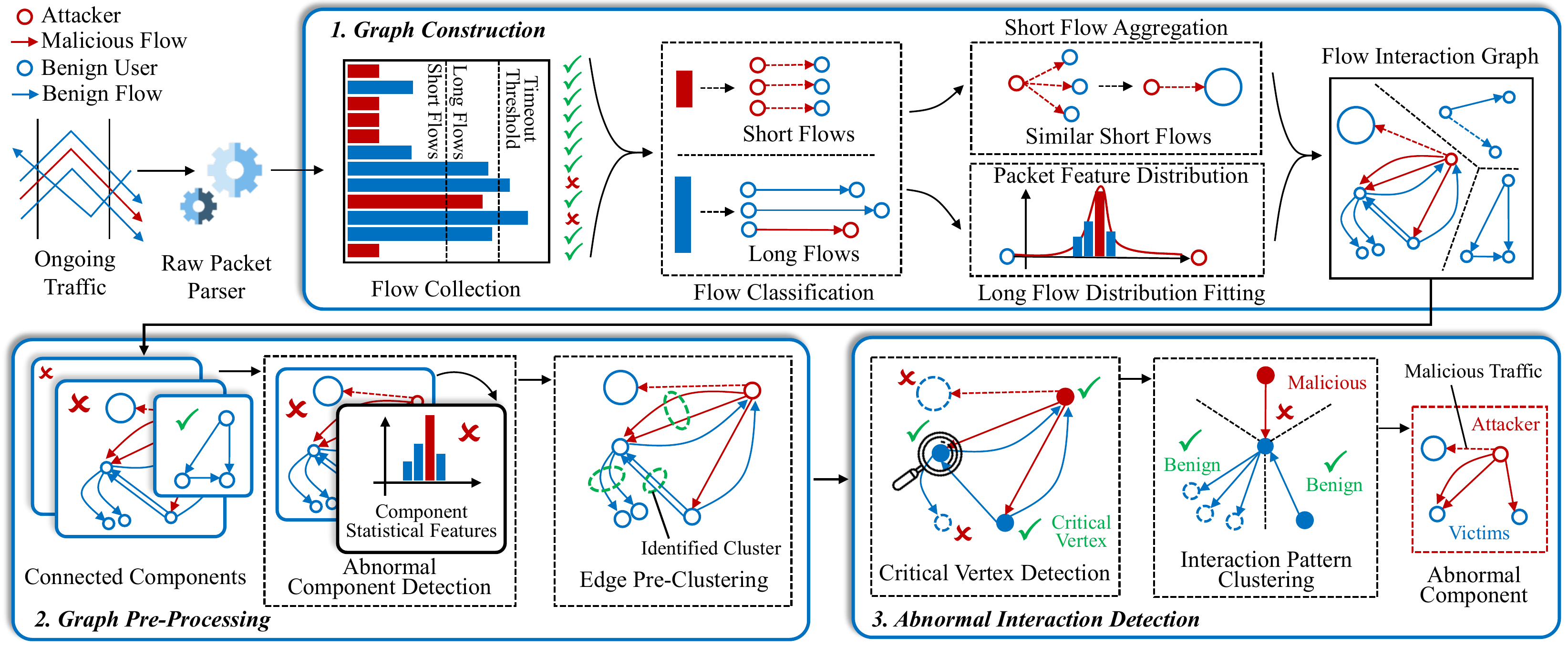}
	\vspace{-3mm}
	\caption{The overview of \name.}
	\label{diagram:high-level}
	\end{center}
	\vspace{-6mm}
\end{figure*}

The rest of the paper is organized as follows: Section~\ref{section:statement} introduces the threat model of \name. Section~\ref{section:overview} presents the high-level design of \name. In section~\ref{section:construction},~\ref{section:splitting}, and~\ref{section:detection}, we describe the detailed designs. In Section~\ref{section:analysis}, we conduct the theoretical analysis. In Section~\ref{section:evaluation}, we experimentally evaluate the performances. Section~\ref{section:related} reviews related works and Section~\ref{section:conclusion} concludes this paper. Finally, we present details in Appendix.

\section{Threat Model And Design Goals}\label{section:statement}
We aim to develop a realtime system (i.e., \name) to detect encrypted malicious traffic. It performs detection according to the traffic replicated by routers through port mirroring~\cite{SPAN}, which ensures that the system will not interfere with the traffic forwarding. After identifying encrypted malicious traffic, it can cooperate with the existing on-path malicious traffic defenses~\cite{NDSS20-Poseidon,TIFS22-Qi,TPDS22-Qi} to throttle the detected traffic. To perform detection on encrypted traffic, we cannot parse and analyze application layer headers and payloads.

In this paper, we focus on detecting active attacks constructed with encrypted traffic. We do not consider passive attacks that do not generate traffic to victims, e.g., traffic eavesdropping~\cite{USEC20-LTE} and passive traffic analysis~\cite{NDSS20-DNS-FP}. According to the existing studies~\cite{IMC17-DoS,SP17-lustrum,DSN17-DynaMiner,ToN22-Qib,ToN22-Qia,NDSS22-PMTUD}, attackers utilize reconnaissance steps to probe the information of victims, e.g., the password of a victim~\cite{CCS13-weakpwd}, the TCP sequence number of a TLS connection~\cite{CCS20-MySC,USEC22-Redirect}, and the randomized memory layout of a web server~\cite{INFO20-ZeroWall}, which cannot be accessed directly by attackers due to lack of prior knowledge. Note that, these attacks are normally constructed with many addresses owned or faked by attackers.

The design goals of \name are as follows: First, it should be able to achieve generic detection, i.e., detect attacks constructed with encrypted or non-encrypted traffic, which ensures that the attacks cannot evade detection by traffic encryption~\cite{AISEC16-Encrypted,IMC17-DataStolen}. Second, it is able to achieve realtime high-speed traffic processing, which means that it can identify whether the passing through encrypted traffic is malicious, while incurring low detection latency. Third, the performed detection by \name is unsupervised, which means that it does not require any prior knowledge of encrypted malicious traffic. That is, it should be able to deal with attacks with unknown patterns, i.e., zero-day attacks, which have not been disclosed~\cite{CCS21-Whisper}. Thus, we do not use any labeled traffic datasets for ML training. These issues cannot be well addressed by the existing detection methods~\cite{C&S21-EncryptedTraffic}.

\section{Overview of \name} \label{section:overview}
In this section, we develop \name that is an unsupervised detection system to capture malicious traffic in real time, in particular, encrypted malicious traffic. Normally, patterns of each flow in the encrypted malicious traffic, i.e., single-flow patterns, may be similar to benign flows, which allow them to evade the existing detection. However, the malicious behaviors appearing in the interaction patterns between the attackers and victims will be more distinct from the benign ones. Thus, in \name, we construct a compact graph to maintain interaction patterns among various flows and detect abnormal interaction patterns by learning the features of the graph. \name analyzes the graph structural features representing the interaction patterns without prior knowledge of known attack traffic and thus can achieve unsupervised detection against various attacks. It realizes generic detection by analyzing flows regardless of the traffic type and can detect encrypted and non-encrypted malicious traffic. Figure~\ref{diagram:high-level} shows three key parts of \name, i.e., graph construction, graph pre-processing, and abnormal interaction detection.

\noindent \textbf{Graph Construction.} \name collects network flows for graph construction. Meanwhile, it classifies the flows into short and long ones and records their interaction patterns separately for the purpose of reducing the density of the graph. In the graph, it uses different addresses as vertices that connect the edges associated with short and long flows, respectively. It aggregates the massive similar short flows to construct one edge for a group of short flows, and thus reduces the overhead for maintaining flow interaction patterns. Moreover, it fits the distributions of the packet features in the long flows to construct the edges associated with long flows, which ensures high-fidelity recorded flow interaction patterns, while addressing the issue of coarse-grained flow features in the traditional methods~\cite{IPFIX}. We will detail how \name maintains the high-fidelity flow interaction patterns in the in-memory graph in Section~\ref{section:construction}.

\noindent \textbf{Graph Pre-Processing.} We pre-process the built interaction graph to reduce the overhead of processing the graph by extracting connected components and cluster the components using high-level statistics. In particular, the clustering can detect the components with only benign interaction patterns accurately and thus filters these benign components to reduce the scale of the graph. Moreover, we perform a pre-clustering and use the generated cluster centers to represent the edges in the identified clusters. We will detail the graph pre-processing in Section~\ref{section:splitting}.

\noindent \textbf{Malicious Traffic Detection Based on the Graph.} We achieve unsupervised encrypted malicious traffic detection by analyzing the graph features. We identify critical vertices in the graph by solving a vertex cover problem, which ensures that the clustering based graph learning processes all edges with the minimum number of clustering. For each selected vertex, we cluster all connected edges according to their flow features and structural features that represent the flow interaction patterns. \name can identify abnormal edges in real time by computing the loss function of the clustering. We will describe the details of graph learning based detection in Section~\ref{section:detection}.

\section{Graph Construction} \label{section:construction}

In this section, we present the design details of constructing the flow interaction graph that maintains interaction patterns among various flows. In particular, we classify different flows, i.e., short and long flows, and aggregate short flows, and perform the distribution fitting for long flows, respectively, for efficient graph construction. In Section~\ref{section:analysis}, we will show that the graph retains the near-optimal information for detection.

\subsection{Flow Classification} \label{section:construction:classification}
\begin{figure}[t]
    \subfigcapskip=-1mm
    \vspace{-4mm}
    \begin{center}
    \subfigure [FCT distribution.]{
        \label{graph:distribution:fct}
		\includegraphics[width=0.23\textwidth]{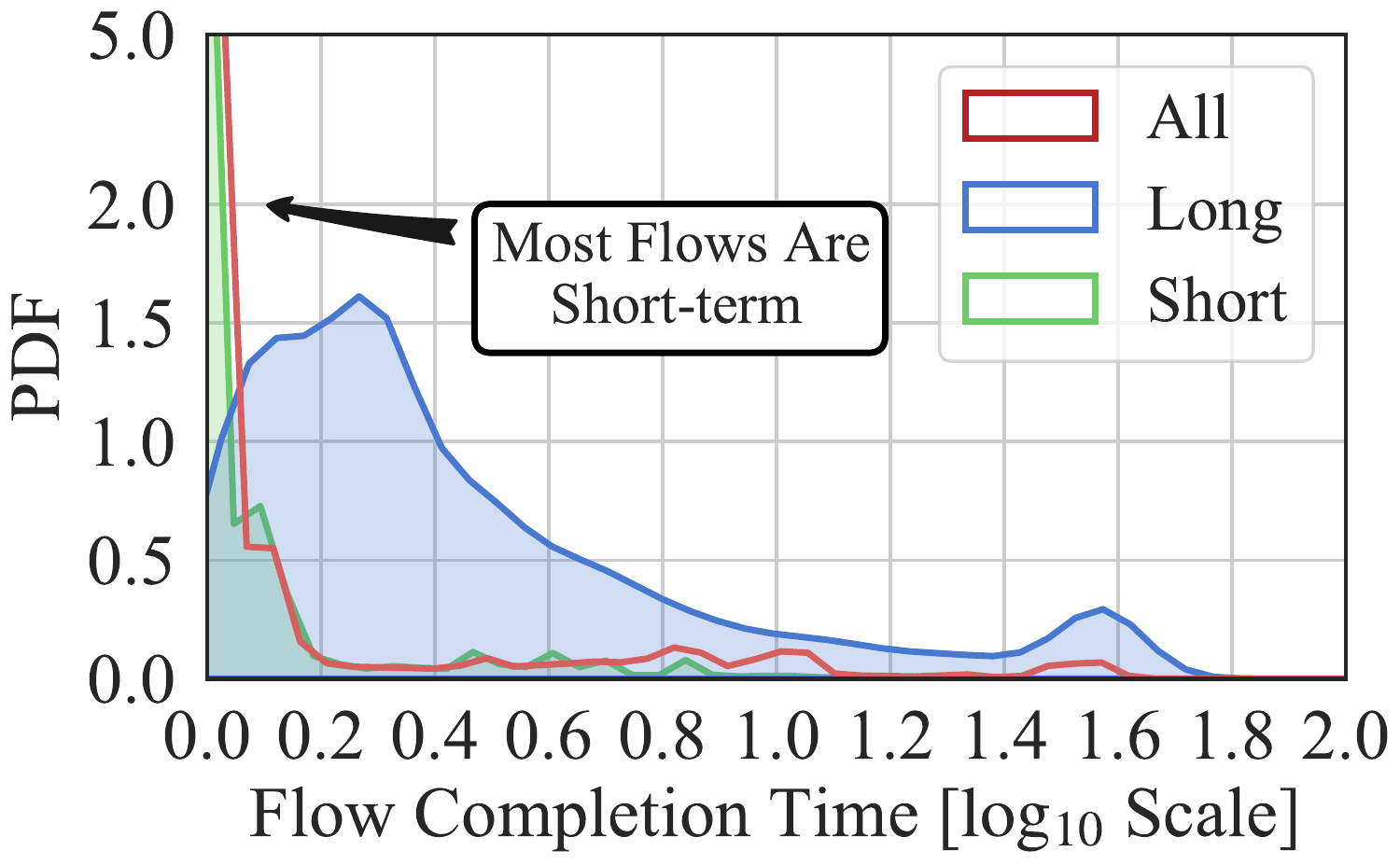}
	}
    \hspace{-3mm}
    \subfigure [Flow length distribution.]{
        \label{graph:distribution:len}
		\includegraphics[width=0.23\textwidth]{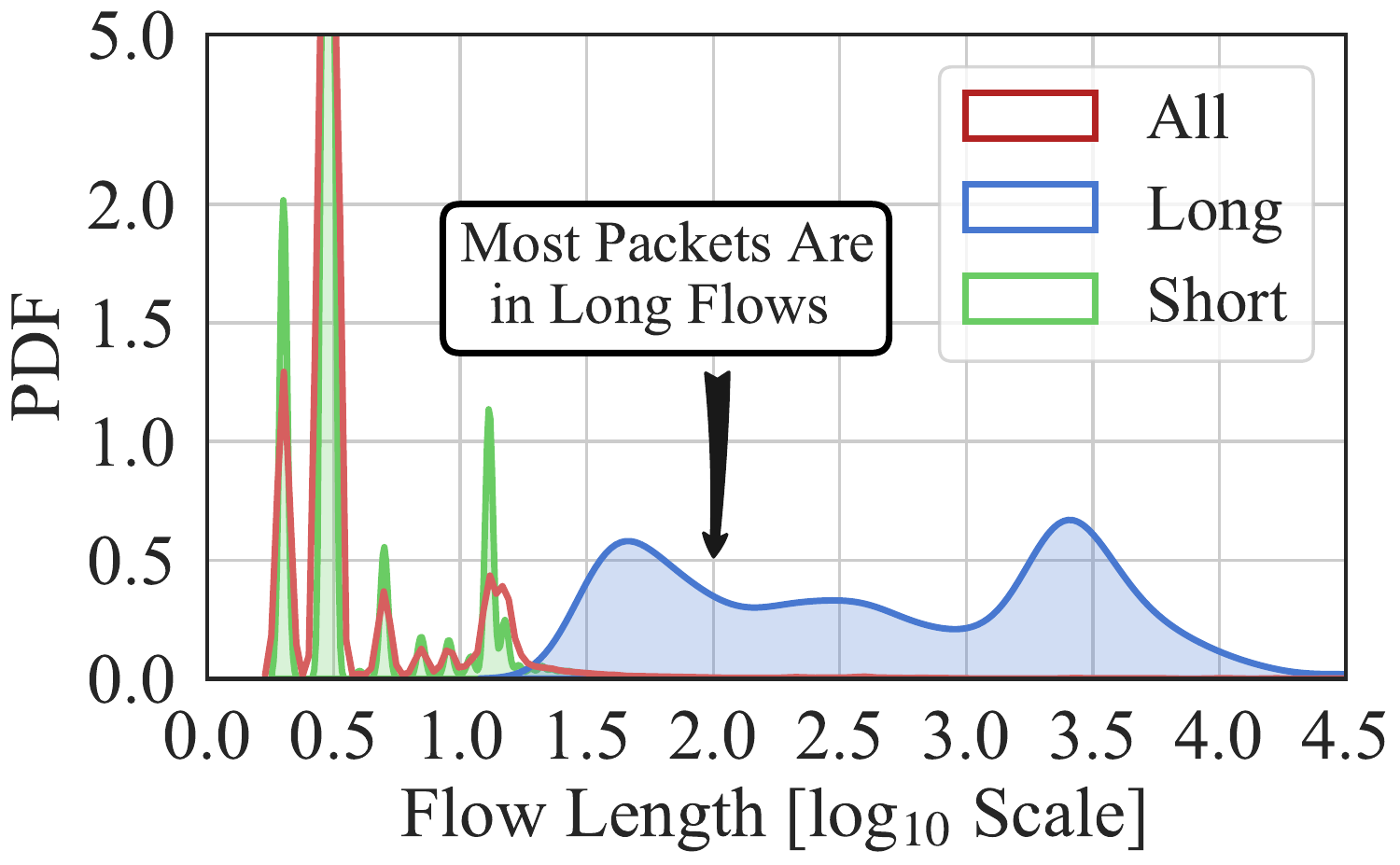}
	}
    \end{center}
    \vspace{-4mm}
    \caption{The real-world flow features distribution of short and long flows.} 
    \vspace{-2mm}
\label{graph:distribution}
\end{figure}

\begin{figure}[t]
    \subfigcapskip=-1mm
    \begin{center}
    \subfigure [Traditional flows as edges.] {
        \label{graph:aggregation:before}
		\includegraphics[width=0.19\textwidth]{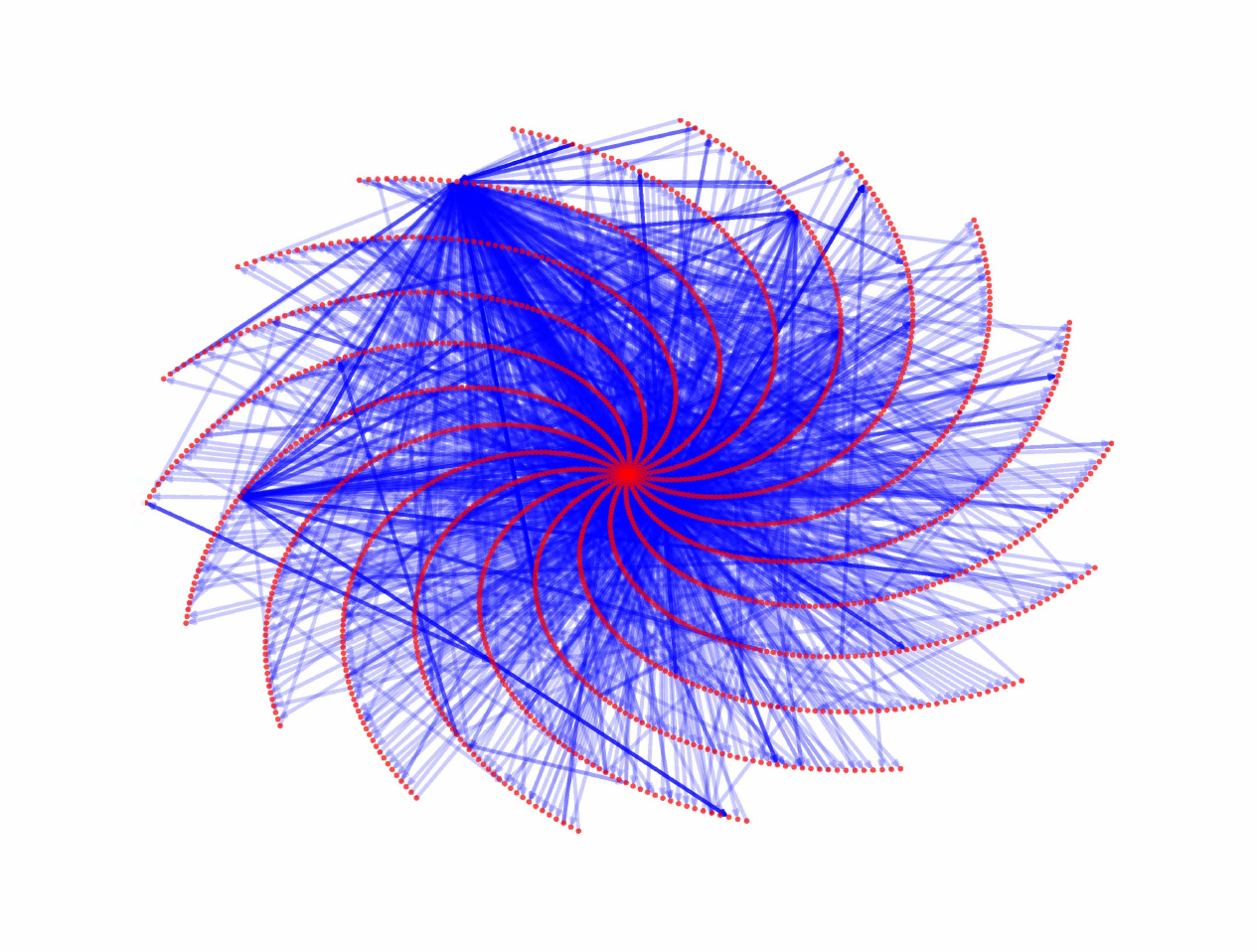}
	}
    \hspace{1mm}
	\subfigure[Short flow aggregation.] {
        \label{graph:aggregation:after}
		\includegraphics[width=0.19\textwidth]{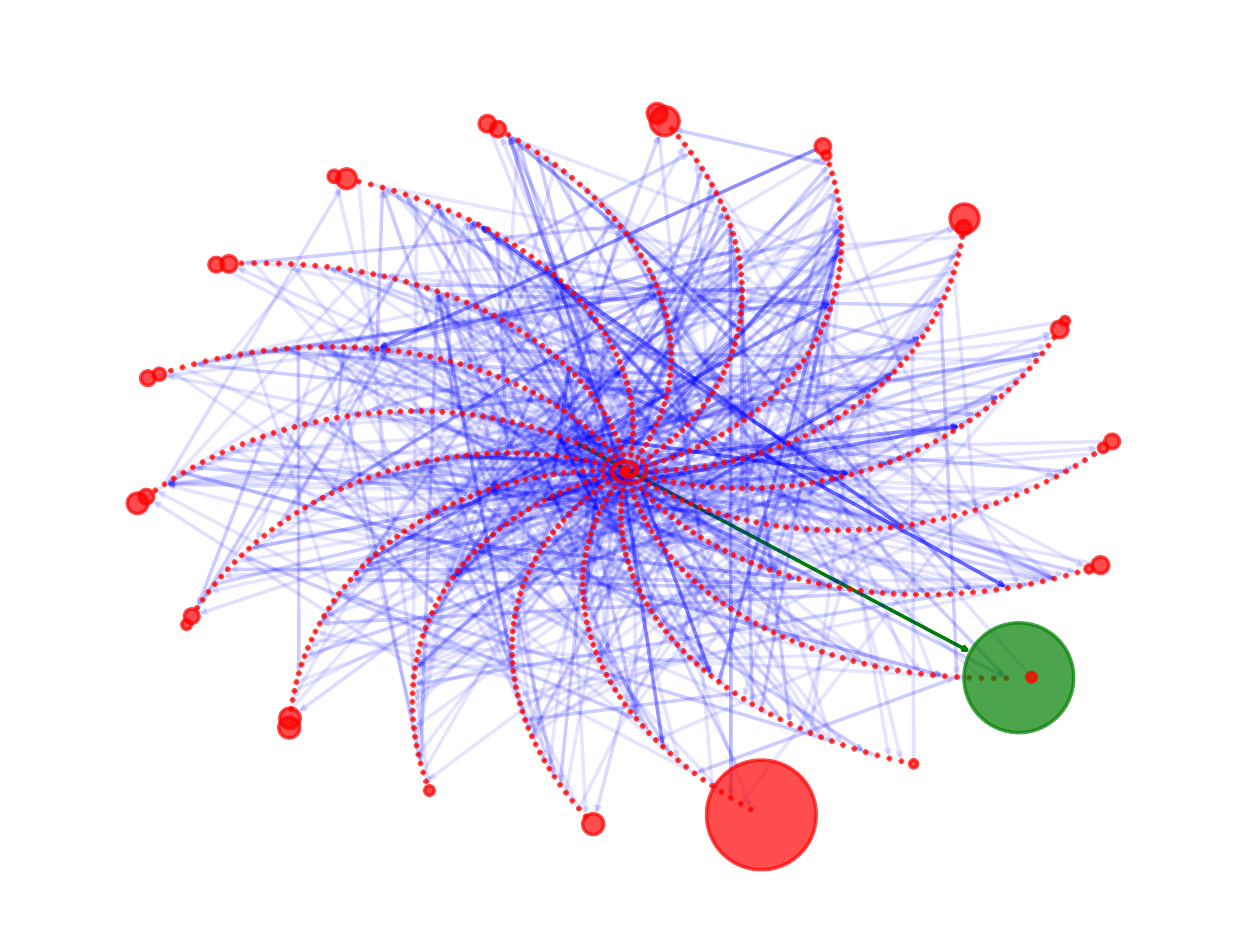}
	}
	\vspace{-4mm}
    \end{center}
    \caption{\name aggregates short flows to reduce the dense graph.} 
    \vspace{-4mm}
    \label{graph:aggregation}
\end{figure}

In order to efficiently analyze flows captured on the Internet, we need to avoid the dependency explosion among flows during the graph construction. We classify the collected flows into short and long flows, according to the flow size distribution~\cite{TOCS03-FlowDistribution} (see Figure~\ref{graph:distribution}), and then reduce the density of the graph (shown in Figure~\ref{graph:aggregation}). Figure~\ref{graph:distribution} shows the distribution of flow completion time (FCT) and flow length of the MAWI Internet traffic dataset~\cite{WIDE} in Jan. 2020. For simplicity, we use the first $13 \times 10^{6}$ packets to plot the figure. According to the figure, we observe that only 5.52\% flows have FCT $>$ 2.0s. However, 93.70\% packets in the dataset are long flows with only 2.36\% proportion. Inspired by the observation, we apply different flow collection strategies for the short and long flows. 

We poll the per-packet information from a data-plane high-speed packet parsing engine and obtain their source and destination addresses, port numbers, and per-packet features, including protocols, lengths, and arrival intervals. These features can be extracted from both encrypted and plain-text traffic for generic detection. We develop a flow classification algorithm to classify the traffic (see Algorithm~\ref{algorithm:classification} in Appendix~\ref{section:appendix:algorithm}). It maintains a timer \textsc{time\_now}, a hash table that uses \textsc{Hash(src, dst, src\_port, dst\_port)} as key and the collected flows indicated by the sequences of their per-packet features as values. It traverses the hash table every \textsc{judge\_interval} second according to \textsc{time\_now} and judges the flow completion when the last packet arrived before \textsc{pkt\_timeout} second of \textsc{time\_now}. When the flows are completed, we classify them as long flows if the flows have more than \textsc{flow\_line} packets. Otherwise, we classify them as short flows. As shown in Figure~\ref{graph:distribution:len}, we can accurately classify short and long flows. The definitions of the hyper-parameters can be found in Table~\ref{table:configure} (see Appendix~\ref{section:appendix:algorithm}). Note that, we poll the state-less per-packet information from data-plane, while not maintaining flow states (e.g., a state machine~\cite{OSDI20-FPGADPI}) on the data-plane to prevent attackers manipulating the states, e.g., side-channel attack~\cite{SP12-SC} and evading detection~\cite{NDSS20-SYMTCP}.

\subsection{Short Flow Aggregation} \label{section:construction:aggregation}
We need to reduce the density of the graph for analysis. As shown in Figure~\ref{graph:aggregation:before}, the graph will be very dense for analysis if we use traditional four-tuple flows as edges, which is similar to the dependency explosion problem in provenance analysis~\cite{NDSS21-WATSON,NDSS21-UIScope}. We observe that most short flows have almost the same per-packet feature sequences. For instance, the encrypted flows of repetitive SSH cracking attempts originated from specific attackers~\cite{CCS13-weakpwd}. Thus, we perform the short flow aggregation to represent similar flows using one edge after the classification. 

We design an algorithm to aggregate short flows (see Algorithm~\ref{algorithm:aggregation} in Appendix~\ref{section:appendix:algorithm}). A set of flows can be aggregated when all the following requirements are satisfied: (i) the flows have the same source and/or destination addresses, which implies similar behaviors generated from the addresses; (ii) the flows have the same protocol type;  (iii) the number of the flows is large enough, i.e., when the number of the short flows reaches the threshold \textsc{agg\_line}, which ensures that the flows are repetitive enough. Next, we construct an edge for the short flows, which preserves one feature sequence (i.e., protocols, lengths, and arrival intervals) for all the flows, and their four-tuples. As a result, four types of edges associated with short flows exist on the graph, i.e., source address aggregated, destination address aggregated, both addresses aggregated, and without aggregation. Thus, a vertex connected to the edge can denote a group of addresses or a single address.

\begin{figure}[t]
    \subfigcapskip=-1mm
    \vspace{-3.8mm}
    \vspace{-0.1mm}
    \begin{center}
	\subfigure[Number of packet length buckets.]{
        \label{graph:fitting:num_len}
		\includegraphics[width=0.23\textwidth]{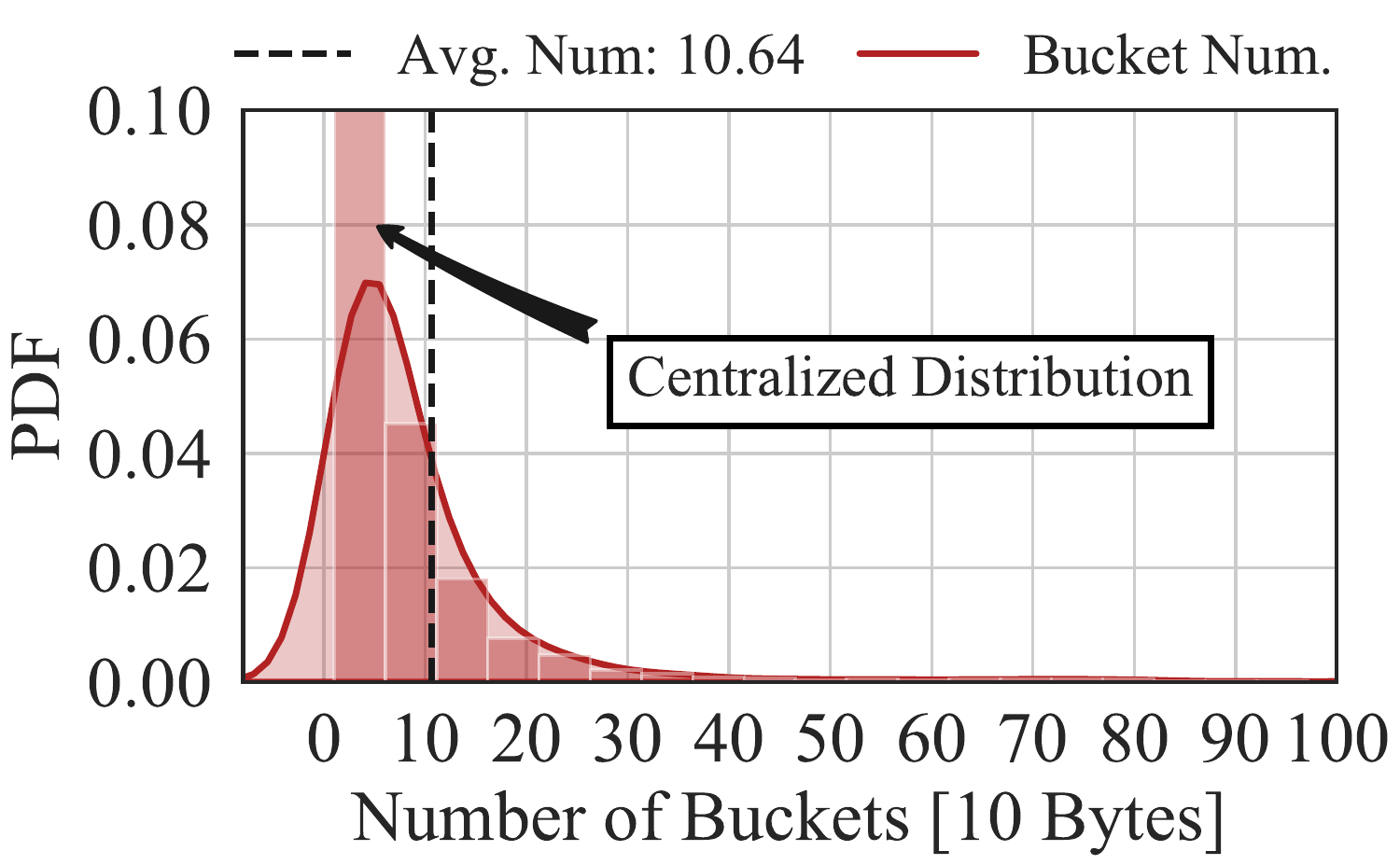}
	}
    \hspace{-4mm}
	\subfigure [Maximum bucket size.]{
        \label{graph:fitting:max_len}
		\includegraphics[width=0.23\textwidth]{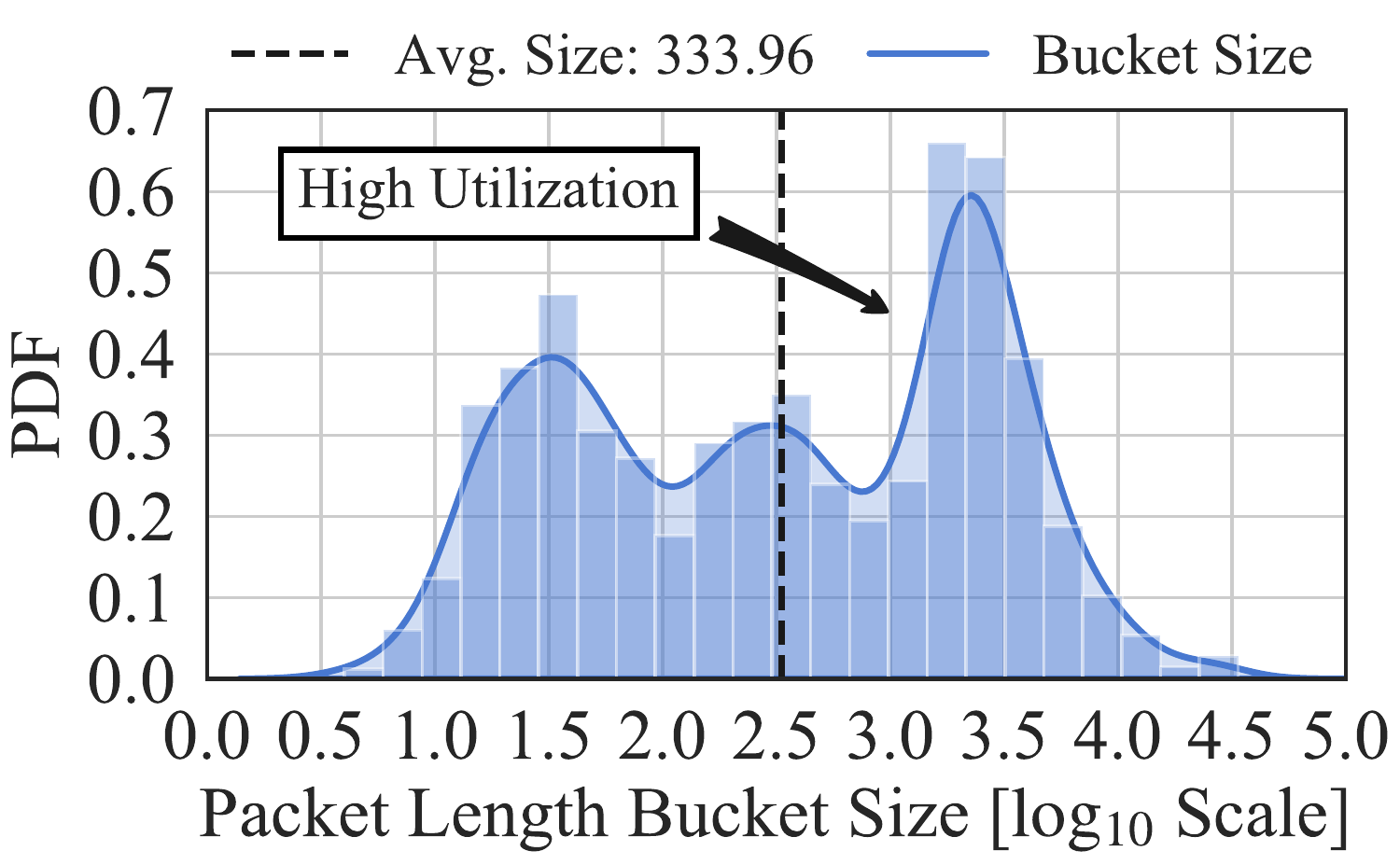}
	}
	\vspace{-2mm}
	\caption{The number and size of the buckets for feature distribution fitting.}
    \label{graph:fitting}
    \end{center}
    \vspace{-6mm}
\end{figure}

Figure~\ref{graph:aggregation} compares the graph using traditional flows as edges and our aggregated graph by using the real-world backbone traffic dataset, which is same to that used in Figure~\ref{graph:distribution}. The diameter of a vertex indicates the number of addresses denoted by the vertex and the depth of the color indicates the repeated edges. In Figure~\ref{graph:aggregation:after}, we observe that the algorithm reduces 93.94\% vertices and 94.04\% edges. The edge highlighted in green indicates short flows (i.e., 2.38 Kpps, from PH) exploiting a vulnerability. Note that, the flow aggregation reduces the storage overhead, which makes it feasible to maintain the in-memory graph for realtime detection.

\subsection{Feature Distribution Fitting for Long Flows} \label{section:construction:fitting}
Now we use histograms to represent the per-packet feature distributions of a long flow which avoid preserving their long per-packet feature sequences, since the features in long flows are centrally distributed. Specifically, we maintain a hash table to construct the histogram for each per-packet feature sequence in each long flow. According to our empirical study, we set the buckets widths for packet-length and arrival interval as 10 bytes and 1 ms, respectively, to trade off between the fitting accuracy and overhead. We calculate the hash code by dividing the per-packet features by the bucket width and increase the counter indexed by the hash code. Finally, we record the hash codes and the associated counters as the histograms. Note that, the coarse-grained flow statistics, e.g., numbers of packets~\cite{IPFIX}, are insufficient for encrypted malicious traffic detection~\cite{CONEXT12-BotFinder}, which also lose the flow interaction information~\cite{NetflowThreat}.

Figure~\ref{graph:fitting} shows the number of the used buckets and the maximum bucket size for the long flows in the same dataset shown in Figure~\ref{graph:distribution}. We confirm the centralized feature distribution, i.e., most packets in the long flows have similar packet lengths and arrival intervals. Specifically, in Figure~\ref{graph:fitting:num_len}, we fit the distribution of packet length using only 11 buckets on average, and most of the buckets collect more than 200 packets (see Figure~\ref{graph:fitting:max_len}), which demonstrate that the histogram based fitting is effective with low storage overhead. Similarly, the fitting for arrival interval uses 121 buckets on average and realizes 71 packets per bucket high utilization. Besides, we use the same method for protocol. We use the mask of protocols as the hash code and use smaller numbers of buckets to realize more efficient fitting due to the limited number of protocol types. Note that, Flowlens~\cite{NDSS21-Flowlens} used a similar histogram to efficiently utilize hardware flow tables on P4 switches. Instead, we construct the histograms to accurately analyze long flows. 

\section{Graph Pre-Processing} \label{section:splitting}
In this section, we pre-process the flow interaction graph to identify key components and pre-cluster the edges, which can enable realtime graph learning based detection against encrypted malicious traffic with unknown patterns.

\begin{figure}[t]
    \subfigcapskip=-1mm
    \vspace{-4mm}
    \begin{center}
	\subfigure[Component size distribution.]{
        \label{graph:component:distribution}
		\includegraphics[width=0.22\textwidth]{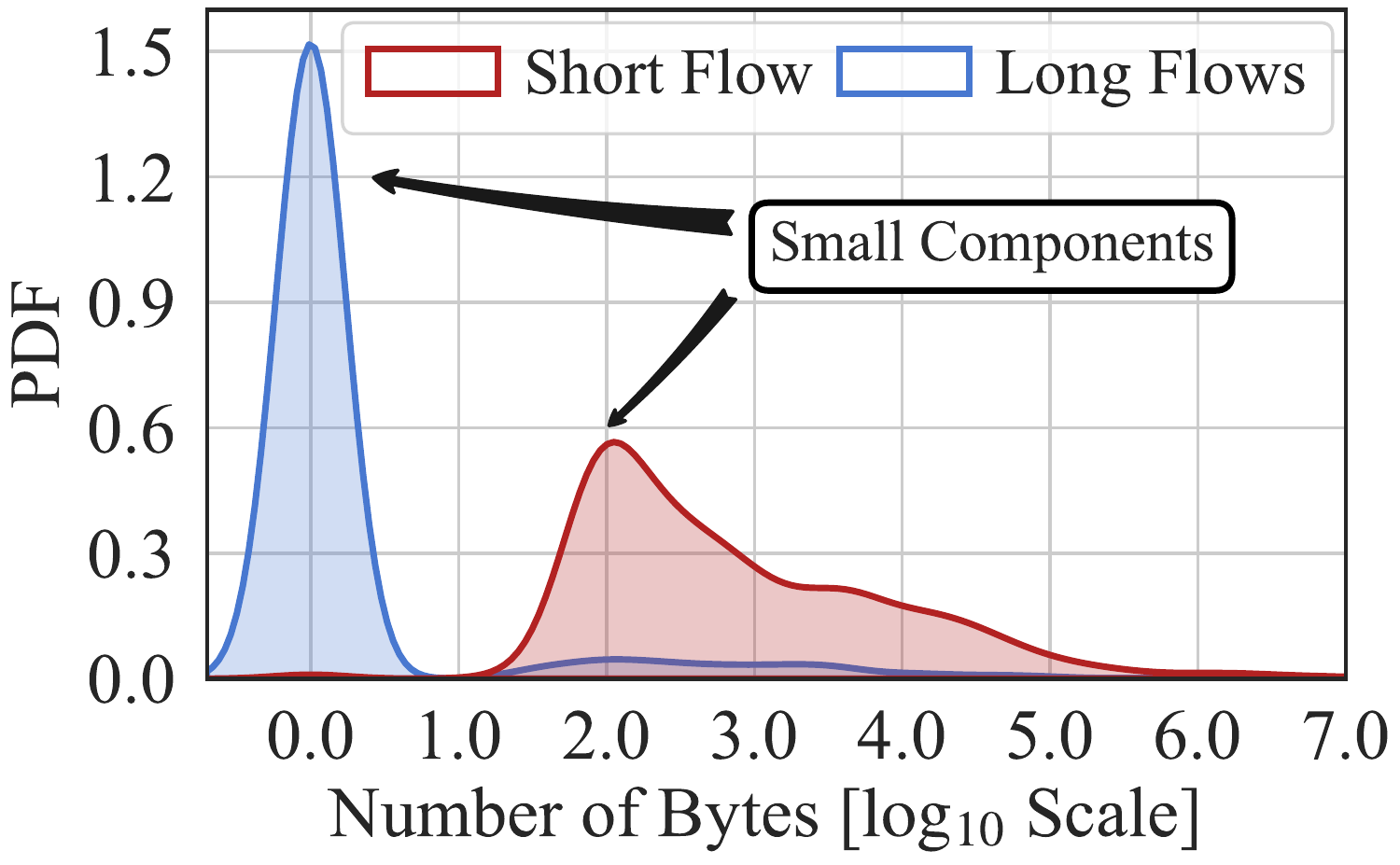}
	}
	\hspace{-2mm}
	\subfigure [Scatter of the components.]{
        \label{graph:component:cluster}
		\includegraphics[width=0.22\textwidth]{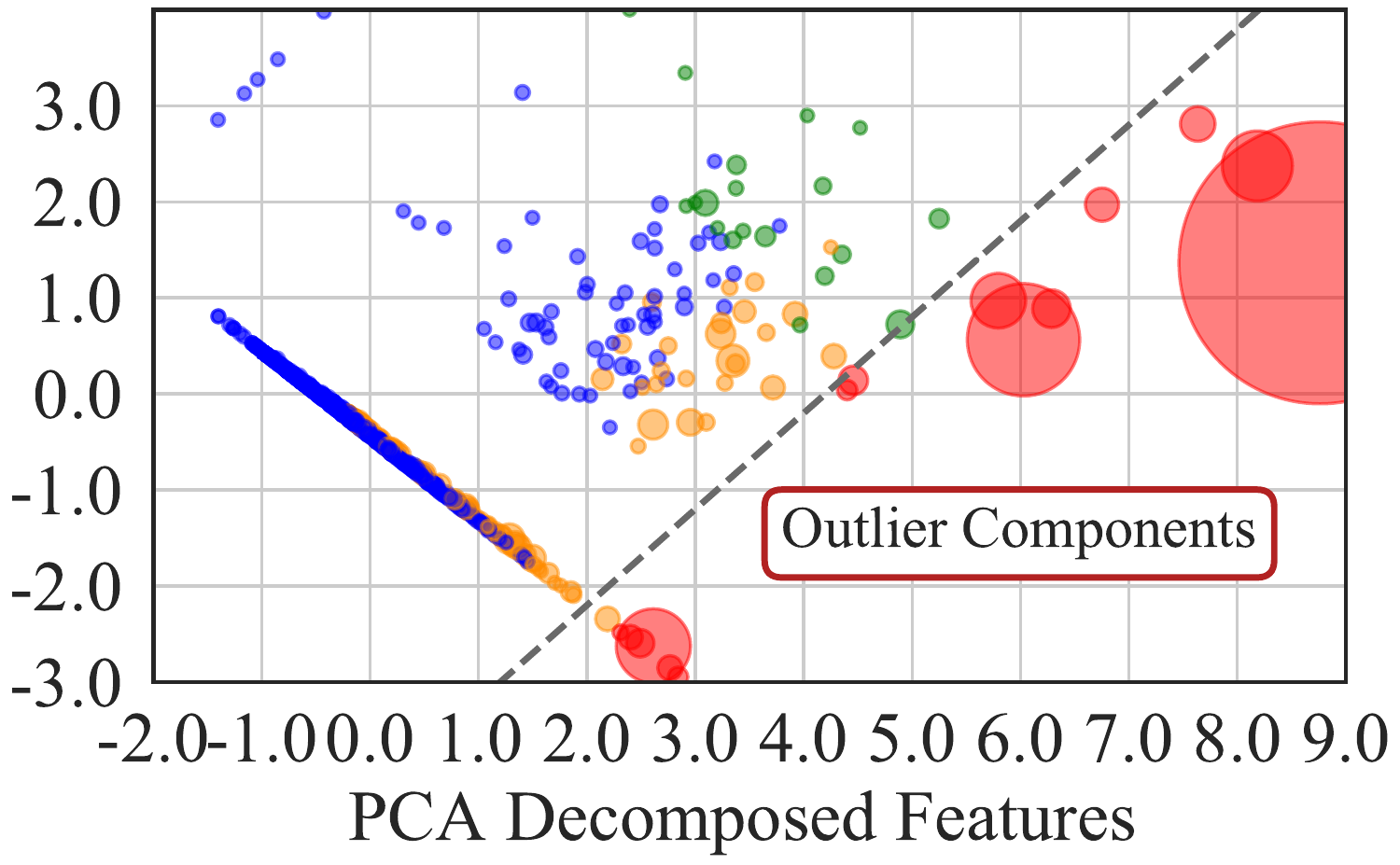}
	}
	\vspace{-2mm}
    \caption{The statistical features of the components.} 
    \label{graph:component}
    \end{center}
    \vspace{-6mm}
\end{figure}

\subsection{Connectivity Analysis} \label{section:splitting:connectivity}
To perform the connectivity analysis of the graph, we obtain the connected components by using depth-first search (DFS) and split the graph by the components. Figure~\ref{graph:component:distribution} presents the size distribution of the identified components of the MAWI traffic dataset~\cite{WIDE} collected in Jan. 2020. We observe that most components contain few edges with similar interaction patterns. Thus, we perform a clustering on the high-level statistics for the connected components to capture the abnormal components that have over one order of magnitude clustering loss than normal components as clustering outliers. Specifically, we extract five features to profile the components, including: (i) the number of long flows; (ii) the number of short flows; (iii) the number of edges denoting short flows; (iv) the number of bytes in long flows; and (v) the number of bytes in short flows. We perform a min-max normalization and acquire the centers using the density based clustering, i.e., DBSCAN~\cite{DBSCAN}. For each component, we calculate the Euclidean distance to its nearest center. We detect an abnormal component when its distance is over the $\mathrm{99^{th}}$ percentile of all the distances based on our empirical study. 

Figure~\ref{graph:component:cluster} shows an instance of the clustering, where the diameters indicate the scale of the traffic on the components (in the unit of bytes). We observe that most components are small, and a high ratio of huge components is classified as abnormal. All edges associated with the normal components are labeled as benign traffic, and the edges associated with the abnormal components will be further processed by the following steps. 

\subsection{Edge Pre-Clustering} \label{section:splitting:preclustering}
\begin{figure}[t]
    \subfigcapskip=-1mm
    \vspace{-4mm}
    \begin{center}
	\subfigure[Adjacent long flows.]{
        \label{graph:features:longlive}
		\includegraphics[width=0.22\textwidth]{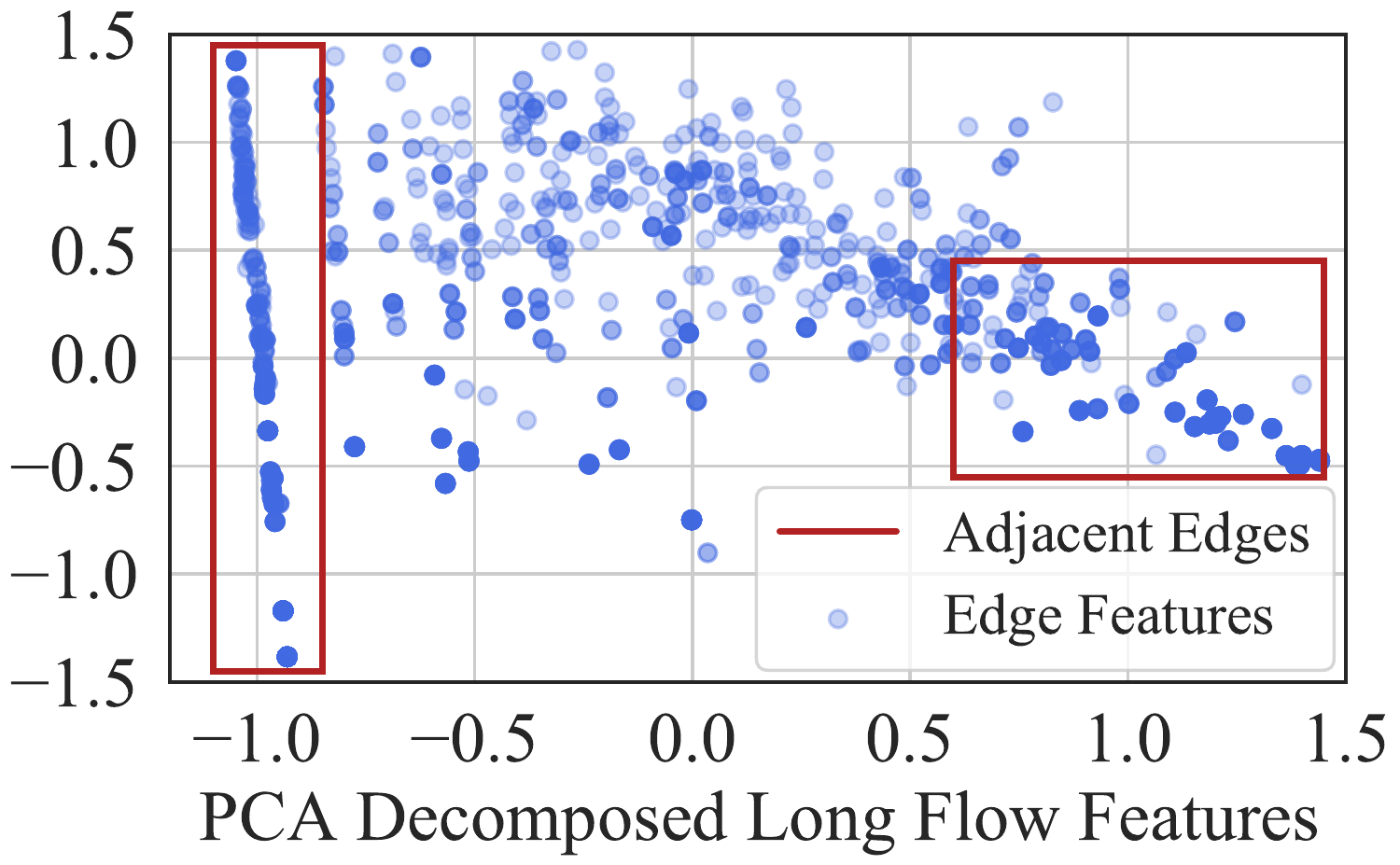}
	}
	\hspace{-2mm}
	\subfigure[Adjacent short flows.]{
        \label{graph:features:ephemeral}
		\includegraphics[width=0.22\textwidth]{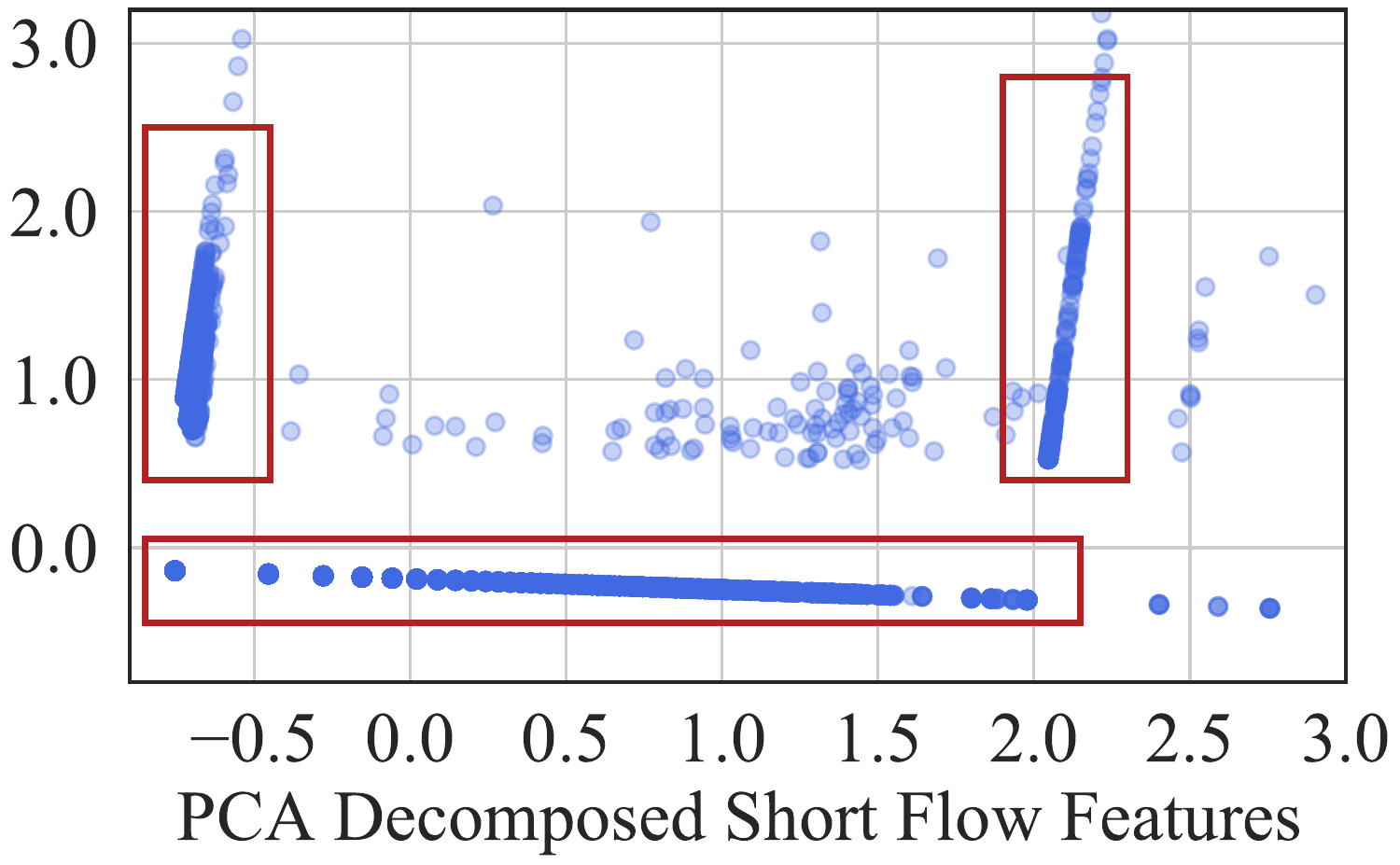}
	}
	\vspace{-2mm}
    \caption{The sparsity of edges in the graph feature space.} 
    \label{graph:features}
    \end{center}
    \vspace{-2mm}
\end{figure}

Now we further need to process and pre-cluster the graph for efficient detection. As shown in Figure~\ref{graph:component}, the abnormal components in the graph have massive vertices and edges. In particular, we cannot directly apply graph representation learning, e.g., graph neural network (GNN), for realtime detection. Figure~\ref{graph:features} shows the edges from the components in the graph structural feature space. We observe that the distribution of the edges is sparse, i.e., most edges are adjacent to massive similar edges in the feature space. To utilize the sparsity, we perform a pre-clustering using DBSCAN~\cite{DBSCAN} that leverages KD-Tree for efficient local search and select the cluster centers of the identified clusters to represent all edges in each cluster to reduce the overhead for graph processing.

\begin{figure}[t]
    \vspace{-2mm}
    \begin{center}
    \includegraphics[width=0.47\textwidth]{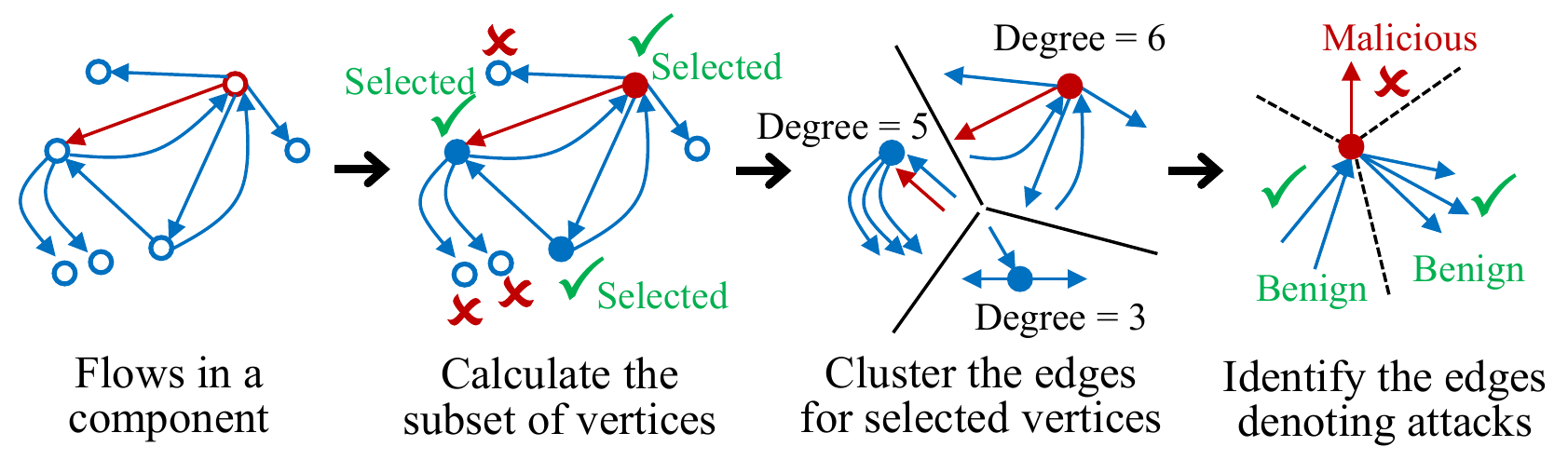}
    \end{center}
    \vspace{-5mm}
    \caption{Critical vertices identification via solving the vertex cover problem.} 
    \vspace{-4mm}
    \label{graph:vertex-cover}
\end{figure}

Specifically, we extract eight and four graph structural features (see Table~\ref{table:feature} in Appendix~\ref{section:appendix:tables}) for the edges associated with short and long flow, respectively, e.g., the in-degree of the source vertex of an edge associated with a long flow. These degree features of malicious traffic are significantly distinct from the benign ones, e.g., the vertices denoting spam bots have higher out-degrees than benign clients due to their frequent interactions with servers. Then, we perform a min-max normalization for the features, and adopt a small search range $\epsilon$ and a large minimum number of points for DBSCAN clustering (see Section~\ref{section:evaluation:setup} for the setting of hyper-parameters) to avoid including irrelevant edges in the clusters, which may incur false positives. Moreover, some edges cannot be clustered and should be treated as outliers, which will be processed as clusters with only one edge.

\section{Malicious Traffic Detection} \label{section:detection}
In this section, we detect encrypted malicious traffic by identifying abnormal interaction patterns on the graph. In particular, we cluster edges connected to the same critical vertex and detects outliers as malicious traffic (see Figure~\ref{graph:vertex-cover}). 

\subsection{Identifying Critical Vertices} \label{section:detection:vertex}
To efficiently learn the interaction patterns of the traffic, we do not perform clustering for all edges directly but cluster edges connected to critical vertices. For each connected component, we select a subset of all vertices in the connected component as the critical vertices according to the following conditions: (i) the source and/or destination vertices of each edge in the component are in the subset, which ensures that all the edges are connected to more than one critical vertices and clustered at least once; and (ii) the number of selected vertices in the subset is minimized, which aims to minimize the number of clustering to reduce the overhead of graph learning. Finding such a subset of vertices is an optimization problem and equivalent to the \textit{vertex cover problem}~\cite{STOC74-npc}, which was proved to be NP Complete (NPC). We select all edges and all vertices on each component to solve the problem. And we reformulate the problem to a Satisfiability Modulo Theories (SMT) problem that can be effectively solved by using Z3 SMT solver~\cite{Z3}. Since we pre-cluster the massive edges and reduce the scale of the problem (see Section~\ref{section:splitting:preclustering}), the NPC problem can be solved in real time.

\subsection{Edge Feature Clustering for Detection} \label{section:detection:clustering}
Now we cluster the edges connected to each critical vertex to identify abnormal interaction patterns. In this step, we use the structural features in Section~\ref{section:splitting:preclustering}, and the flow features extracted from the per-packet feature sequences of short flows or the fitted feature distributions of long flows. All features are shown in Table~\ref{table:feature} (see Appendix~\ref{section:appendix:tables}). We use the lightweight K-Means algorithm to cluster the edges associated with short and long flows, respectively, and calculate the clustering loss that indicates the degree of maliciousness for malicious flow detection. 

\begin{small}
\vspace{-4mm}
\begin{align}
        \mathsf{loss}_{\mathrm{center}}(\mathsf{edge}) &= \min_{C_i\in \lbrace C_1,...,C_K\rbrace} ||C_i - f(\mathsf{edge}) ||_2, \label{equaltion:loss:center} \\
       \mathsf{loss}_{\mathrm{cluster}}(\mathsf{edge}) &= \mathsf{TimeRange(\mathcal{C}(\mathsf{edge}))}, \label{equaltion:loss:cluster} \\
       \mathsf{loss}_{\mathrm{count}}(\mathsf{edge}) &= \log_2(\mathsf{Size(\mathcal{C}(\mathsf{edge}))} + 1), \label{equaltion:loss:aggregation}
\end{align}
\begin{equation} \label{equaltion:loss:all}
    \begin{aligned}
        \mathsf{loss}(\mathsf{edge}) =& \alpha\mathsf{loss}_{\mathrm{center}}(\mathsf{edge}) \\ 
        -& \beta\mathsf{loss}_{\mathrm{cluster}}(\mathsf{edge}) + \gamma\mathsf{loss}_{\mathrm{count}}(\mathsf{edge}),
    \end{aligned}
\end{equation}
\vspace{-2mm}
\end{small}

\noindent where $K$ is the number of obtained cluster centers, $C_i$ is the $i^\mathrm{th}$ center, $f(\mathsf{edge})$ is the feature vector, $\mathcal{C}(\mathsf{edge})$ contains all edges in the cluster of $\mathsf{edge}$ produced by pre-clustering, and $\mathsf{TimeRange}$ calculates the time range covered by the flows denoted by the edges. 

According to Equation~\eqref{equaltion:loss:all}, the loss has three parts: (i) $\mathsf{loss}_{\mathrm{center}}$ in \eqref{equaltion:loss:center} is the Euclidean distance to the cluster centers which indicates the difference from other edges connected to the critical vertex; (ii) $\mathsf{loss}_{\mathrm{cluster}}$ in \eqref{equaltion:loss:cluster} indicates the time range covered by the cluster identified by the pre-clustering in Section~\ref{section:splitting:preclustering} which implies long lasting interaction patterns tend to be benign; (iii) $\mathsf{loss}_{\mathrm{count}}$ in \eqref{equaltion:loss:aggregation} is the number of flows denoted by the edges, which means a burst of massive flows implies malicious behaviors. Moreover, we used weights: $\alpha, \beta, \gamma$ to balance the loss terms. Finally, it detects the associated flows as malicious when the loss function of the edge is larger than a threshold.

\section{Theoretical Analysis}\label{section:analysis}
In this section, we develop a theoretical analysis framework, i.e., \textit{flow recording entropy model}, to analyze the information preserved in the graph of \name for graph learning based detection. The detailed analysis can be found in Appendix~\ref{section:appendix:theory}.

\subsection{Information Entropy Based Analysis}
We develop the framework that aims to quantitatively evaluate the information retained by the exiting traffic recording modes, which decide the data representations for malicious traffic detection, by using three metrics: (i) the amount of information, i.e., the average Shannon entropy obtained by recording one packet; (ii) the scale of data, i.e., the space used to store the information; (iii) the density of information, i.e., the amount of information on a unit of storage. By using this framework, we model the graph based traffic recording mode used by \name as well as three typical types of flow recording modes, i.e., (i) idealized mode that records and stores the whole per-packet feature sequence; (ii) event based mode (e.g., Zeek) that records specific events~\cite{CCS17-Deeplog,AISEC16-Encrypted}; and (iii) sampling based mode (e.g., NetFlow) that records coarse-grained flow information~\cite{ACSAC12-Disclose,USEC21-Jaqen}.

We model a flow, i.e., a sequence of per-packet features, as a sequence of random variables represented by an aperiodic irreducible discrete-time Markov chain (DTMC). Let $\mathcal{G}=\lbrace \mathcal{V}, \mathcal{E}\rbrace$ denote the state diagram of the DTMC, where $\mathcal{V}$ is the set of states (i.e., the values of the variables) and $\mathcal{E}$ denotes the edges. We define $s=|\mathcal{V}|$ as the number of different states and use $\mathcal{W}=[w_{ij}]_{s \times s}$ to denote the weight matrix of $\mathcal{G}$. All of the weights are equal and normalized:

\begin{small}
\vspace{-2mm}
\begin{equation}
    \begin{aligned}
        \forall \hspace{2mm} 1 \leq i,j,m,n \leq s, (w_{ij} = &w_{mn}) \lor  ( w_{ij} = 0 \lor w_{mn} = 0), \\
        w_i = \sum_{j=1}^{s} w_{ij},&\quad 1 = \sum_{i=1}^{s} w_{i}.
    \end{aligned}
\end{equation}
\end{small}

The state transition is performed based on the weights, i.e., the transition probability matrix $P=[P_{ij}]$, $P_{ij}=w_{ij}/{w_i}$. Therefore, the DTMC has a stationary distribution $\mu$:

\begin{small}
\vspace{-4mm}
\begin{equation}
    \begin{cases}
        \mu P = \mu,\\
        1 = \sum_{j=1}^{s} \mu_j
    \end{cases}
    \Rightarrow \quad \mu_j = w_j, \quad \forall \hspace{2mm} 1 \leq j \leq s.
\end{equation}
\end{small}

Assume that the stationary distribution is a binomial distribution with the parameter: $0.1 \le p \le 0.9$ to approach Gaussian distribution with low skewness:

\begin{small}
\vspace{-3mm}
\begin{equation}
    \mu \sim B(s, p) \stackrel{App.}{\longrightarrow} \mathcal{N}(sp, sp(1-p)).
\end{equation}
\vspace{-4mm}
\end{small}

Based on the distribution, we obtain the entropy rate of the DTMC which is the expected Shannon entropy increase for each step in the state transition, i.e., the expected Shannon entropy of each random variable in the sequence, (using \textit{nat} as unit, 1~\textit{nat} $\approx$ 1.44~\textit{bit}):

\begin{small}
\vspace{-5mm}
\begin{equation}
    \begin{aligned}
        \mathcal{H}[\mathcal{G}]=\sum_{i=1}^{s}\mu_i\sum_{j=1}^{s} p_{ij}& \ln \frac{1}{p_{ij}} = -\sum_{i=1}^{s}\sum_{j=1}^{s}w_{ij}\ln w_{ij} + \sum_{j=1}^{s}w_j\ln w_j \\
        &= \ln |\mathcal{E}| - \frac{1}{2} \ln 2\pi sep(1-p).
    \end{aligned}
\end{equation}
\end{small}

Moreover, for the real-world flow size distribution, we assume that the length of the sequence of random variables obeys a geometric distribution with high skewness, i.e., $L\sim G(q)$ with a parameter: $0.5 \le q \le 0.9$. $\mathcal{H}$, $\mathcal{L}$, and $\mathcal{D}$ denote the expectation of the metrics, i.e., the amount of information, the scale of data, and the density, respectively.

\begin{figure*}[t]
    \subfigcapskip=-1mm
    \vspace{-4mm}
    \begin{center}
	\subfigure[The entropy of the modes.]{
        \label{graph:3d:entropy}
		\includegraphics[width=0.23\textwidth]{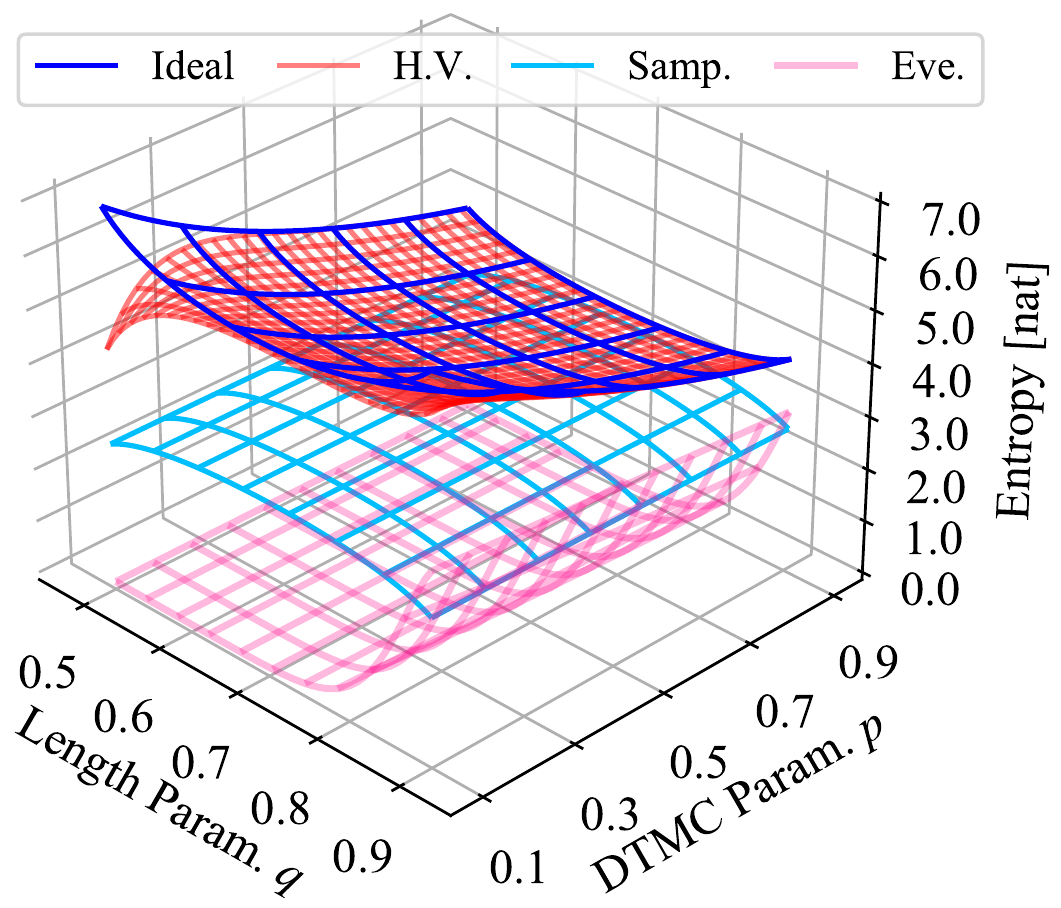}
	}
	\hspace{-2mm}
	\subfigure [The data scale of the modes.]{
        \label{graph:3d:length}
        \includegraphics[width=0.23\textwidth]{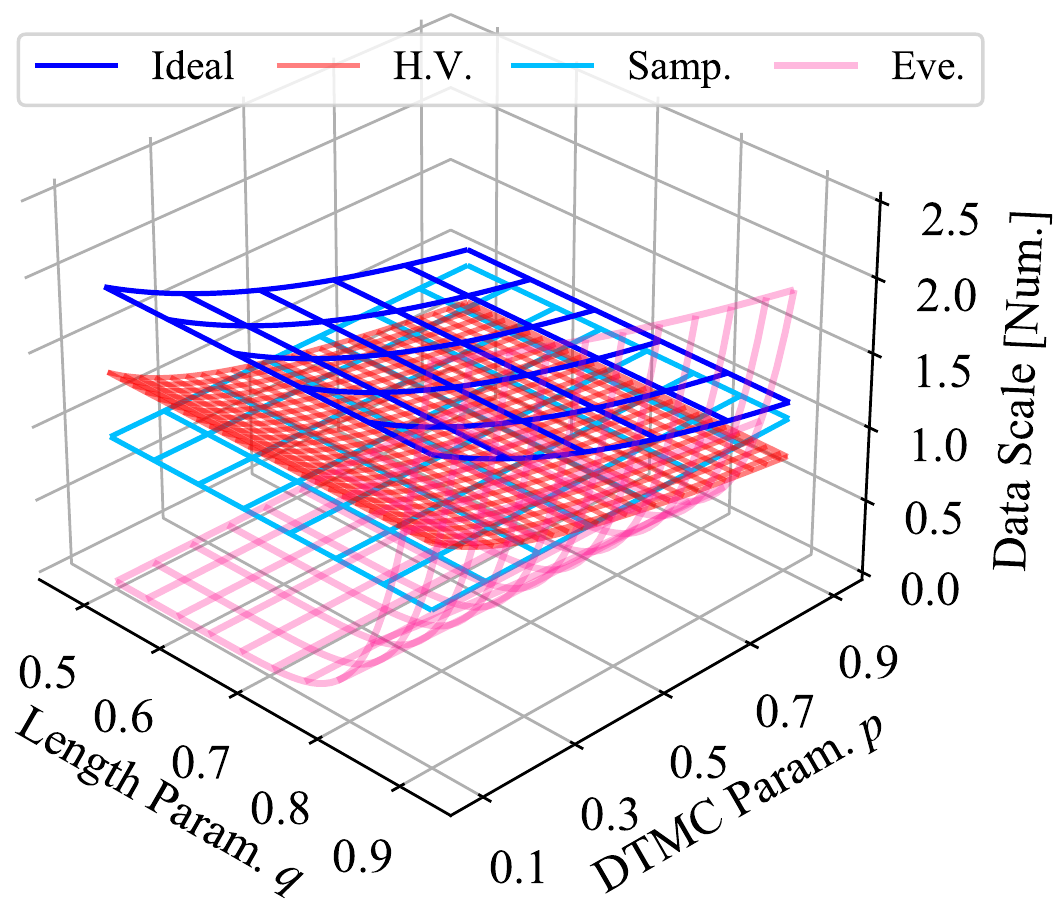}
	}
    \hspace{-2mm}
    \subfigure[The density of the modes.]{
        \label{graph:3d:density}
		\includegraphics[width=0.23\textwidth]{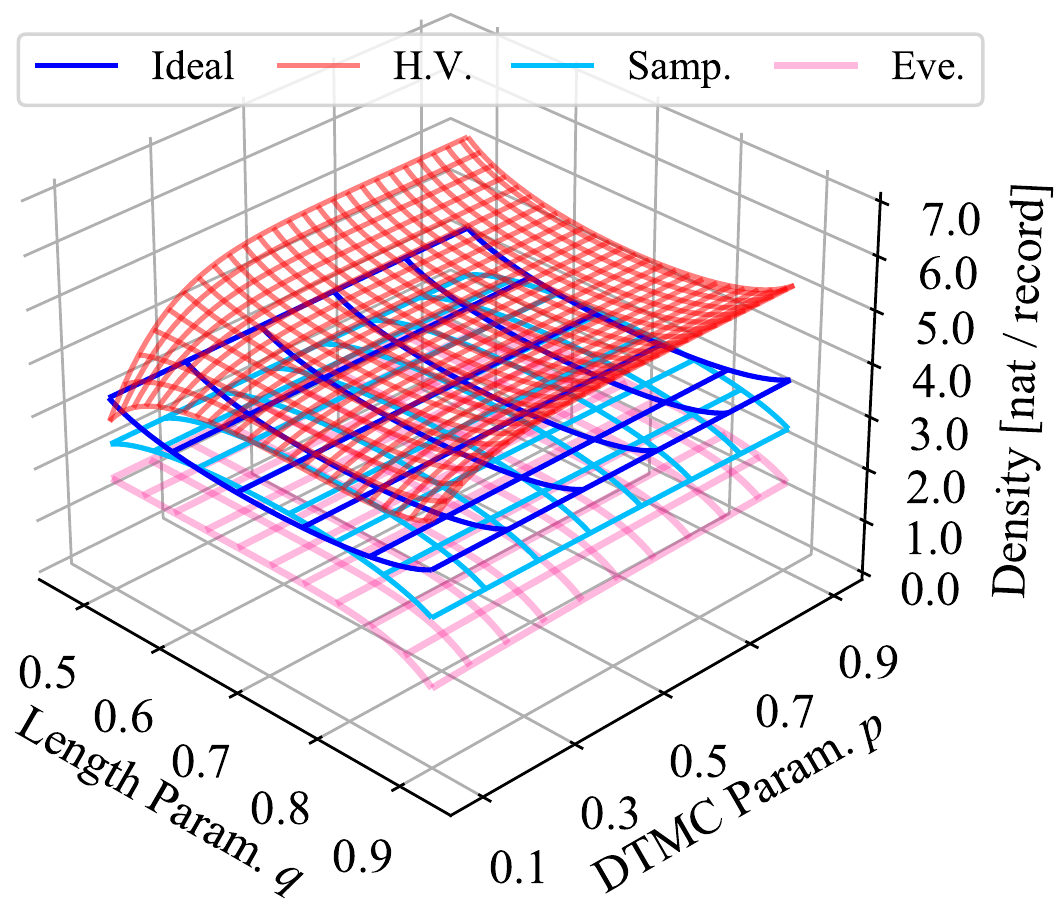}
	}
    \hspace{-2mm}
	\subfigure [The density improvement.]{
        \label{graph:3d:difference}
		\includegraphics[width=0.23\textwidth]{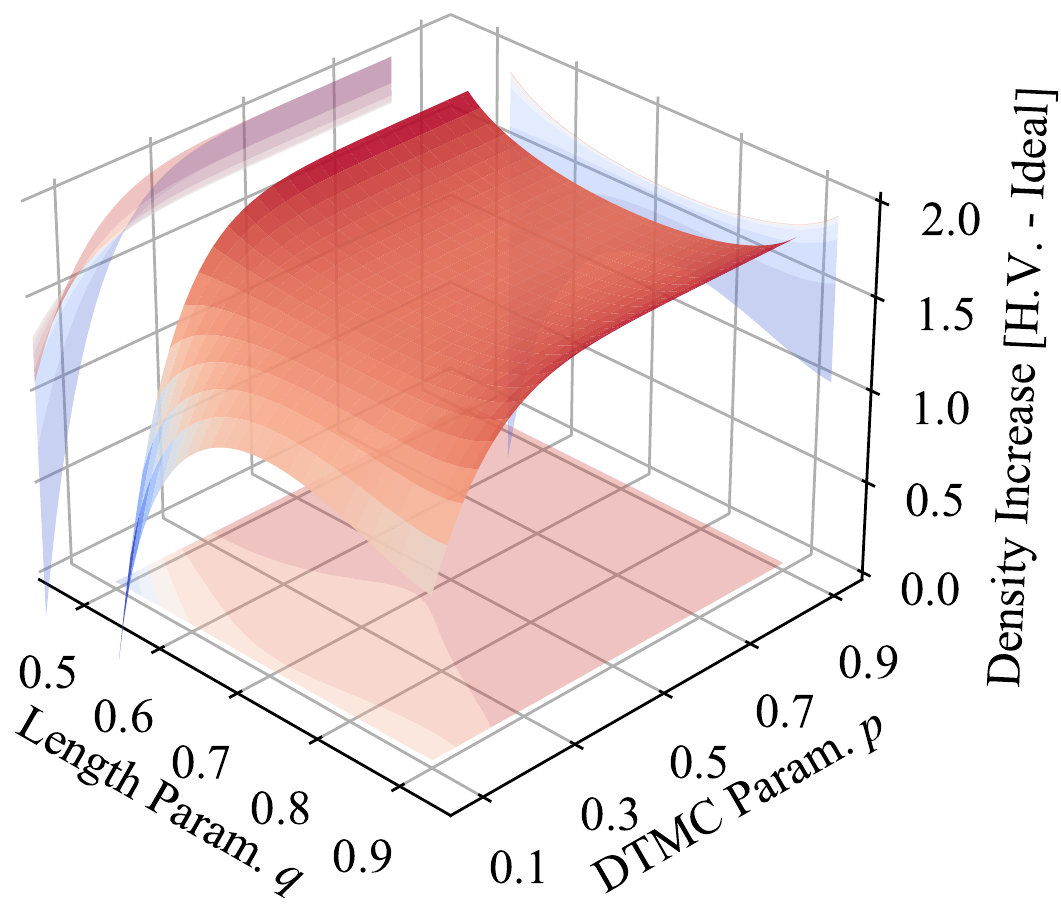}
	}
	\vspace{-2mm}
    \caption{The traffic information retained by different recording modes on the feasible region of the parameters.}
    \label{graph:3d}
    \end{center}
    \vspace{-5mm}
\end{figure*}

\noindent \textbf{Idealized Recording Mode.} The idealized recording mode has infinite storage and captures optimal fidelity traffic information by recording each random variable from the sequence without any processing. Thus, the obtained information entropy of the idealized mode grows at the entropy rate of the DTMC: 

\begin{small}
\vspace{-4mm}
\begin{equation}
    \mathcal{H}_{\mathrm{Ideal}}=\mathrm{E}[L\mathcal{H}[G]]=\frac{1}{q}\ln|\mathcal{E}| - \frac{1}{2q}\ln2\pi sep(1-p).
\end{equation}
\vspace{-3mm}
\end{small}

According to data processing inequality~\cite{TIT98-DPI}, the information retained in the idealized recording mode reaches the optimal value. It implies that processing of the observed per-packet features denoted by the random variables may incur information loss. In the following sections, we will show that the other mode incurs information loss.

We can obtain the scale of data and the density of information for the idealized recording mode as follows:

\begin{small}
\vspace{-3mm}
\begin{equation}
    \mathcal{L}_{\mathrm{Ideal}} = \mathrm{E}[L] = \frac{1}{q}.
\end{equation}
\begin{equation}
    \mathcal{D}_{\mathrm{Ideal}} = \frac{\mathcal{H}_{\mathrm{Ideal}}}{\mathcal{L}_{\mathrm{Ideal}}} = \mathcal{H}[G].  
\end{equation}
\end{small}

\noindent \textbf{Graph Based Recording Mode of \name.}
\name applies different strategies to process short and long flows for the graph construction. Let $K$ denote the threshold for classifying the flows. When $L < K$, it collects all random variables from the sequence for short flows. Otherwise, it collects the histogram to fit the distribution for long flows. Then, we can obtain the lower bound to estimate the information entropy in the graph of \name:

\begin{small}
\vspace{-3mm}
\begin{equation}
    \begin{aligned}
        \mathcal{H}_{\mathrm{\Name}}& = \frac{1-(Kq+1)(1-q)^K}{q}\mathcal{H}[G] + \frac{1}{4}s(1-q)^K\\ [(1+s)\ln & ps + 2\ln2\pi e + 2q\ln K - 2s(1 + p + \gamma)].
    \end{aligned}
\end{equation}
\end{small}

We can also obtain the expected data scale and the density:

\begin{small}
\vspace{-3mm}
    \begin{equation}
    \begin{aligned}
        \mathcal{L}_{\mathrm{\Name}} = s(1-q)^K &+ \frac{1-(Kq+1)(1-q)^K}{Cq},
         \end{aligned}
    \end{equation}
\vspace{-6mm}
\end{small}

\noindent where $C$ is the average number of flows denoted by an edge associated with short flows.

\begin{small}
\vspace{-2mm}
\begin{equation}
    \begin{aligned}
        \mathcal{D}_{\mathrm{\Name}} &= \frac{\mathcal{H}_{\mathrm{\Name}}}{\mathcal{L}_{\mathrm{\Name}}}.
    \end{aligned}
    \end{equation}
\vspace{-3mm}
\end{small}

\noindent \textbf{Sampling Based Recording Mode.}
Similarly, the sampling based mode extracts and records flow statistics for the detection. We analyze the accumulative statistics (e.g. the total number of bytes) that are widely adopted~\cite{Netflow,IPFIX}. Let $\langle s_1, s_2, ..., s_L\rangle$ denote the sequence of random variables, and $X_{\mathrm{Samp.}}=\sum_{i=1}^{L}s_i$ indicates the flow statistic to be recorded. We can obtain a tight lower bound as an estimation for the amount of information and the other metrics as follows:

\begin{small}
\vspace{-4mm}
\begin{equation}
    \mathcal{H}_{\mathrm{Samp.}} = \mathcal{H}[X_{\mathrm{Samp.}}]= \frac{1}{2}\ln 2\pi esp(1-p) + \frac{\ln 2}{2} q (1 - q).
\end{equation}
\begin{equation}    
    \mathcal{L}_{\mathrm{Samp.}} = 1.
\end{equation}   
\begin{equation}
    \mathcal{D}_{\mathrm{Samp.}} = \mathcal{H}_{\mathrm{Samp.}}
\end{equation}
\end{small}

\begin{figure}[t]
    \subfigcapskip=-1mm
    \vspace{-2mm}
    \begin{center}
	\subfigure[Fix $q$ and leave $p$ as variable.]{
        \label{graph:approach:paramp}
		\includegraphics[width=0.22\textwidth]{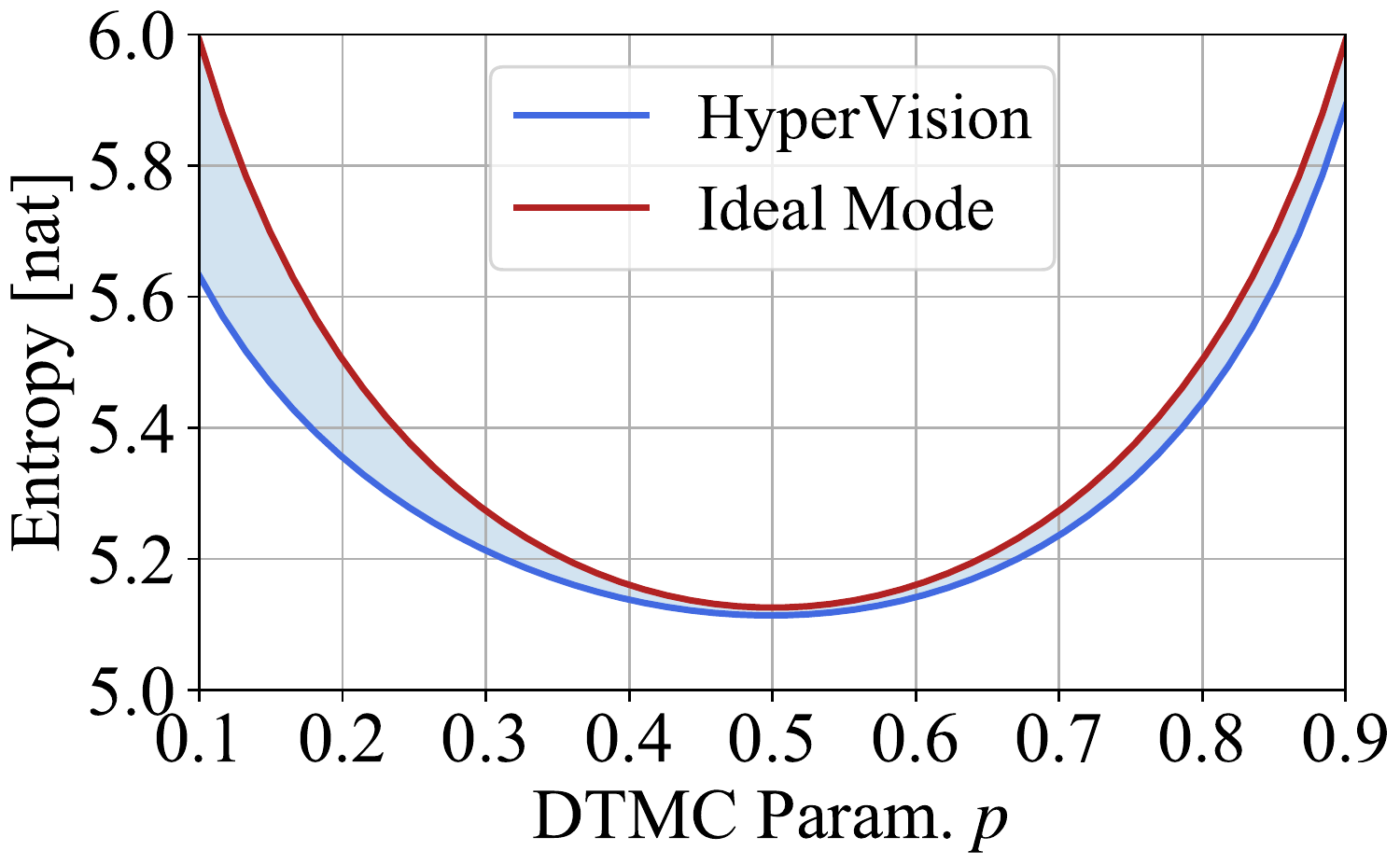}
	}
	\hspace{-2mm}
	\subfigure [Fix $p$ and leave $q$ as variable.]{
        \label{graph:approach:paramq}
		\includegraphics[width=0.22\textwidth]{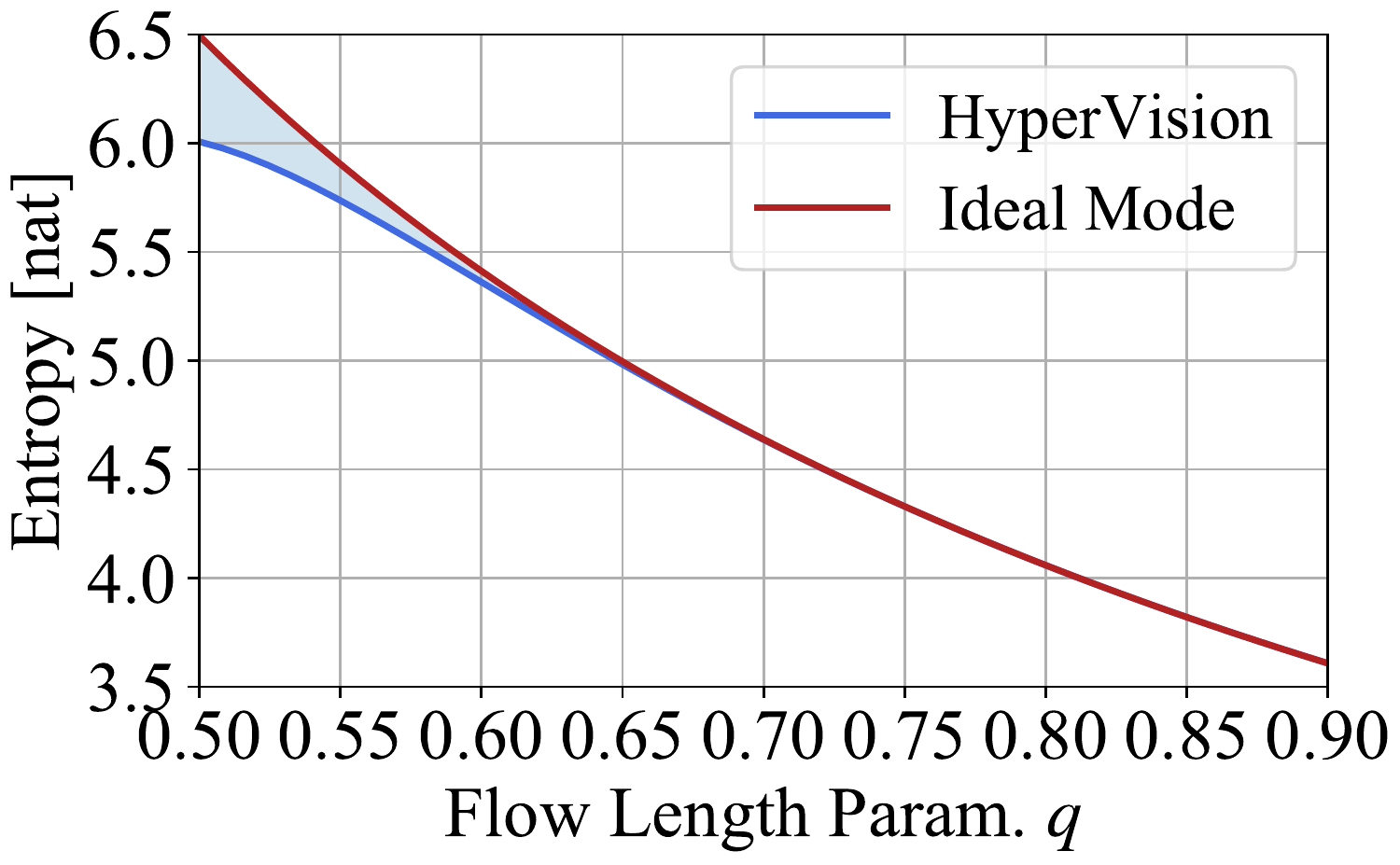}
	}
	\vspace{-2mm}
    \caption{\name approaches the idealized flow recording mode on information entropy.} 
    \label{graph:approach}
    \end{center}
    \vspace{-3mm}
\end{figure}

\noindent \textbf{Event Based Recording Mode.}
The event based recording mode inspects each random variable in the sequence and records events with a small probability. Since the observation that the event based methods do not generate repetitive events for a long flow with a larger $s$, for simplicity, we assume that the probability is $p^s \varpropto 1/s$. Then, we can obtain the concise closed-form solution of the amount of information, the scale of data, and the density of information for event based recording mode as follows:

\begin{small}
\vspace{-2mm}
    \begin{equation}
        \mathcal{H}_{\mathrm{Eve.}} = -2\theta \ln \theta, 
    \end{equation} 
\end{small}

\noindent where $\theta=\frac{\zeta}{\eta}$, $\zeta=q-qp^s$, and $\eta=q-p^s(q-1)$.
\begin{small}
    \vspace{-2mm}
    \begin{equation}
        \mathcal{L}_{\mathrm{Eve.}} = -\frac{p^s}{\eta}.
    \end{equation}
    \vspace{-2mm}
    \begin{equation} 
        \mathcal{D}_{\mathrm{Eve.}} = \frac{2\zeta}{p^s}\ln \theta.
    \end{equation}
\end{small}

\setlength\tabcolsep{2.8pt}
\def\d{\mathrm{d}}
\newcommand{\upp}[1]{\color[rgb]{0.117, 0.447, 0.999}#1}
\newcommand{\dow}[1]{\color[rgb]{0.753,0,0}#1}
\newcommand{\inc}[1]{\upp\mathbf{\blacktriangle #1\%}}
\newcommand{\dec}[1]{\dow\mathbf{\blacktriangledown #1\%}}

\subsection{Analysis Results}
We perform numerical studies to compare the flow recording modes in real-world setting. We select three per-packet features: protocol, length, and the arrival interval (in ms) as the instances of the DTMC, then we measure the parameters of the DTMC, i.e., $|\mathcal{E}|$ and $|\mathcal{V}|$ according to the first $10^{6}$ packets in the MAWI dataset on Jan. 2020~\cite{WIDE}. We also measure $K$, $C$, and estimate the geometric distribution parameter $q$ via the second moment. We have the following three key results.

\noindent\textit{(1) \name maintains more information using the graph than the existing methods.} Figure~\ref{graph:3d} shows the results on the feasible region ($\mathcal{F} = \lbrace 0.1 \le p \le 0.9$, $0.5 \le q \le 0.9 \rbrace$). We observe that \name maintains at least 2.37 and 1.34 times information entropy than traditional flow sampling and event based flow recording. Thus, the traditional detection methods cannot retain high-fidelity flow interaction information. Actually, they only analyze the features of a single flow, which can be evaded by encrypted traffic. According to Figure~\ref{graph:3d:length}, \name has 69.69\% data scale of the sampling based mode. It implies that the data scale is the key challenge for the existing methods to utilize flow interaction patterns. We well address this issue by using the compact graph for maintaining the interactions among flows. 

\noindent\textit{(2) \name maintains near-optimal information using the graph.} According to Figure~\ref{graph:3d:entropy}, we observe that the information maintained by the graph almost equals to the theoretical optimum, with the difference ranging from $4.6 \times 10^{-9}$ to $2.6$ \textit{nat}. When the parameter of the geometric distribution of $L$ approaches 0.9, the flow information loss is larger because of the increasing ratio of long flows that incur more information loss. Figure~\ref{graph:approach} compares the information in \name and the idealized system when $q=0.59$ and $p=0.8$. We have similar results. The gaps between the graph mode and the optimal mode are only 0.056 and 0.021.

\renewcommand{\arraystretch}{1.2}
\begin{table}[t]
    \footnotesize
    \caption{The integral of the density in the feasible region.}
    \vspace{-4mm}
    \begin{center}
    \begin{threeparttable}
        \begin{tabular}{c|ccc}
        \toprule
        \textbf{\tabincell{c}{Per-Packet Features}} & \textbf{\tabincell{c}{Packet Length}} & \textbf{\tabincell{c}{Time Interval}} & \textbf{\tabincell{c}{Protocol Type}} \\
        \midrule
        $\iint_{\mathcal{F}}\mathcal{D}_{\mathrm{Ideal}}(p, q) \d p \d q$ & $1.011_{\dec{32.10}}$ & $0.918_{\dec{32.00}}$ & $0.795_{\dec{32.51}}$ \\
        $\iint_{\mathcal{F}}\mathcal{D}_{\mathrm{Samp.}}(p, q) \d p \d q$ & $0.965_{\dec{35.17}}$ & $0.963_{\dec{28.66}}$ & $0.800_{\dec{32.08}}$ \\
        $\iint_{\mathcal{F}}\mathcal{D}_{\mathrm{Eve.}}(p, q) \d p \d q$ & $0.588_{\dec{60.51}}$ & $0.588_{\dec{56.44}}$ & $0.588_{\dec{50.08}}$ \\
        \midrule
        $\iint_{\mathcal{F}}\mathcal{D}_{\mathrm{\Name}}(p, q) \d p \d q$ & $1.489_{\inc{47.27}}$ & $1.350_{\inc{35.51}}$ & $1.178_{\inc{48.18}}$\\
        \bottomrule
        \end{tabular}
    \end{threeparttable}
    \label{table:integral}
    \end{center}
    \vspace{-5mm}
\end{table}

\noindent\textit{(3) \name has higher information density than the existing methods.}  Figure~\ref{graph:3d:density} shows that \name realizes 1.46, 1.54, and 2.39 times information density than the existing methods, respectively. Although the idealized system realizes the optimal amount of traffic information, the density is only 78.55\% of \name in the worst case, as shown in Figure~\ref{graph:3d:difference}. From Table~\ref{table:integral}, we find that, for all kinds of per-packet features, \name can increase the density ranging between 35.51\% and 47.27\% due to the different recording strategies for short and long flows. 

In summary, the flow interaction graph provides high-fidelity and low-redundancy traffic information with obvious flow interaction patterns, which ensures that \name achieves realtime and unsupervised detection, particularly, detecting encrypted malicious traffic with unknown patterns.

\section{Experimental Evaluation} \label{section:evaluation}
\subsection{Experiment Setup} \label{section:evaluation:setup}
\noindent\textbf{Implementation.} We prototype \name with more than 8,000 Line of Code (LOC). The prototype is compiled by gcc 9.3.0 and cmake 3.16.3. We use DPDK~\cite{DPDK} version 19.11.9 encapsulated by libpcap++~\cite{pcapplusplus} version 21.05 to implement the high-speed data-plane module. The graph construction module maintains the graph in memory for realtime detection. The graph learning module detects the encrypted malicious traffic on the interaction graph. It uses DBSCAN and K-Means in mlpack~\cite{mlpack} (version 3.4.2) for clustering and Z3 SMT Solver~\cite{Z3} (version 4.8) to identify the critical vertices.

\noindent\textbf{Testbed.} We deploy \name on a testbed built upon DELL servers (PowerEdge R410, produced in 2012) with two Intel Xeon E5645 CPUs (2 $\times$ 12 cores), Ubuntu 20.04.2 (Linux 5.11.0), Docker 20.10.7, 24GB memory, one Intel 82599ES 10 Gb/s NIC, and two Intel 850nm SFP+ laser ports for optical fiber connections. We configure 6GB huge page memory for DPDK (3GB/NUMA Node) and bind 8 threads on 8 physical cores for 16 NIC RX queues to parse the per-packet features from high-speed traffic. We use 8 cores for in-memory graph construction, and 7 cores are used for graph learning, the rest one core is used as DPDK master core.

\noindent\textbf{Datasets.} We use real-world backbone network traffic datasets from the vantage-G of WIDE MAWI project~\cite{WIDE} in AS2500, Tokyo Japan, Jan. $\sim$ Jun. 2020 as background traffic. The vantage transits traffic from/to its BGP peers and providers using 10 Gb/s fiber linked to its IXP (DIX-IE), and the traffic is collected using port mirroring, which is consistent with our threat model and the physical testbed described above. We remove the attack traffic with obvious patterns in the background traffic dataset according to the rules defined by the existing studies~\cite{IMC19-ScanScan,IMC19-ScanDoS,USEC14-ScanScan}, e.g., traffic will be detected as scanning traffic if it has scanned over 10\% IPv4 addresses~\cite{USEC14-ScanScan}. We generate the malicious traffic by constructing real attacks or replaying the existing traces in our testbed. Specifically, we collect malicious traffic in our virtual private cloud (VPC) with more than 1,500 instances. We manipulate the instances to perform attacks according to the real-world measurements~\cite{IMC15-Cloud,IMC19-ScanDoS,IMC19-ScanScan,IMC17-DoS,USEC14-ScanScan,DSN17-DynaMiner,Kaspersky} and the same settings in the existing studies~\cite{CCS20-MySC,SIGCOMM03-LRTCPDOS,USEC16-ACKSC,SP13-Crossfire}. We classify 80 new datasets used in our experiments (see Table~\ref{table:datasetdetail} for details) into four groups, three of which are encrypted malicious traffic:
\begin{itemize}[leftmargin=*]
    \vspace{-2mm}
    
    \item \textit{Traditional brute force attack.} Although \name\ focuses on encrypted traffic, we generate 28 kinds of traditional flooding attacks to verify its generic detection and the correctness of baselines including 18 high-rate and 10 low-rate attacks: (i) the brute scanning with the real packet rates~\cite{USEC14-ScanScan}; (ii) the source spoofing DDoS with various rates~\cite{IMC17-DoS}; (iii) the amplification attacks~\cite{IMC19-ScanDoS}; (iv) probing vulnerable applications~\cite{USEC14-ScanScan,USEC13-zmap}. We collected the traffic in our VPC to avoid interference with real services.

    \item \textit{Encrypted flooding traffic.} Different from the brute force flooding, the encrypted flooding is generated by repetitive attack behaviors which target specific applications: (i) the link flooding generates encrypted low-rate flows, e.g., the low-rate TCP attacks~\cite{SIGCOMM03-LRTCPDOS,NDSS05-PDoS} and the Crossfire attack~\cite{SP13-Crossfire}, to congest links; (ii) injecting encrypted flows that exploits protocol vulnerabilities by flooding attack traffic and inject packets into the channel~\cite{CCS20-MySC,USEC16-ACKSC,TON22-IPIDSC}; (iii) the password cracking performs slow attempts to hijack the encrypted communication protocols~\cite{SP19-PasswordCracking,CCS13-weakpwd}. We perform SSH cracking in the VPC with the scale of SSH servers in the ASes reachable to AS2500. 

    \item \textit{Encrypted web malicious traffic.} Web malicious traffic is normally encrypted by HTTPS. We collect the traffic generated by seven widely used web attacks including automatic vulnerabilities discovery (including XSS, CSRF, various injections)~\cite{CCS17-Deemon}, SSL vulnerabilities detection~\cite{USEC19-SCAN}, and crawlers. We also collect the SMTP-over-TLS spam traffic that lures victims to visit the phishing sites~\cite{USEC20-Sunrise}.

    \item \textit{Malware generated encrypted traffic.} The traffic of malware campaigns is low-rate and encrypted, e.g., malware component update or delivery~\cite{USEC11-PPI}, command and control (C\&C) channel~\cite{ACSAC12-Disclose}, and data exfiltration~\cite{IMC17-DataStolen}. We use the malware infection statistics published in 2020~\cite{Kaspersky} and probed active addresses from the adopted vantage~\cite{hurricane,ripe} to estimate the number of visible victims. We use the same number of instances to replay public malware traffic datasets~\cite{CIC-dataset,Stratosphere-dataset} to mimic malware campaigns, which is similar to the existing study~\cite{NDSS14-Cyberprobe}. 

\end{itemize}

\renewcommand{\arraystretch}{1.2}
\begin{table}[t]
    \scriptsize
    \vspace{1.2mm}
    \caption{The average accuracy on the groups of datasets.}
    \vspace{-4mm}
    \setlength\tabcolsep{1.9pt}
    \begin{center}
    \begin{threeparttable}
        \begin{tabular}{@{}c|c|c|c|c|c|c@{}}
        \toprule
        \tabincell{c}{Method} & \tabincell{c}{Metric} & \tabincell{c}{Traditional\\Attacks} & \tabincell{c}{Flooding\\Enc. Traffic} & \tabincell{c}{Enc. Web\\Attacks} & \tabincell{c}{Malware\\Traffic} & Overall \\
        \midrule
        \multirow{2}{*}{Jaqen} & AUC & $0.913_{\dec{7}}$ & $0.782_{\dec{19}}$ & N/A\tnote{1} & N/A & $0.867_{\dec{12}}$ \\
        & F1 & $0.819_{\dec{16}}$ & $0.495_{\dec{46}}$ & N/A & N/A & $0.705_{\dec{26}}$ \\
        \cline{1-2}
        \multirow{2}{*}{FlowLens} & AUC & $0.939_{\dec{4}}$ & $0.757_{\dec{22}}$ & $0.685_{\dec{30}}$ & $0.768_{\dec{22}}$ & $0.752_{\dec{36}}$ \\
        & F1 & $0.799_{\dec{18}}$ & $0.651_{\dec{29}}$ & $0.384_{\dec{59}}$ & $0.411_{\dec{57}}$ & $0.451_{\dec{41}}$ \\
        \cline{1-2}
        \multirow{2}{*}{Whisper} & AUC & $0.951_{\dec{3}}$ & $0.932_{\dec{4}}$ & $0.958_{\dec{2}}$ & $0.648_{\dec{34}}$ & $0.752_{\dec{23}}$ \\
        & F1 & $0.705_{\dec{27}}$ & $0.461_{\dec{50}}$ & $0.546_{\dec{42}}$ & $0.357_{\dec{62}}$ & $0.407_{\dec{57}}$ \\
        \cline{1-2}
        \multirow{2}{*}{Kitsune} & AUC & $0.748_{\dec{24}}$ & -~\tnote{2} & $0.759_{\dec{22}}$ & - & $0.751_{\dec{23}}$ \\
        & F1 & $0.419_{\dec{57}}$ & - & $0.366_{\dec{61}}$ & - & $0.402_{\dec{58}}$ \\
        \cline{1-2}
        \multirow{2}{*}{DeepLog} & AUC & $0.716_{\dec{27}}$ & $0.621_{\dec{26}}$ & $0.767_{\dec{22}}$ & $0.653_{\dec{34}}$ & $0.666_{\dec{32}}$ \\
        & F1 & $0.513_{\dec{47}}$ & $0.508_{\dec{45}}$ & $0.572_{\dec{40}}$ & $0.628_{\dec{34}}$ & $0.597_{\dec{37}}$ \\
        \midrule
        \multirow{2}{*}{\Name} & AUC & $0.988_{\inc{8}}$ & $0.974_{\inc{4}}$ & $0.985_{\inc{2}}$ & $0.993_{\inc{29}}$ & $0.988_{\inc{13}}$ \\
        & F1 & $0.978_{\inc{19}}$ & $0.927_{\inc{42}}$ & $0.957_{\inc{67}}$ & $0.970_{\inc{54}}$ & $0.960_{\inc{36}}$ \\
        \bottomrule
        \end{tabular}
    \begin{tablenotes}
        \scriptsize
        \item[1] The results are N/A  because Jaqen is designed for detection of volumetric attacks. 
        \item[2] - means that the average AUC is lower than 0.60, which is nearly the result of random guessing.
    \end{tablenotes}
    \end{threeparttable}
    \label{table:average}
    \end{center}
    \vspace{-5mm}
\end{table}

\newcommand{\bs}[1]{\color[rgb]{0.117, 0.447, 0.999}#1}
\newcommand{\ws}[1]{\color[rgb]{0.753,0,0}#1}
\newcommand{\Inc}[1]{$\upp{^\blacktriangle}$\upp{#1}\%}
\newcommand{\Dec}[1]{$\dow{^\blacktriangledown}$#1\%}

\renewcommand{\arraystretch}{1.2}
\begin{table*}[!t]
    \footnotesize
    \vspace{-4mm}
    \centering
    \setlength\tabcolsep{1.7pt}
    \caption{Detection accuracy of \name and the baselines on traditional brute force attacks.}
    \vspace{-4.2mm}
    \begin{center}
    \begin{threeparttable}
    \begin{tabular}{@{}c|c|ccccccc|ccccccc|cccc@{}}
    \toprule
        \multirow{2}{*}{\tabincell{c}{Method}} & \multirow{2}{*}{\tabincell{c}{Metric}} & \multicolumn{7}{c|}{\tabincell{c}{Brute Scanning}} & \multicolumn{7}{c|}{\tabincell{c}{Amplification Attack}} & \multicolumn{4}{c}{\tabincell{c}{Source Spoofing DDoS}} \\\cline{3-20}
         &  & ICMP & NTP & SSH & SQL & DNS & HTTP &  HTTPS &
        NTP & DNS & CharG. & SSDP & RIPv1 & Mem. & CLDAP & SYN & RST & UDP & ICMP \\
        \midrule
        \multirow{2}{*}{\tabincell{c}{Jaqen}} & AUC & 0.9478 & 0.9989 & 0.9706 & 0.9851 & 0.9989 & 0.9774 & 0.9988 & 0.9822 & 0.9590 & 0.9860 & 0.9907 & 0.9011 & 0.9586 & 0.9537 & 0.9976 & 0.9985 & 0.9682 & \bs 0.9995 \\
        & F1 & 0.9710 & 0.9356 & 0.9835 & 0.9924 & 0.9965 & 0.9884 & 0.9299 & 0.9457 & 0.8816 & 0.7986 & 0.7054 & 0.6549 & 0.8500 & 0.7931 & 0.9614 & 0.9236 & 0.5603 & \bs 0.9861 \\
        \hline
        \multirow{2}{*}{\tabincell{c}{FlowLens}} & AUC & 0.9906 & 0.9021 & 0.9961 & 0.9993 & 0.9985 & 0.9874 & 0.9226 & 0.9784 & 0.8001 & 0.9998 & 0.9907 & 0.9833 & 0.9786 & 0.9993 & 0.9912 & 0.9918 & \bs 0.9999 & \ws 0.6351 \\
        & F1 & 0.9181 & 0.6528 & 0.8899 & \bs 0.9996 & \bs 0.9992 & \bs 0.9936 & 0.9572 & 0.9794 & 0.7127 & 0.9991 & 0.8918 & \bs 0.9889 & 0.9691 & 0.9986 & 0.8638 & 0.8173 & \bs 0.9990 & \ws 0.2632 \\
        \hline
        \multirow{2}{*}{\tabincell{c}{Whisper}} & AUC & 0.9499 & 0.9796 & 0.9562 & 0.9811 & 0.9832 & 0.9658 & 0.9827 & 0.9125 & 0.9645 & 0.8489 & 0.9662 & 0.9761 & 0.8954 & 0.9402 & 0.9563 & 0.9658 & 0.8956 & 0.9489 \\
        & F1 & 0.7004 & 0.7585 & 0.8869 & 0.7022 & 0.6748 & 0.7182 & 0.7489 & 0.8248 & 0.8435 & 0.4686 & 0.6195 & 0.6396 & 0.6956 & 0.8620 & 0.7587 & 0.8778 & 0.4857 & 0.4192 \\
        \hline
        \multirow{2}{*}{\tabincell{c}{Kitsune}} & AUC & \ws 0.4522 & \ws 0.7252 & \ws{-}~\tnote{2} & 0.7439 & \ws 0.7228 & 0.7380 & 0.9614 & \ws 0.7340 & 0.9994 & \bs 0.9998 & \bs 0.9989 & \ws 0.4343 & \ws 0.3993 & 0.7592 & \ws 0.6210 & 0.4086 & 0.8534 & 0.7913 \\
        & F1 & \ws{-}~\tnote{1} & \ws 0.3459 & \ws - & 0.5033 & 0.4923 & 0.4798 & 0.4878 & \ws 0.4461 & \ws 0.5031 & \ws 0.4609 & \ws 0.4360 & \ws - & \ws - & 0.3838 & \ws 0.3361 & \ws - & 0.4539 & 0.4153 \\
        \hline
        \multirow{2}{*}{\tabincell{c}{DeepLog}} & AUC & 0.6717 & 0.8232 & 0.8377 & \ws 0.6518 & 0.8261 & \ws 0.6617 & \ws 0.5545 & 0.7475 & \ws 0.7428 & \ws 0.7462 & \ws 0.7458 & 0.7487 & 0.7480 & \ws 0.7483 & 0.7564 & \ws 0.2470 & \ws 0.7012 & 0.7521 \\
        & F1 & 0.3566 & 0.4178 & 0.5266 & \ws 0.2695 & \ws 0.4050 & \ws 0.2668 & \ws 0.3653 & 0.5108 & 0.7201 & 0.5705 & 0.4313 & 0.3368 & 0.3321 & \ws 0.3424 & 0.6074 & \ws - & \ws 0.4370 & 0.3428 \\
        \midrule
        \multirow{2}{*}{\tabincell{c}{\Name}} & AUC & \bs 0.9999 & \bs 0.9999  & \bs 0.9999 & \bs 0.9999 & \bs 0.9999 & \bs 0.9999 & \bs 0.9999 & \bs 0.9999 & \bs 0.9999 & 0.9998 & 0.9989 & \bs 0.9998 & \bs 0.9969 & \bs 0.9999 & \bs 0.9999 & \bs 0.9999 & 0.9996 & 0.9928 \\
        & F1 & \bs 0.9939 & \bs 0.9928 & \bs 0.9960 & 0.9932 & 0.9831 & 0.9808 & \bs 0.9892 & \bs 0.9998 & \bs 0.9998 & \bs 0.9992 & \bs 0.9956 & 0.9984 & \bs 0.9983 & \bs 0.9996 & \bs 0.9993 & \bs 0.9571 & 0.9981 & 0.9295 \\
        \bottomrule
    \end{tabular}
    \begin{tablenotes}
        \scriptsize
        \item[1] We highlight the best accuracy in {\bs{$\bullet$}} and the worst accuracy in {\ws{$\bullet$}}. We mark - for the F1 when the AUC is lower than 0.50, which is the accuracy of random guessing.
        \item[2] Kitsune did not finish the detection within 90 min (i.e., meaningless for defenses). And \Name is short for \name.
    \end{tablenotes}
    \end{threeparttable}
    \end{center}
    \label{table:traditional}
    \vspace{-5mm}
\end{table*}

The malicious traffic is replayed with the background traffic datasets on the physical testbed simultaneously according to their original packet rates~\cite{WIDE} which is the same as the existing studies~\cite{USEC21-Jaqen,TON21-Poseidon,CCS21-Whisper}. Specifically, each dataset contains 12$\sim$15 million packets and the replay lasts 45s and the first 75\% time does not contain malicious traffic for collecting flow interactions and training the baselines. Note that, the rates of the encrypted attack flows in our datasets are only 0.01 $\sim$ 8.79 Kpps which consume only 0.01\% $\sim$ 0.72\% bandwidth. We will show that these stealthy attacks evade most baselines.

To eliminate the impact of the dataset bias, we also use 12 existing datasets including the Kitsune datasets~\cite{NDSS18-Kitsune}, the CIC-DDoS2019 datasets~\cite{CIC19}, and the CIC-IDS2017 datasets~\cite{CIC17}, which are collected in the real-world. These detailed results can be found in Appendix~\ref{section:appendix:experiment:dataset}. In particular, the traffic in two CIC datasets~\cite{CIC17, CIC19} lasts 6$\sim$8 hours under multiple attacks, which aims to verify the long-run performances of \name (see Appendix~\ref{section:appendix:experiment:longrun}). Moreover, we validate the robustness of \name against evasion attacks with obfuscation techniques, which can be found in Appendix~\ref{section:appendix:experiment:robust}.

\noindent\textbf{Baselines.} We use five state-of-the-art generic malicious traffic detection methods as baselines:
\begin{itemize}[leftmargin=*]
    \vspace{-2mm}
    \item \textbf{Jaqen} \textit{(sampling based recording and signature based detection)}. Jaqen~\cite{USEC21-Jaqen} uses Sketches to obtain flow statistics and applies the threshold based detection. We prototype Jaqen on the testbed, and adjust the signatures for each statistic and each attack to obtain the best accuracy.
    \item \textbf{FlowLens} \textit{(sampling based recording and ML based detection)}. FlowLens~\cite{NDSS21-Flowlens} uses sampled flow distribution and supervised learning, i.e., random forest. We use the hyper-parameter setting with the best accuracy used in the paper to retrain the ML model.
    \item \textbf{Whisper} \textit{(flow-level features and ML based detection)}. Whisper~\cite{CCS21-Whisper,TON23-Whisper} extracts the frequency domain features of flows and uses clustering to learn the features. We deploy Whisper on the physical testbed without modifications and then retrain the clustering model.
    \item \textbf{Kitsune} \textit{(packet-level features and DL based detection)}. Kitsune extracts per-packet features and uses autoencoders to learn the features which is an unsupervised method~\cite{NDSS18-Kitsune}. We use its default hyper-parameters and retrain the model. 
    \item \textbf{DeepLog} \textit{(event based recording and DL based detection)}. DeepLog is a general log analyzer using LSTN RNN~\cite{CCS17-Deeplog}. We use the logs of connections for detection and its original hyper-parameter setting to achieve the best accuracy.
\end{itemize}

Note that, in the baselines above, we do not include DPI-based encrypted malicious traffic detection because they are unable to investigate encrypted payloads~\cite{USEC08-Botsniffer}. Also, we do not compare the task-specific detection methods~\cite{KDD17-EncryptedTraffic,CONEXT12-BotFinder} because they cannot achieve acceptable detection accuracy. Features in FlowLens, Kitsune, and Whisper are similar to them, e.g., flow features~\cite{KDD17-EncryptedTraffic}, packet header features~\cite{AISEC16-Encrypted}, and time-series~\cite{CONEXT12-BotFinder}.

\noindent\textbf{Metrics.} We mainly use AUC and F1 score because they are most widely used in the literature~\cite{NDSS18-Kitsune,CCS21-Whisper,INFO20-ZeroWall,CCS17-Deeplog,Conext20-Qian,ACSAC12-Disclose,CCS21-nPrint}. Also, we use other six metrics to validate the improvements of \name, including precision, recall, F2, ACC, FPR, and EER. 

\noindent\textbf{Hyper-parameter Selection.} We conduct four-fold cross validation to avoid overfitting and hyper-parameter bias. Specifically, the datasets are equally partitioned into four subsets. Each subset is used once as a validation set to tune the hyper-parameters via the empirical study and the remaining three subsets are used as testing sets. Finally, four results are averaged to produce final results. Moreover, our ablation study shows that the different threshold settings incur at most 5.2\% accuracy loss. Therefore, the hyper-parameter selection has limited impacts on the detection results.

\subsection{Accuracy Evaluation} \label{section:evaluation:accuracy}
Table~\ref{table:average} summarizes the detection accuracy and the improvements of \name over the existing methods. In general, \name achieves average F1 ranging between 0.927 and 0.978 and average AUC ranging between 0.974 and 0.993 on the 80 datasets, which are 35\% and 13\% improvements over the best accuracy of the baselines. In 44 datasets, none of the baselines achieves F1 higher than 0.80, which means that they are not effective to detect the attacks. Due to the page limits, we do not show the failed detection results of these baselines.

\begin{figure}[t]
    \subfigcapskip=-1.5mm
    \begin{center}
	\subfigure[ROC of detecting NTP DDoS.]{
        \label{graph:roc-prc:ntprdos-roc}
		\includegraphics[width=0.22\textwidth]{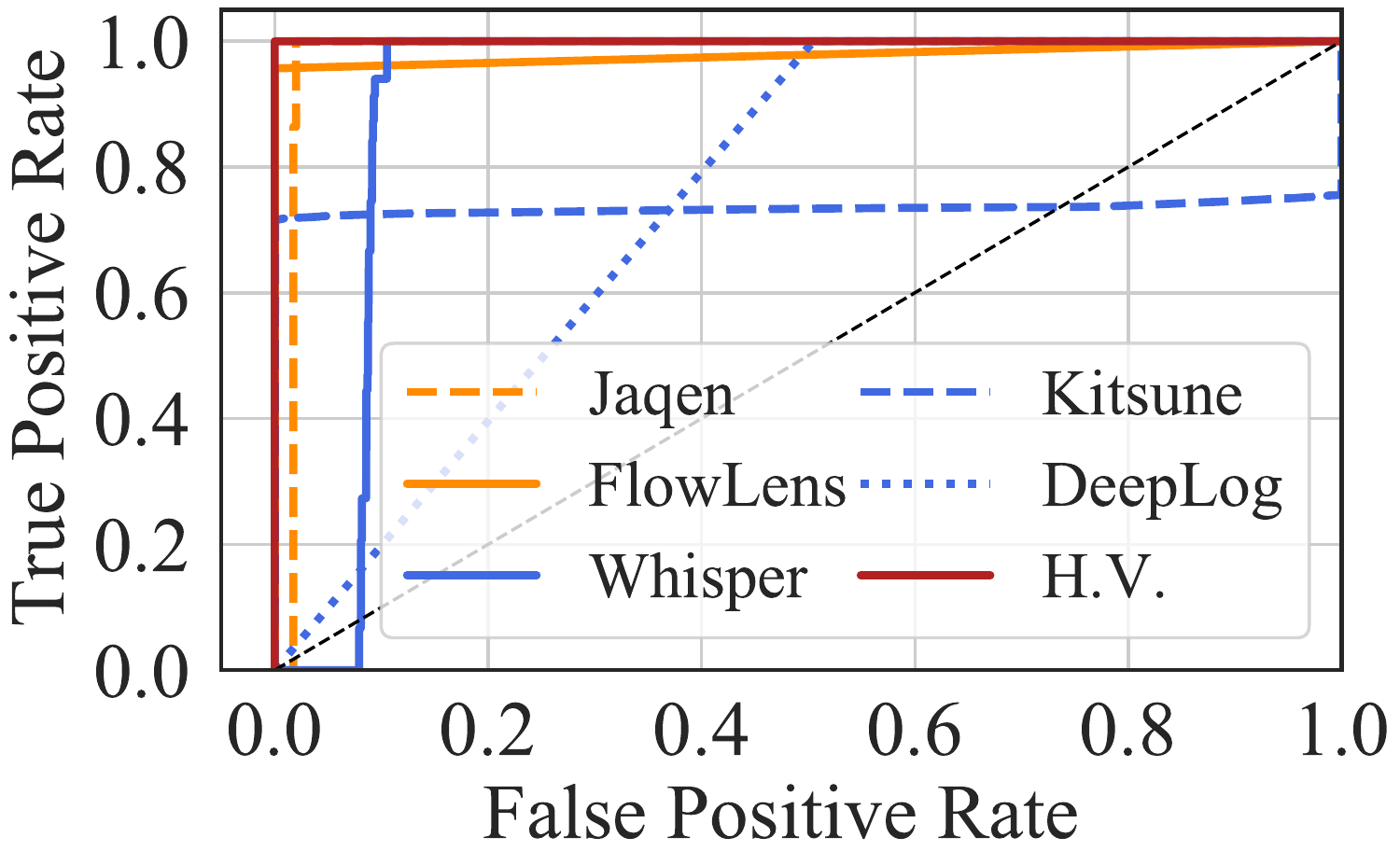}
	}
	\hspace{-2mm}
	\subfigure [ROC of detecting HTTP scan.]{
        \label{graph:roc-prc:httpscan-roc}
		\includegraphics[width=0.22\textwidth]{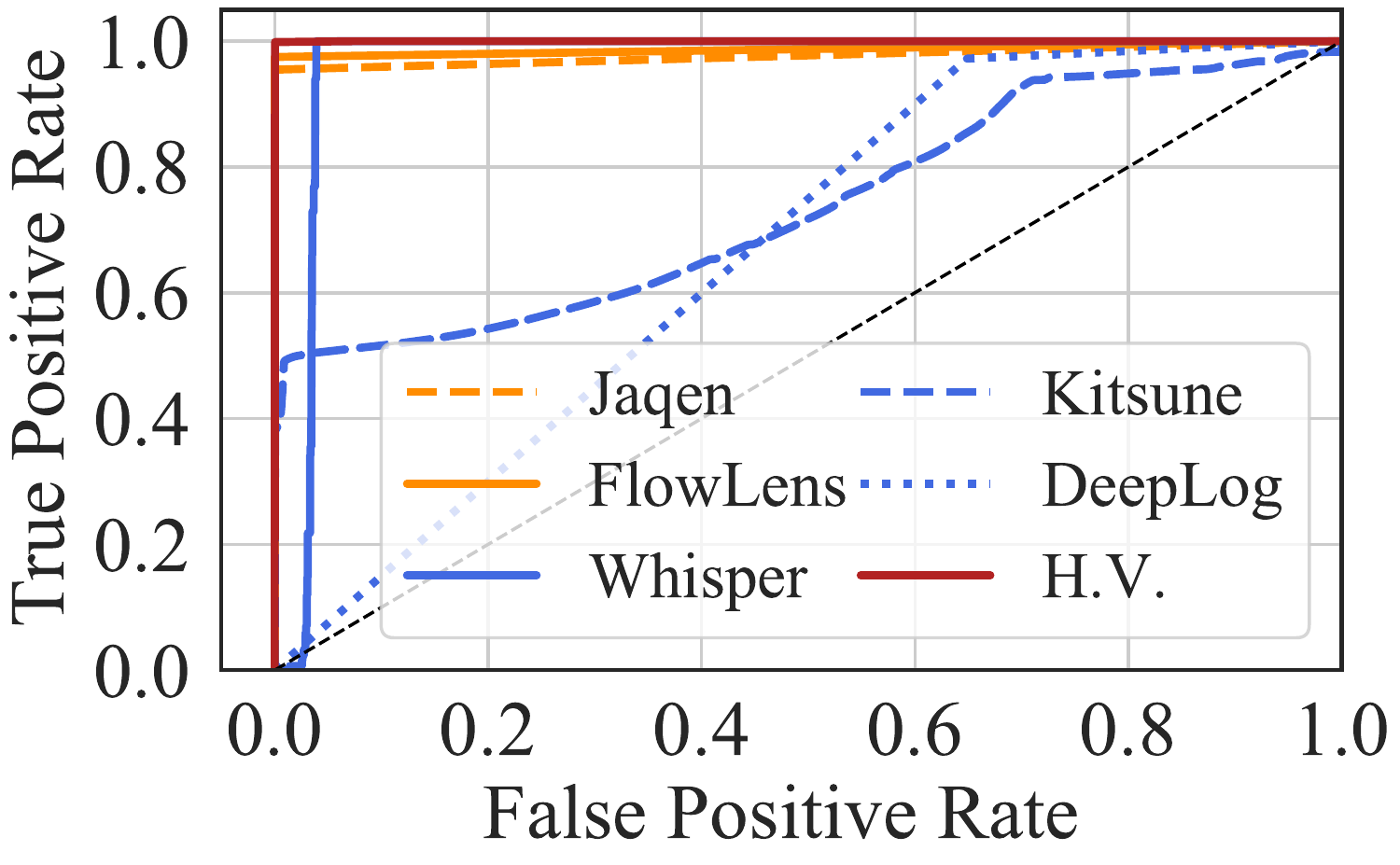}
	}
    \\
    \vspace{-2mm}
    \subfigure[PRC of detecting NTP DDoS.]{
        \label{graph:roc-prc:ntprdos-prc}
		\includegraphics[width=0.21\textwidth]{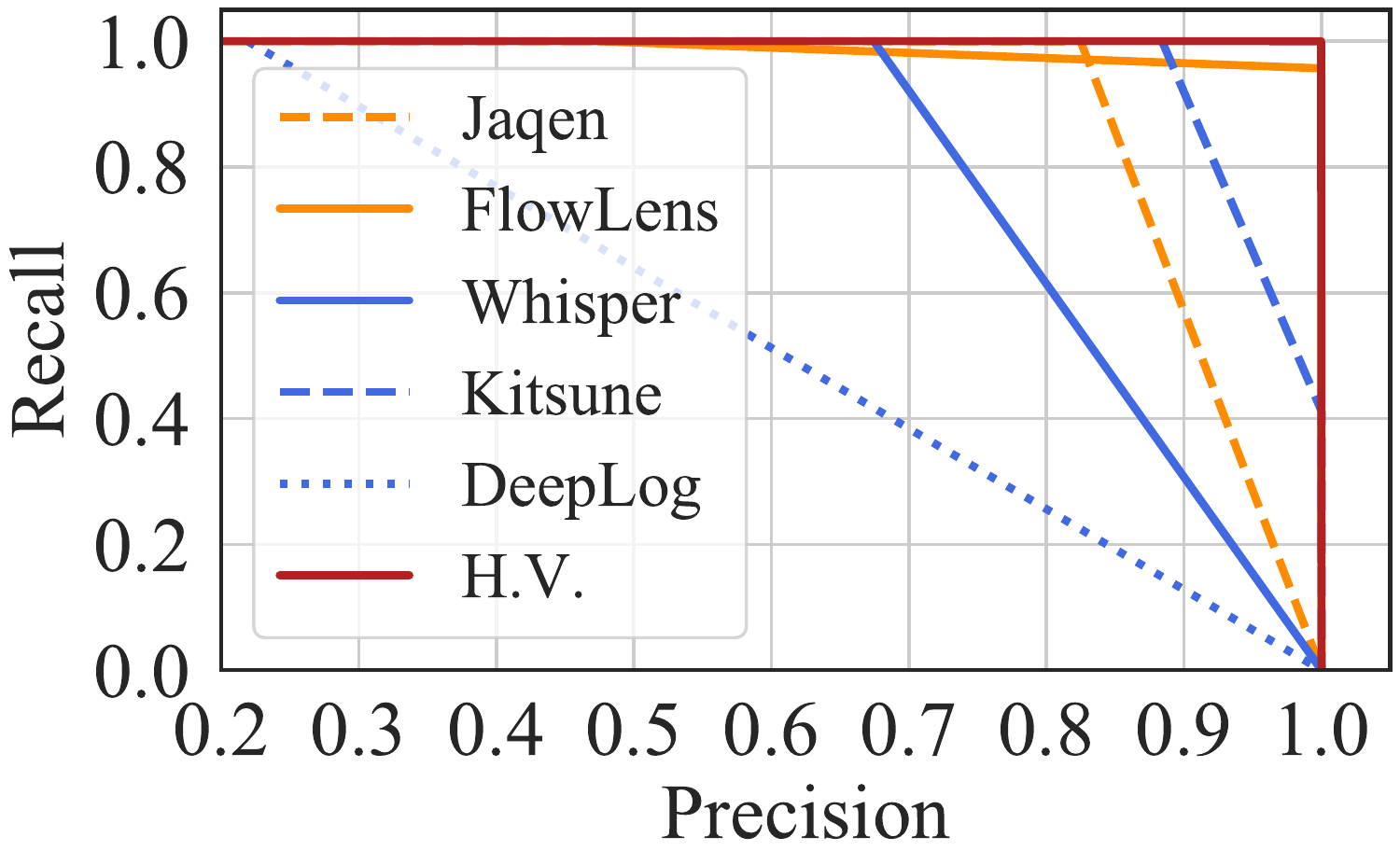}
	}
	\hspace{-2mm}
	\subfigure [PRC of detecting SYN DDoS.]{
        \label{graph:roc-prc:synsdos-prc}
		\includegraphics[width=0.21\textwidth]{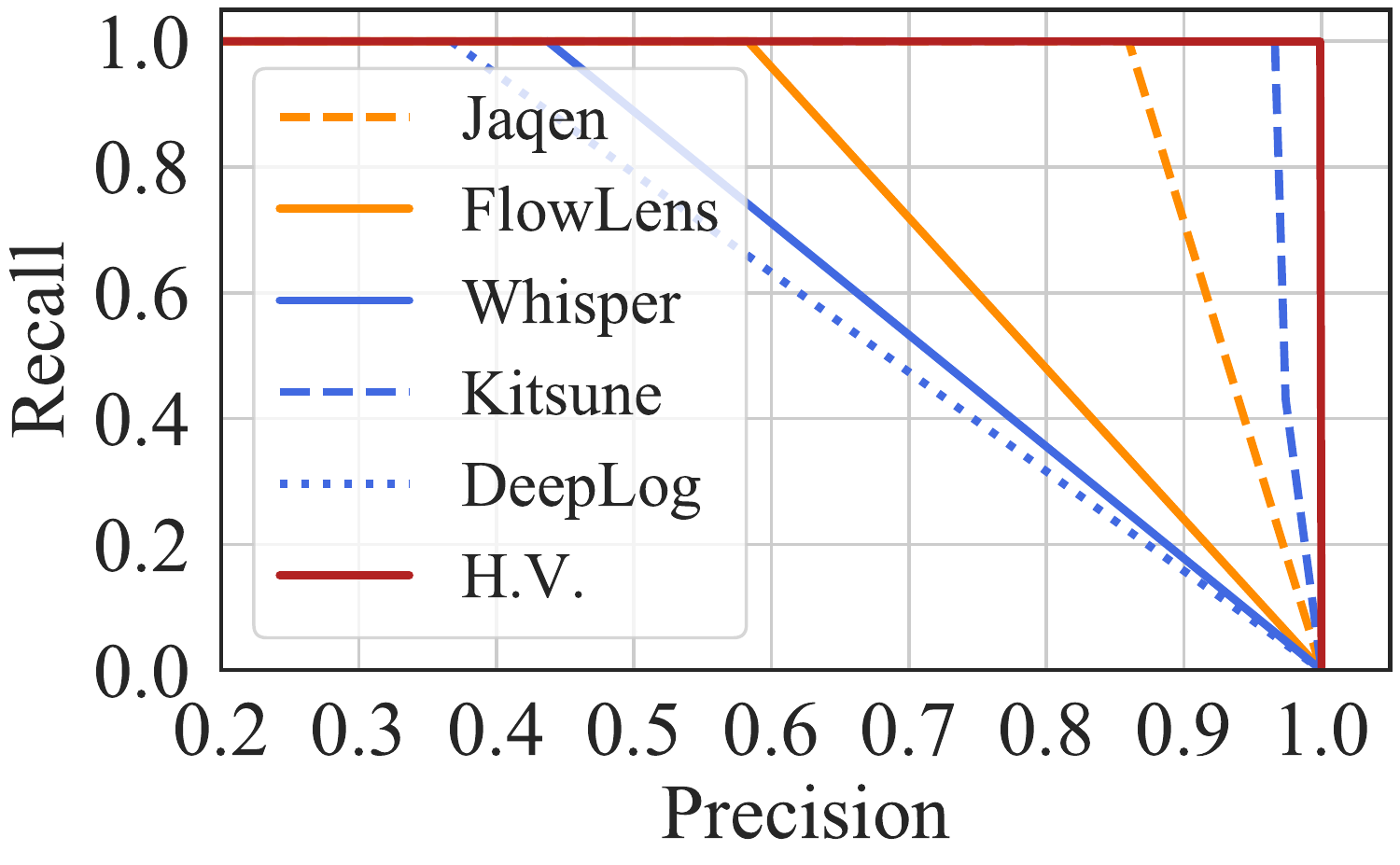}
	}
	\vspace{-2mm}
    \caption{ROC and PRC of \name and all the baselines.} 
    \label{graph:roc-prc}
    \end{center}
    \vspace{-6mm}
\end{figure}

\noindent\textbf{Traditional Brute Force Attacks.}
First, we measure the performance of the baselines by using the flooding attacks with short flows. Although \name is designed for encrypted malicious traffic detection, we find that it can also detect traditional attacks accurately. The results are shown in Table~\ref{table:traditional}. \name has 0.992 $\sim$ 0.999 AUC and 0.929 $\sim$ 0.999 F1, which achieves at most 13.4\% and 1.3\% improvement of F1 and AUC over the best performance of the baselines. The ROC and PRC results are illustrated in Figure~\ref{graph:roc-prc}. According to Figure~\ref{graph:roc-prc:ntprdos-roc} and~\ref{graph:roc-prc:httpscan-roc}, we observe that \name has less false positives while achieving similar accuracy. Figure~\ref{graph:roc-prc:ntprdos-prc} and Figure~\ref{graph:roc-prc:synsdos-prc} show that the PRC of \name is largely better than the baselines, which means that it has a higher precision when all methods reach the same recall.

Second, by comparing \name with Jaqen, we can see that \name can realize higher accuracy (i.e., a 19.4\% F1 improvement) than Jaqen with the best threshold set manually. That is, the unsupervised method allows reducing manual design efforts. Moreover, it has 56.3\% AUC improvement over the typical supervised ML based method (FlowLens). Note that, we assume that \name cannot acquire labeled datasets for training, which is more realistic. Also, it outperforms Whisper with 11.6\% AUC, which is an unsupervised detection in high-speed network. We observe that Kitsune and DeepLog have lower accuracy because they cannot afford high-speed backbone traffic.

\begin{figure}[t]
    \subfigcapskip=-1.5mm
    \vspace{-3.6mm}
    \begin{center}
	\subfigure[AUC of detecting probing vulnerable application.]{
        \hspace{-2mm}
        \includegraphics[width=0.45\textwidth]{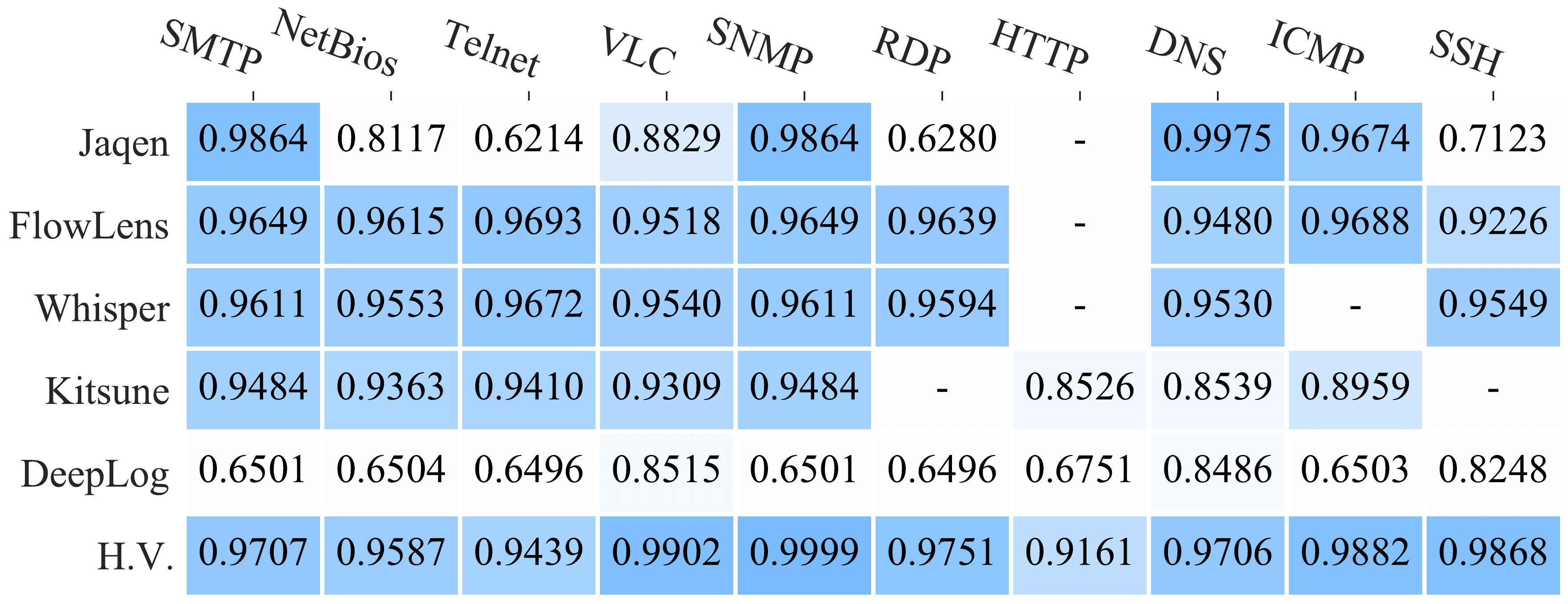}
	}
    \\
    \vspace{-1mm}
    \subfigure [F1 of detecting probing vulnerable application.]{
        \hspace{-2mm}
        \includegraphics[width=0.45\textwidth]{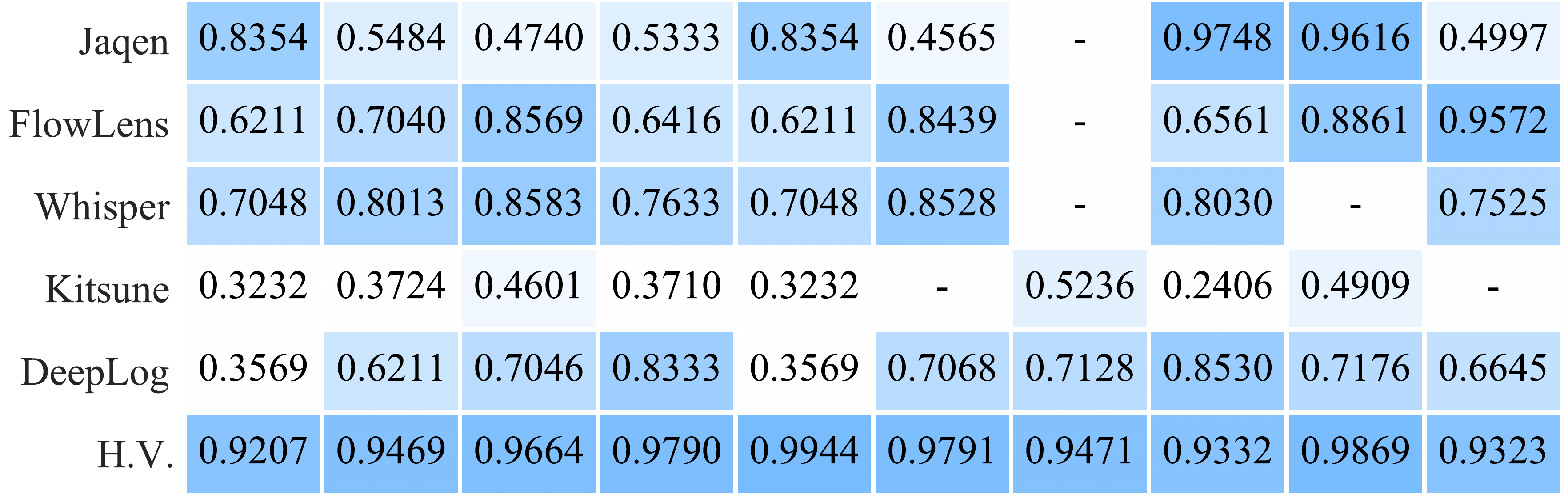}
	}
	\vspace{-2mm}
    \caption{Heatmap of accuracy for probing vulnerabilities.} 
    \label{graph:lrscan}
    \end{center}
    \vspace{-1mm}
\end{figure}

Third, we measure the detection accuracy of probing vulnerable applications. As shown in Figure~\ref{graph:lrscan}, we see that \name can detect the low-rate attacks with 0.920 $\sim$ 0.994 F1 and 0.916 $\sim$ 0.999 AUC under 6 $\sim$ 268 attackers with 17.6 $\sim$ 97.9 Kpps total bandwidth. It also achieves at most 46.8\% F1 and 27.3\% AUC improvements over the baselines that have a more significant accuracy decrease than the high-rate attacks. For example, FlowLens only achieves averagely 0.684 F1, which is only 77\% under the high-rate attacks. Although Jaqen can be deployed on programmable switches, its thresholds are invalided by the low-rate attacks. And Whisper is unable to detect the attacks with two datasets. Moreover, Kitsune and DeepLog cannot detect the attacks because of the low rate of malicious packets ($\le$ 1.2\%). 

The reason why \name can detect the slow probing while maintaining the similar accuracy to the high-rate attacks is that the graph preserves flow interaction patterns. Although the flows from a single attacker are slow, e.g., at least 244 pps, \name can record and analyze their interaction patterns. Specifically, each flow in the stealthy attack traffic can be represented by an edge in the graph, while the vertices in the graph indicate the addresses generating the traffic. Thus, the traffic can be captured by identifying vertices with large out-degrees (i.e., a large number of edges). Moreover, the brute force attacks validate that our method is effective to capture the DDoS traffic because it utilizes the short flow aggregation to construct the edge associated with short flows and avoids inspecting each short spoofing flow. Besides, the experiment results also show that the critical vertices denote the addresses of major active flows, e.g., web servers, DNS servers, and scanners. Note that, we exclude the results of the baselines that cannot detect encrypted traffic with lower rates in the following sections due to the page limits.

\begin{figure}[t]
    \subfigcapskip=-1.5mm
    \vspace{-3mm}
    \begin{center}
    \hspace{-2mm}
	\subfigure[AUC of detecting encrypted link-flooding and encrypted channel injection.]{
        \label{graph:misc:link-sc-auc}
		\includegraphics[width=0.48\textwidth]{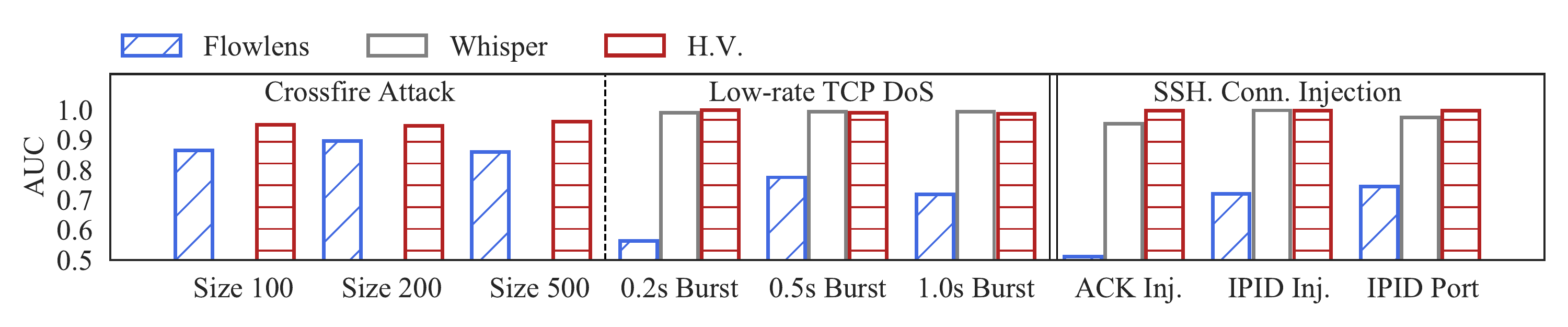}
	}
	\vspace{-4mm}
	\\
    \hspace{-2mm}
	\subfigure [F1 of detecting encrypted link-flooding and encrypted channel injection.]{
        \label{graph:misc:link-sc-f1}
		\includegraphics[width=0.48\textwidth]{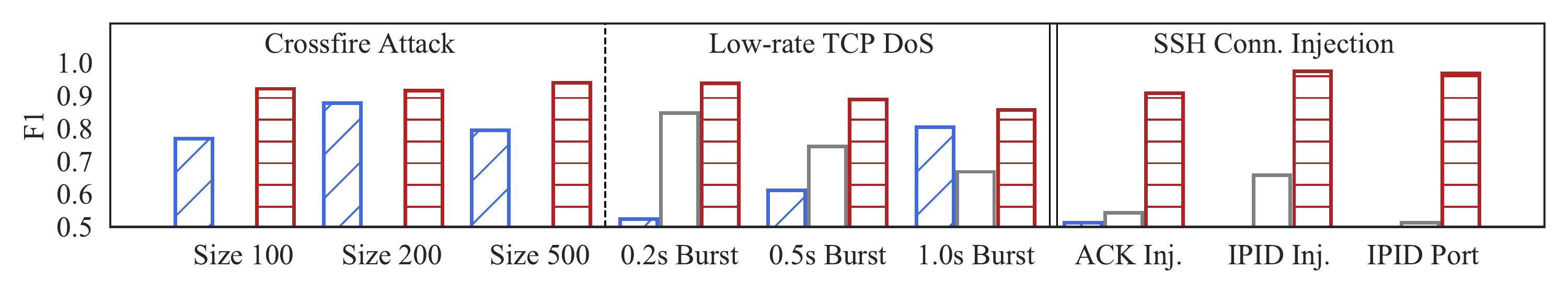}
	}
	\vspace{-2mm}
    \\
    \subfigure[F1 of password cracking.]{
        \label{graph:misc:link-pwd-auc}
		\includegraphics[width=0.23\textwidth]{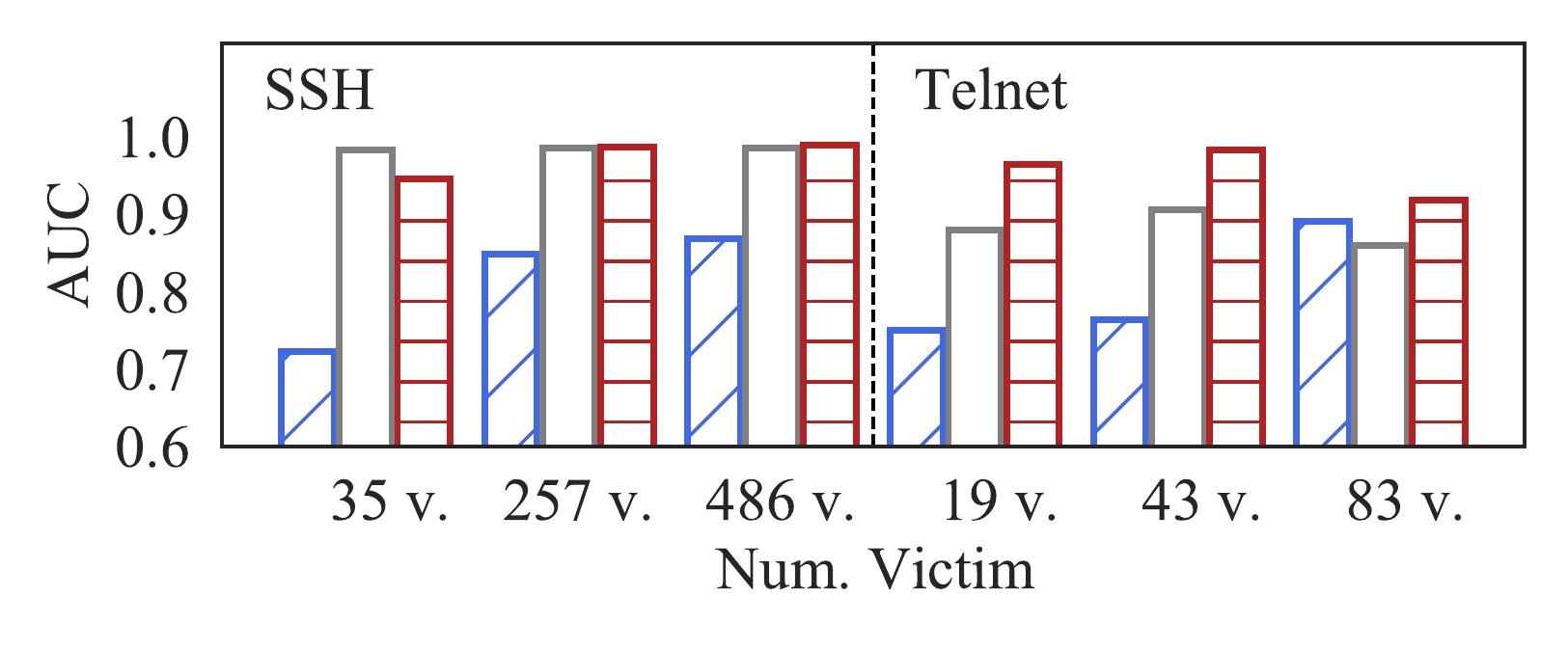}
	} 
	\hspace{-4mm}
	\subfigure [AUC of password cracking.]{
        \label{graph:misc:link-pwd-f1}
		\includegraphics[width=0.23\textwidth]{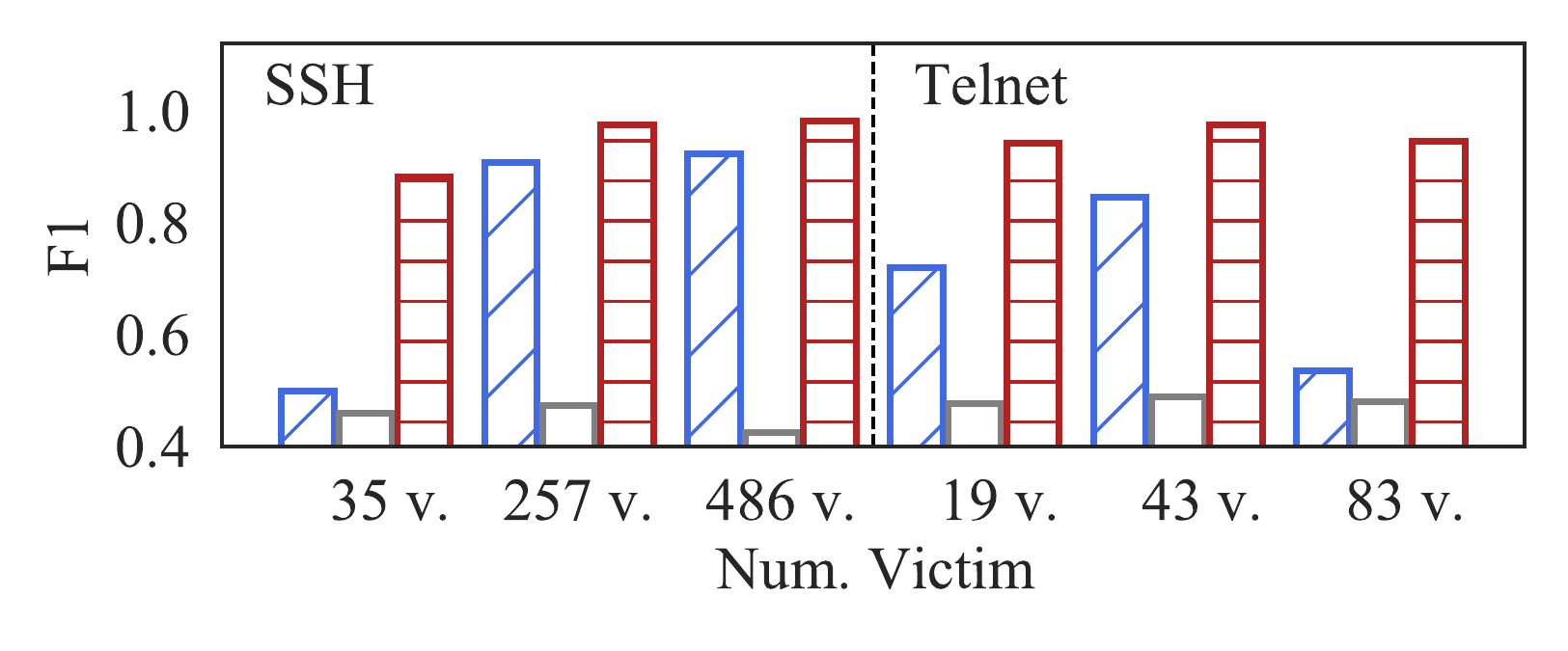}
	}
	\vspace{-2mm}
	\caption{Detection accuracy of encrypted flooding traffic.}
	\label{graph:misc}
	\end{center}
\vspace{-5.2mm}
\end{figure}

\begin{figure}[!t]
    \subfigcapskip=-2.2mm
    \vspace{-0.4mm}
    \begin{center}
	\subfigure[Crossfire.]{
        \label{graph:visual:crossfire}
		\includegraphics[width=0.21\textwidth]{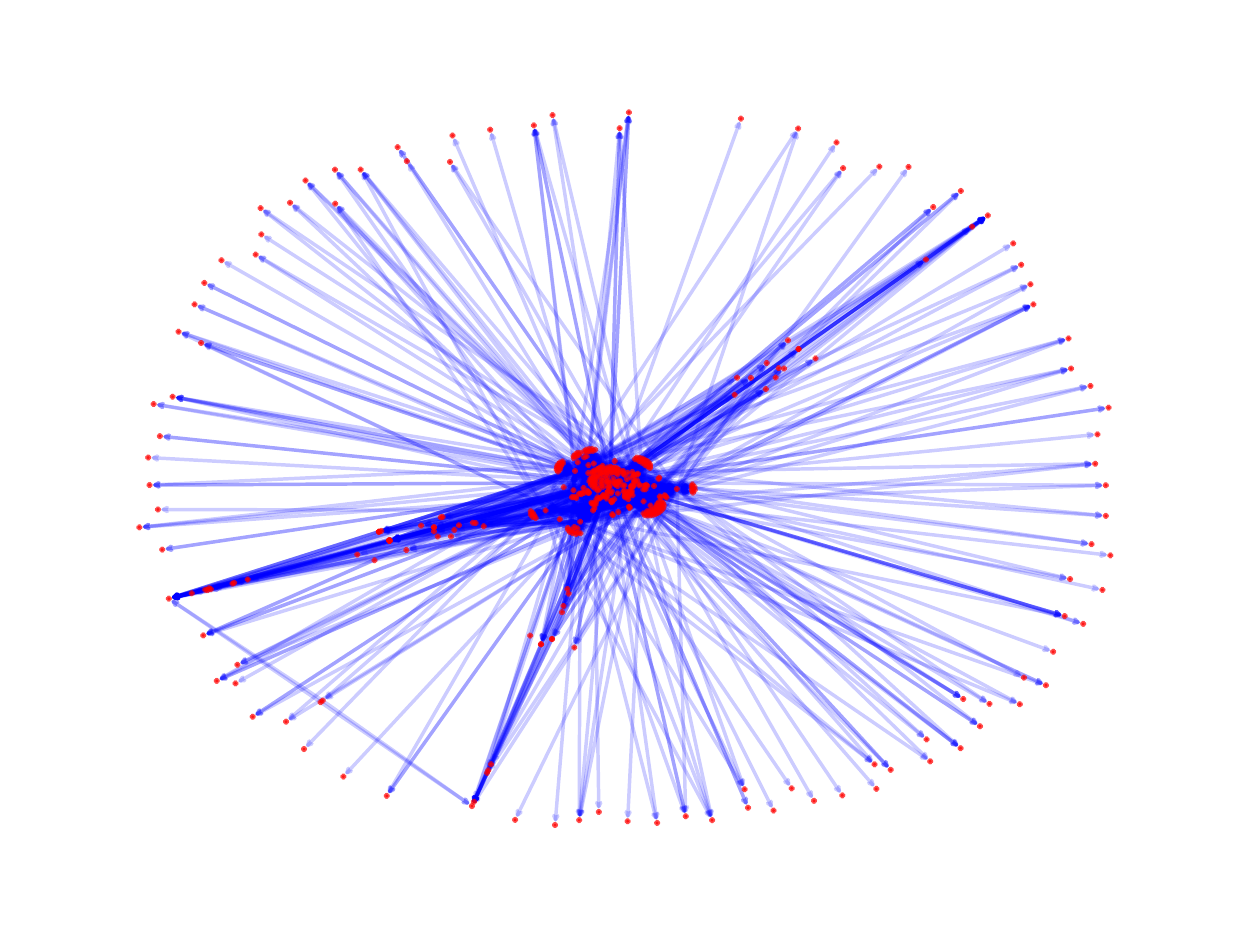}
	}
	\hspace{-2mm}
    \subfigure [SSH cracking.]{
        \label{graph:visual:password}
		\includegraphics[width=0.21\textwidth]{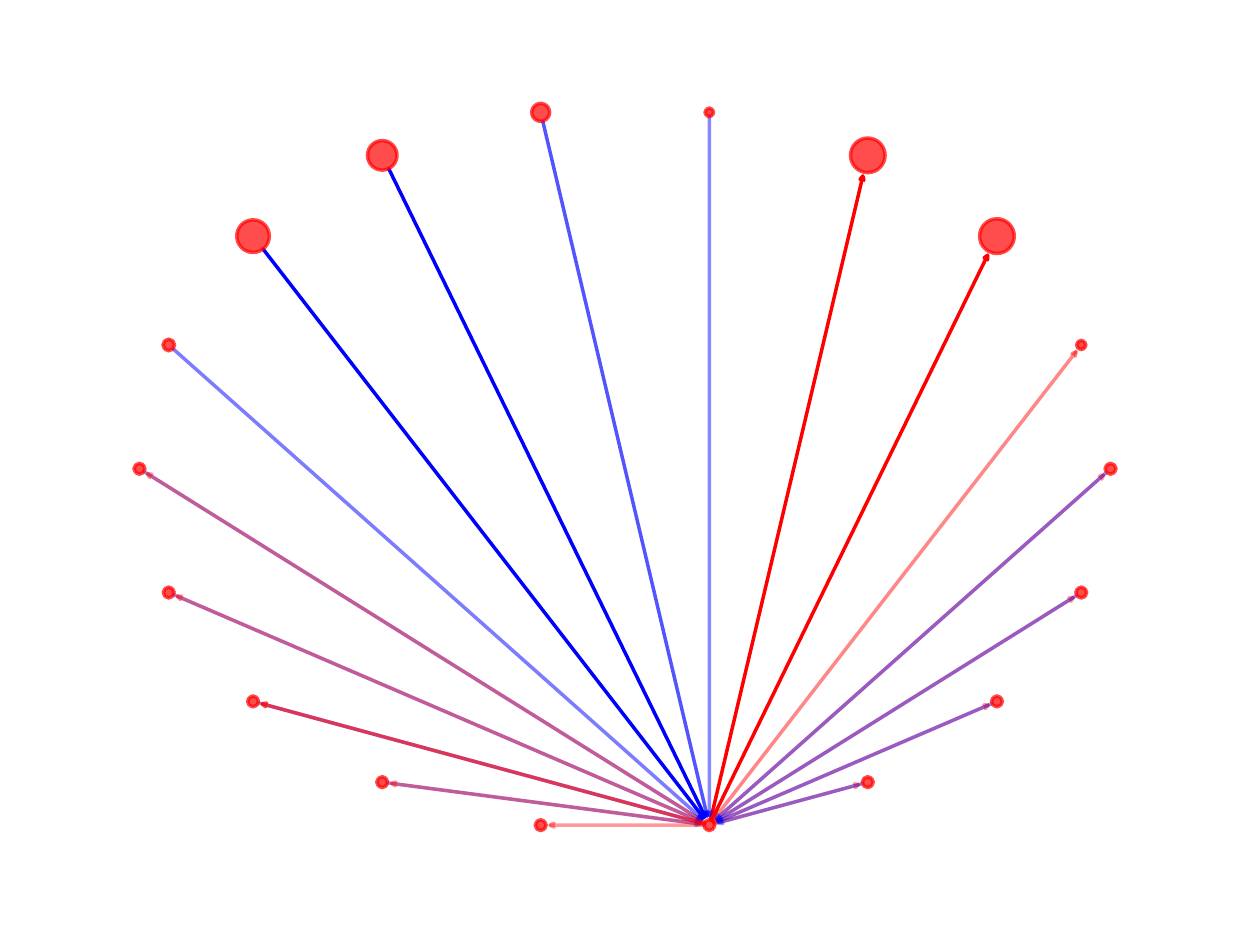}
	}
    \\
    \vspace{-2mm}
    \subfigure [XSS detection.]{
        \label{graph:visual:xss}
		\includegraphics[width=0.22\textwidth]{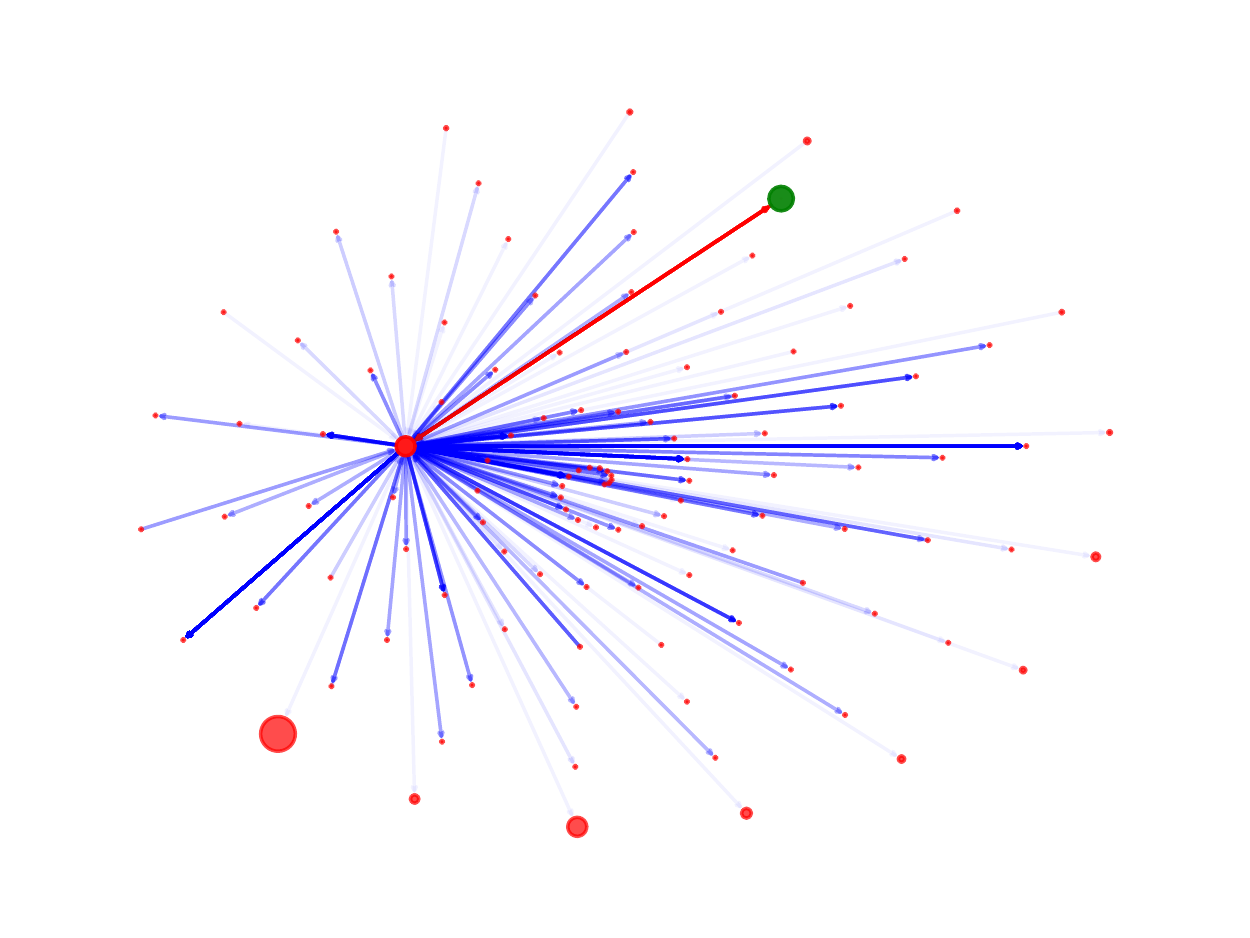}
	}
	\hspace{-2mm}
	\subfigure [P2P botnet.]{
        \label{graph:visual:sality}
		\includegraphics[width=0.22\textwidth]{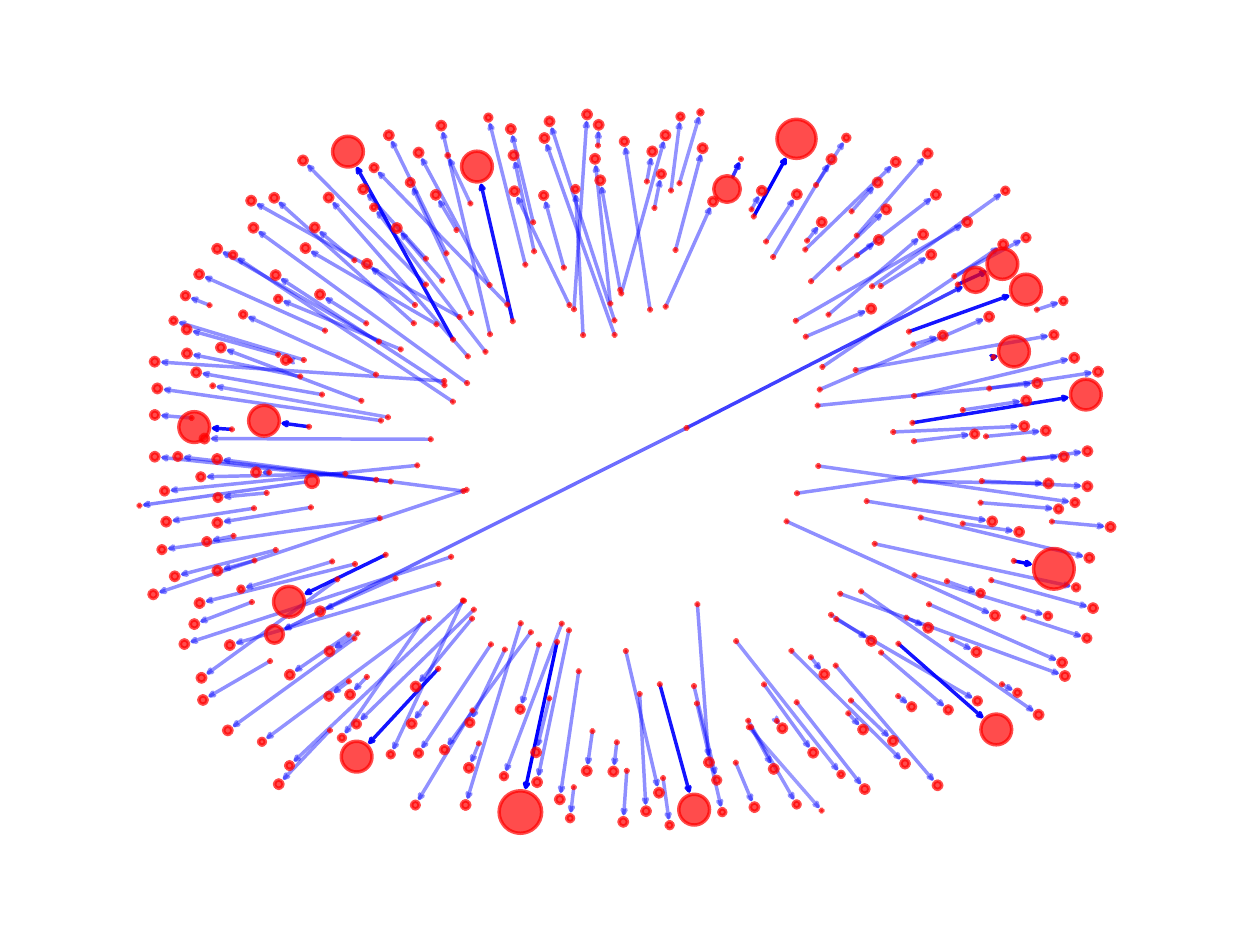}
	}
	\vspace{-1.5mm}
    \caption{Subgraph with various encrypted malicious traffic.} 
    \label{graph:visual}
    \end{center}
    \vspace{-5.6mm}
\end{figure}

\noindent\textbf{Encrypted Flooding Traffic.} Figure~\ref{graph:misc} shows the detection accuracy under flooding attacks using encrypted traffic. Generally, \name achieves 0.856 $\sim$ 0.981 F1 and 0.917 $\sim$ 0.998 AUC, which are 58.7\% and 25.3\% accuracy improvements over the baselines that can detect such attacks. Specifically, as shown in Figure~\ref{graph:misc:link-sc-auc} and~\ref{graph:misc:link-sc-f1}, we observe that \name can accurately detect the link flooding traffic consists of various encrypted traffic with different parameters. For instance, it can detect the Crossfire attack using HTTPS web requests generated by different sizes of botnets~\cite{SP13-Crossfire} with at most 0.939 F1. The massive web traffic generated by bots, which is low-rate ($\le$ 4Kbps) and encrypted, evades the detection of Whisper and FlowLens (F1 $\le$ 0.8). As shown in Figure~\ref{graph:visual:crossfire}, \name can detect the attack efficiently by splitting the botnet clusters into a single connected component to exclude the interference from the similar benign web traffic, where the inner layer denotes botnets and the outer denotes decoy servers. 

Moreover, we find that \name can detect low-rate TCP DoS attacks that use burst encrypted video traffic for at most 0.995 AUC and 0.938 F1. Although Whisper has slightly better AUC in some cases, we find that it cannot achieve high accuracy on all scenarios. As a result, it has only 55.5\% AUC in the worse case. Moreover, \name can aggregate the short flows in the SSH connection injection attacks and achieves more than 0.95 F1. The attacks exploiting protocol vulnerabilities realize low-rate packet injection and evade the detection of FlowLens (i.e., AUC $\le$ 0.774, F1 $\le$ 0.513). Figure~\ref{graph:misc:link-pwd-auc} and~\ref{graph:misc:link-pwd-f1} illustrate that \name can identify slow and persisted password attempts for the channels with over 0.881 F1 and 0.917 AUC, which are 1.19 and 1.28 times improvements over FlowLens and Whisper. The reason is that \name maintains the interaction patterns of attackers using the graph, e.g., the massive short flows for login attempts shown as red edges in Figure~\ref{graph:visual:password}.

\begin{figure}[t]
    \subfigcapskip=-2.4mm
    \vspace{-1.3mm}
    \begin{center}
	\subfigure[AUC of detecting encrypted web attack traffic.]{
        \hspace{-2mm}
        \includegraphics[width=0.48\textwidth]{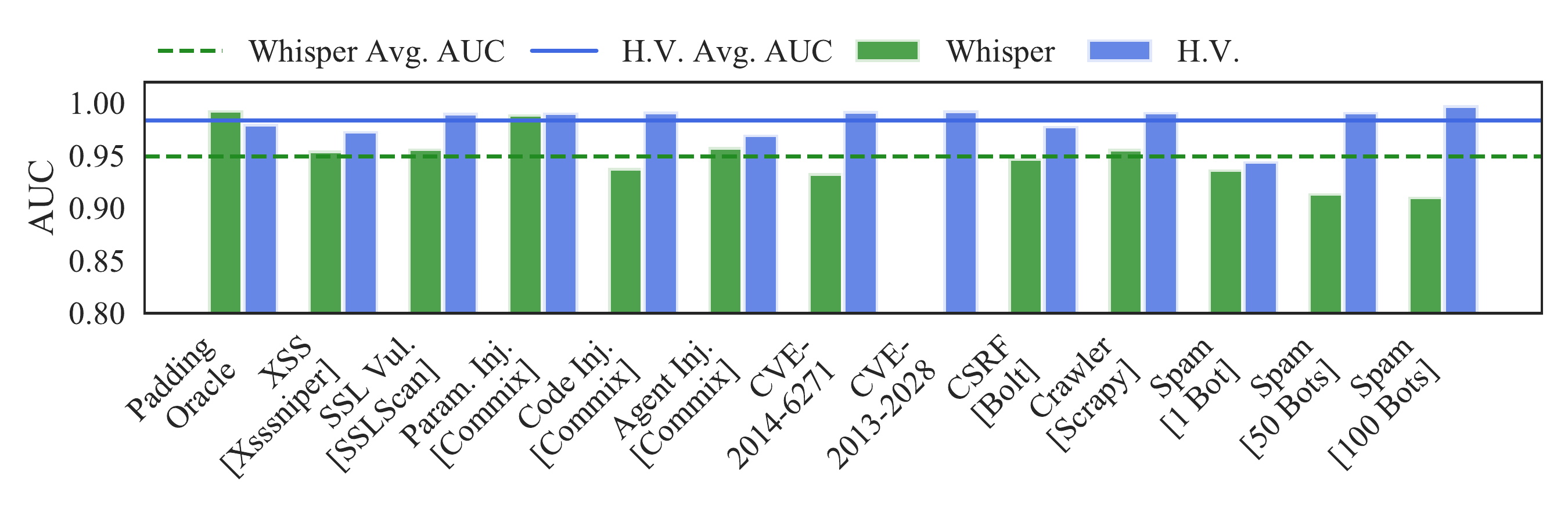}
	}
    \\
    \vspace{-1.5mm}
    \subfigure [F1 of detecting encrypted web attack traffic.]{
        \hspace{-2mm}
        \includegraphics[width=0.48\textwidth]{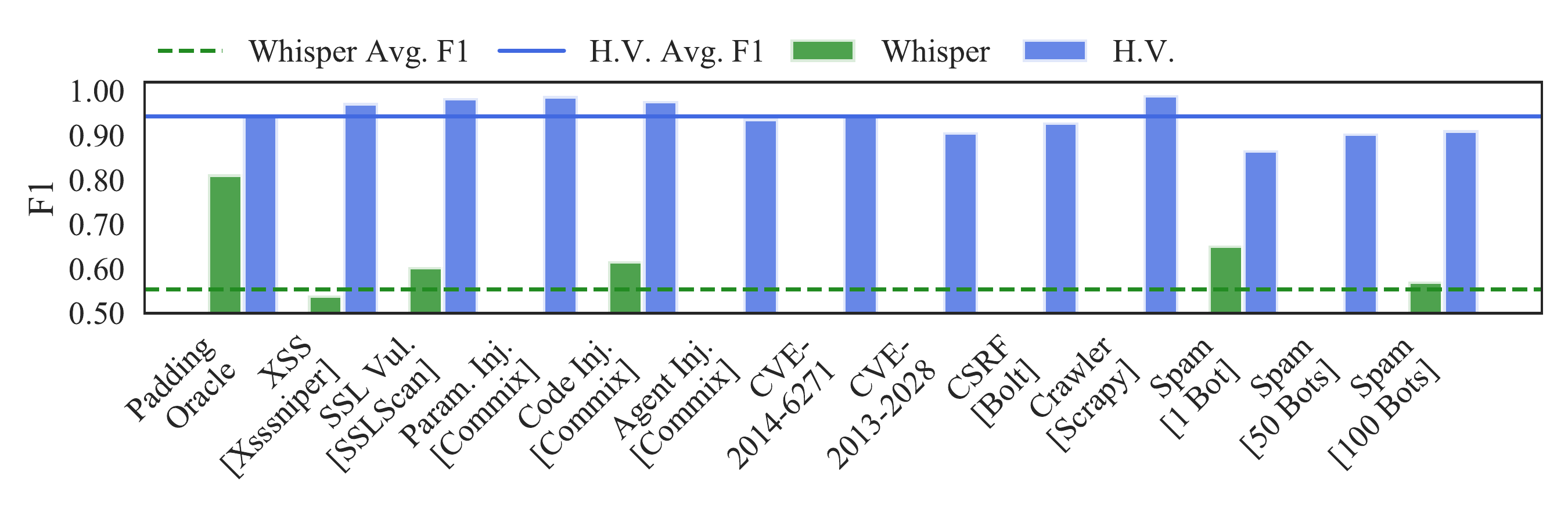}
	}
	\vspace{-3mm}
    \caption{Accuracy of encrypted web attack traffic detection.} 
    \label{graph:web}
    \end{center}
    \vspace{-4mm}
\end{figure}

\begin{figure*}[!t]
    \subfigcapskip=-1mm
    \vspace{-5mm}
    \begin{center}
        \includegraphics[width=0.99\textwidth]{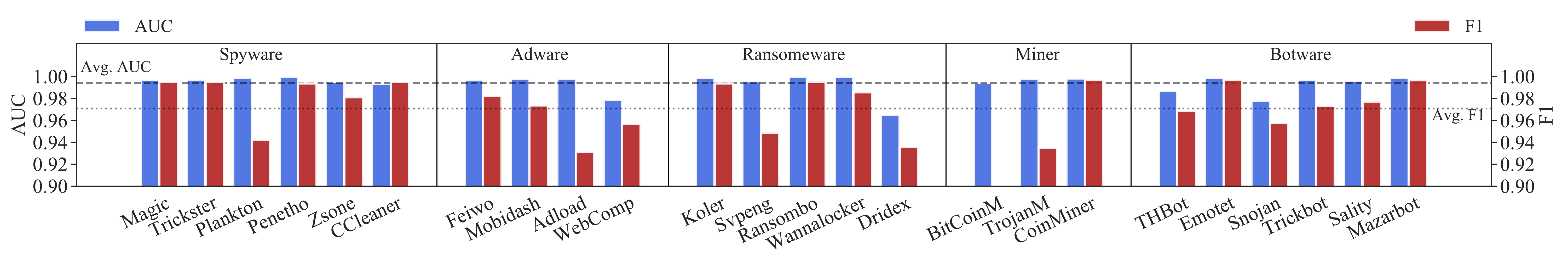}
	\vspace{-6mm}
    \caption{\name can detect various encrypted malware traffic.} 
    \label{graph:malware}
    \end{center}
    \vspace{-6mm}
\end{figure*}

\noindent\textbf{Encrypted Web Malicious Traffic.} Figure~\ref{graph:web} presents the detection accuracy of the encrypted traffic generated by various web vulnerabilities discovery. \name achieves 0.985 average AUC and 0.957 average F1 (i.e., 2.8\% and 75.2\% increase compared to Whisper). The flow based ML detection cannot detect web encrypted malicious traffic because the traffic has single-flow patterns that are almost same to benign web access flows. \name can accurately detect the encrypted web malicious traffic, because, as shown in Figure~\ref{graph:visual:xss}, it captures the traffic from the frequent interactions as the edges associated with long flows, and identifies the malicious traffic (denoted by red edges) generated by the attacker (denoted by the green vertex) by clustering the edges associated with benign web traffic that are connected to the same critical vertex (denoted by the red solid vertex).

\begin{figure}[t]
    \subfigcapskip=-1.5mm
    \begin{center}
	\subfigure[Graph construction throughput.]{
        \label{graph:throughput:audit-avg}
		\includegraphics[width=0.22\textwidth]{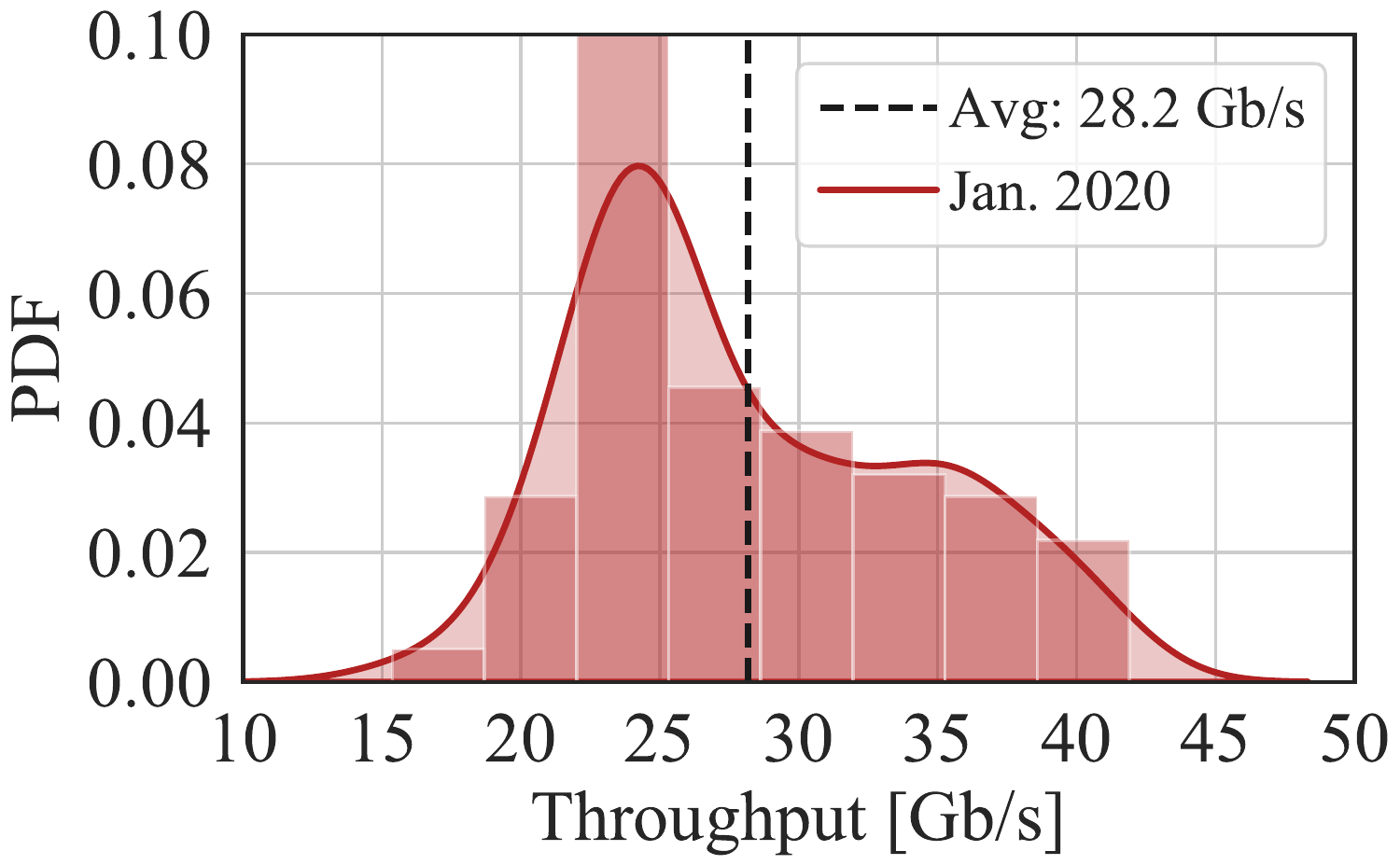}
	}
	\hspace{-2mm}
	\subfigure [Max construction throughput.]{
        \label{graph:throughput:audit-max}
		\includegraphics[width=0.22\textwidth]{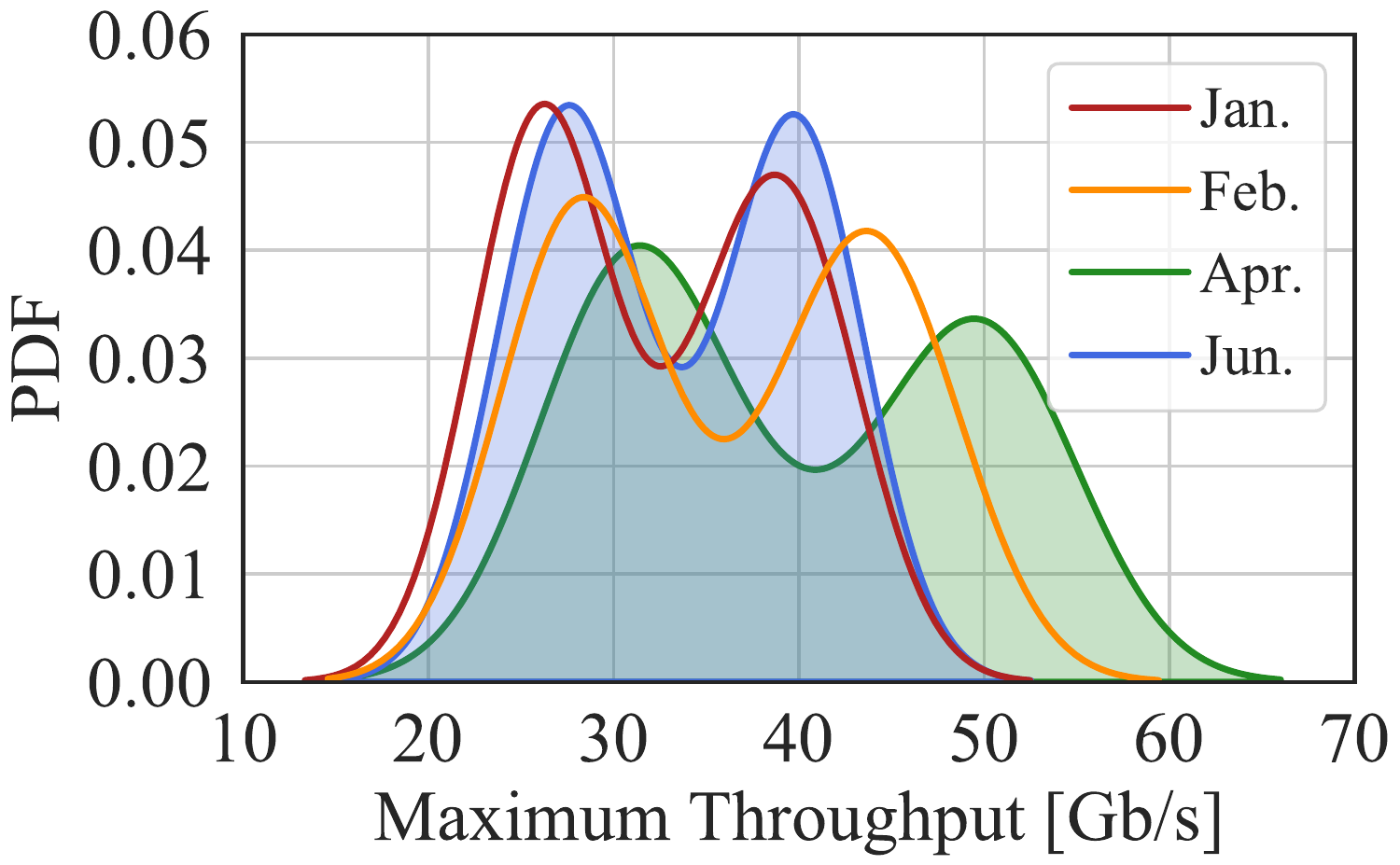}
	}\\
    \vspace{-1mm}
    \subfigure[Graph detection throughput.]{
        \label{graph:throughput:detect-avg}
		\includegraphics[width=0.22\textwidth]{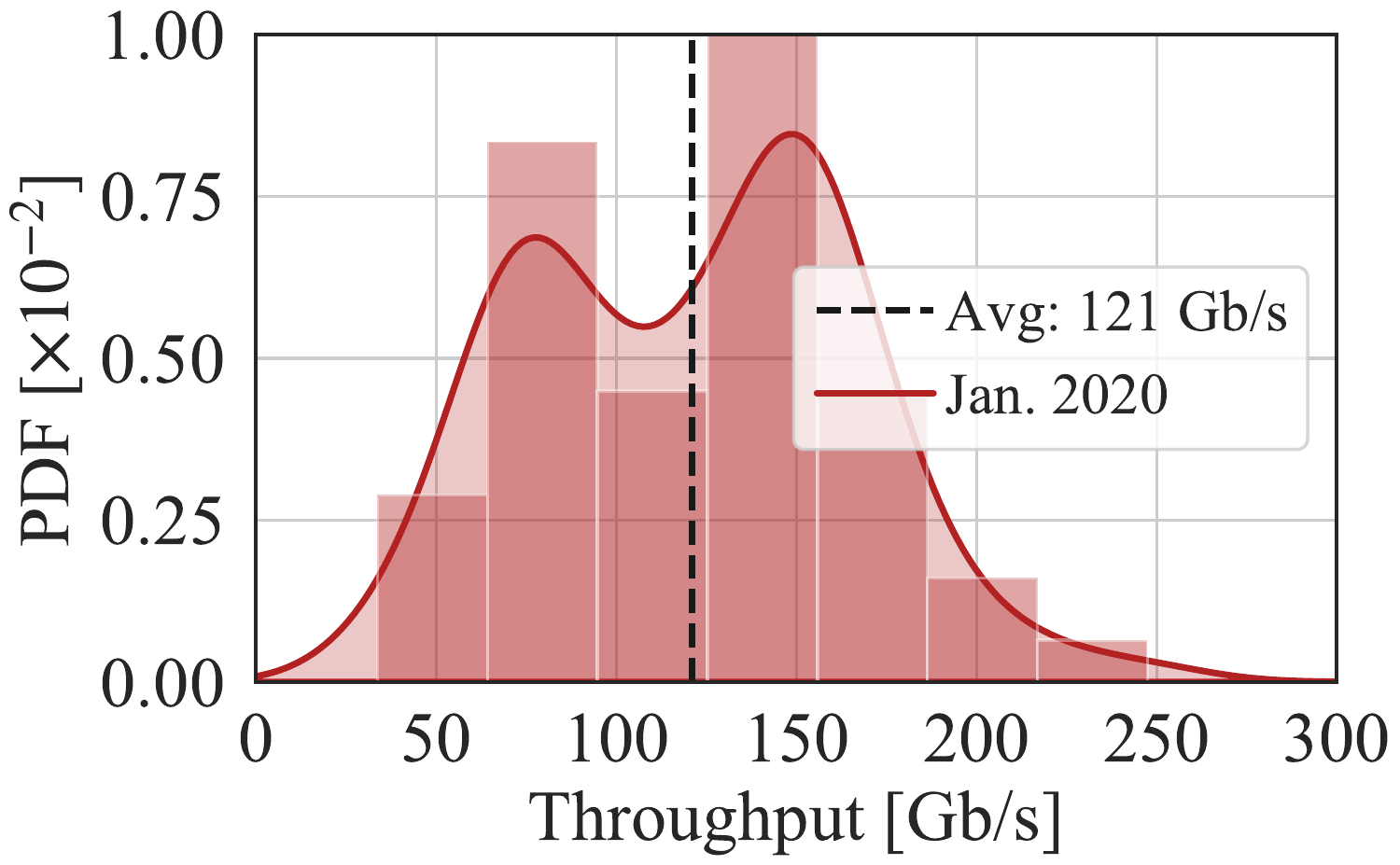}
	}
	\hspace{-2mm}
	\subfigure [Stable detection throughput.]{
        \label{graph:throughput:detect-max}
		\includegraphics[width=0.22\textwidth]{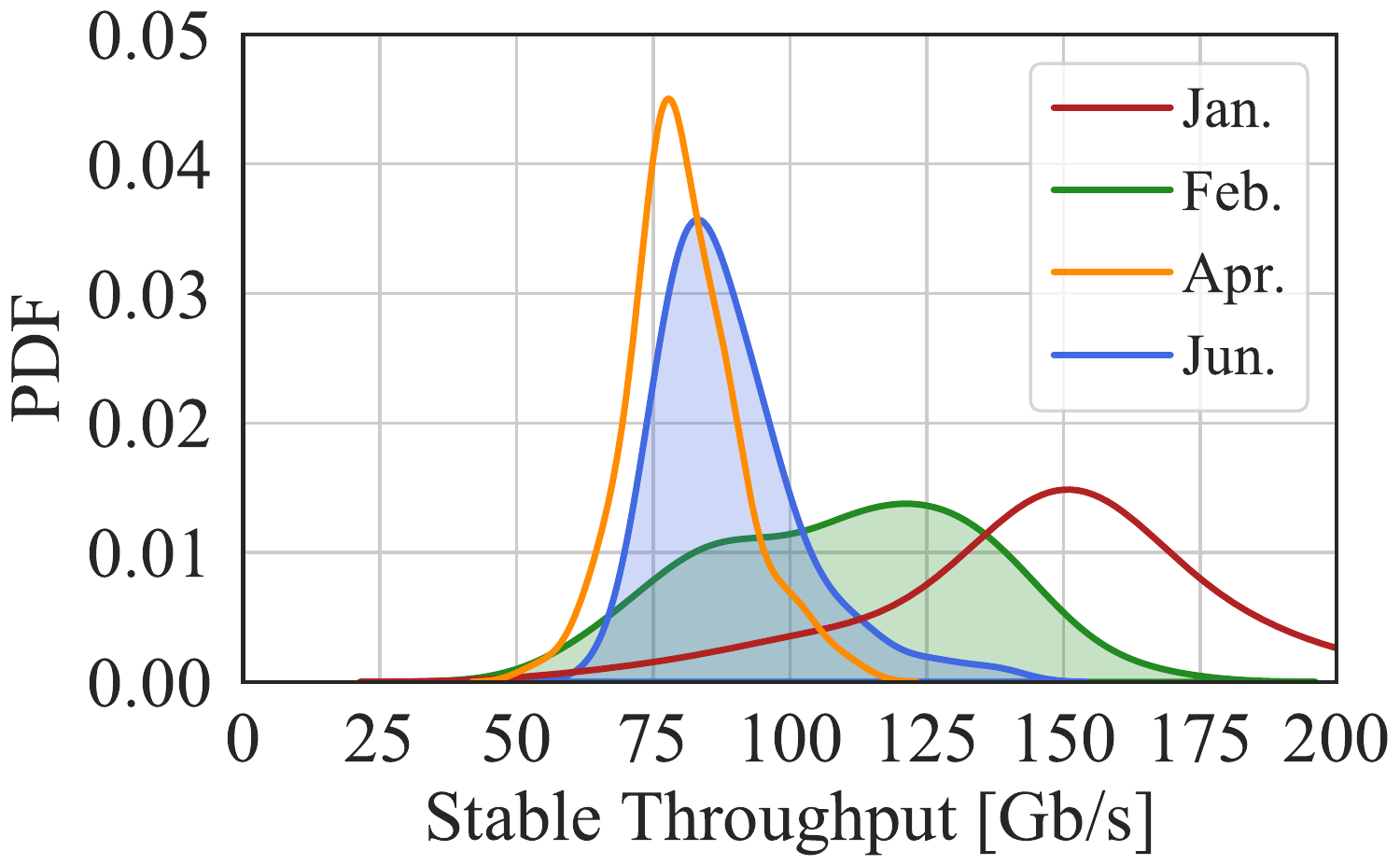}
	}
	\vspace{-2mm}
    \caption{Throughput of graph construction and detection.} 
    \label{graph:throughput}
    \end{center}
    \vspace{-2mm}
\end{figure}

\begin{figure}[t]
    \subfigcapskip=-1.5mm
    \vspace{-1mm}
    \begin{center}
	\subfigure[Graph construction latency.]{
        \label{graph:latency:audit-avg}
		\includegraphics[width=0.22\textwidth]{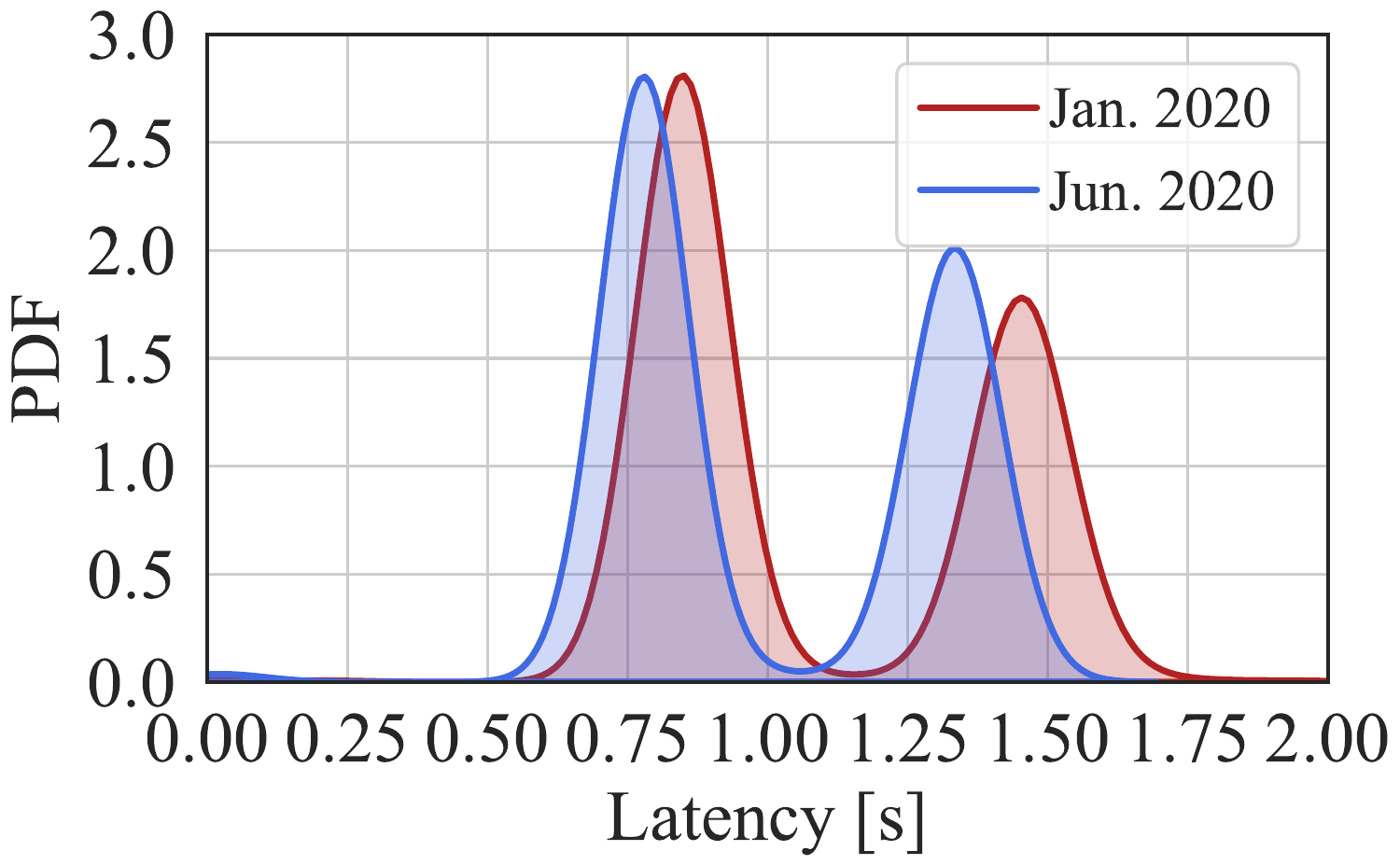}
	}
	\hspace{-2mm}
	\subfigure [Construct latency composition.]{
        \label{graph:latency:audit-comp}
		\includegraphics[width=0.22\textwidth]{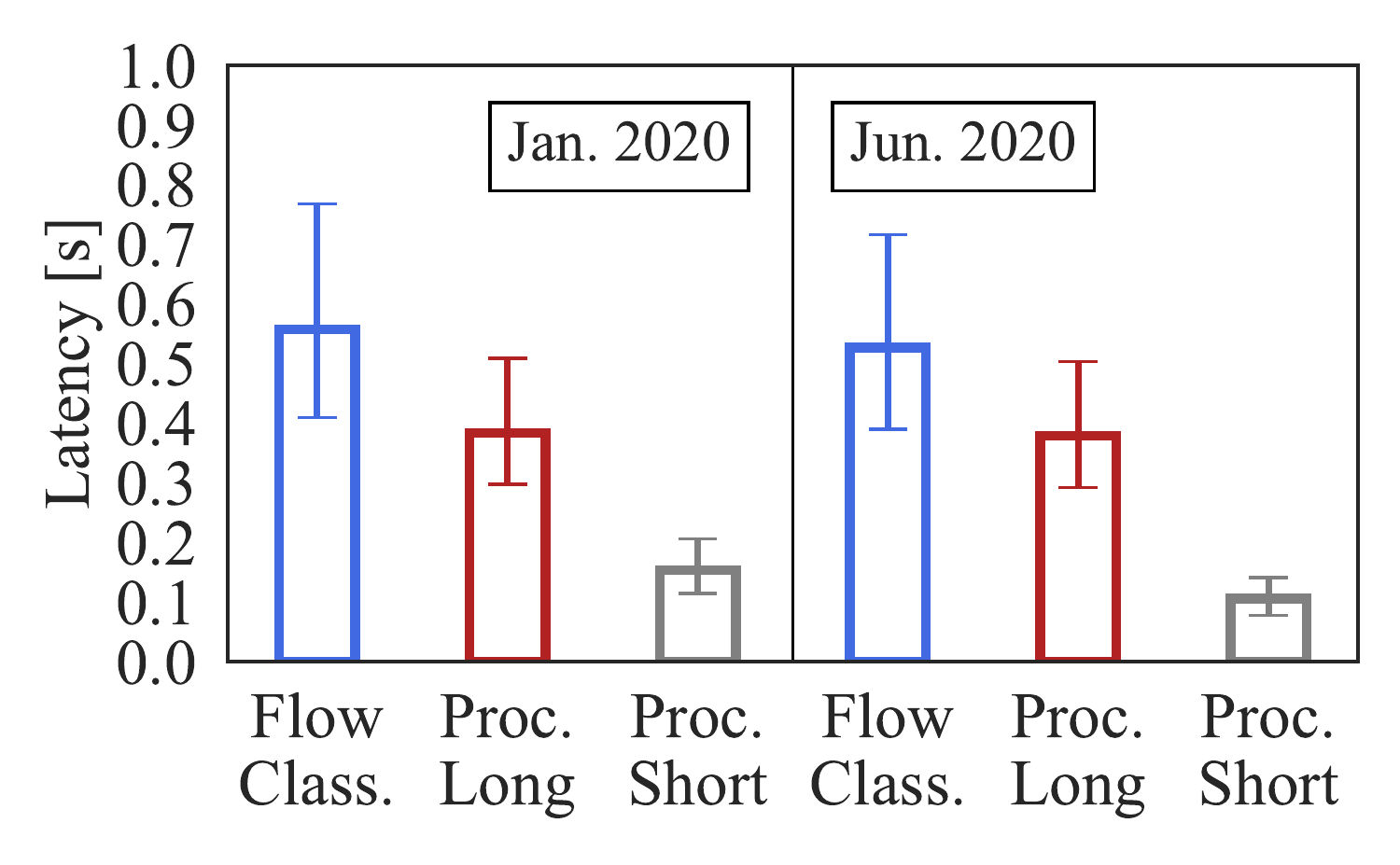}
	}\\
    \vspace{-1mm}
    \subfigure[Graph detection latency.]{
        \label{graph:latency:detect-avg}
		\includegraphics[width=0.22\textwidth]{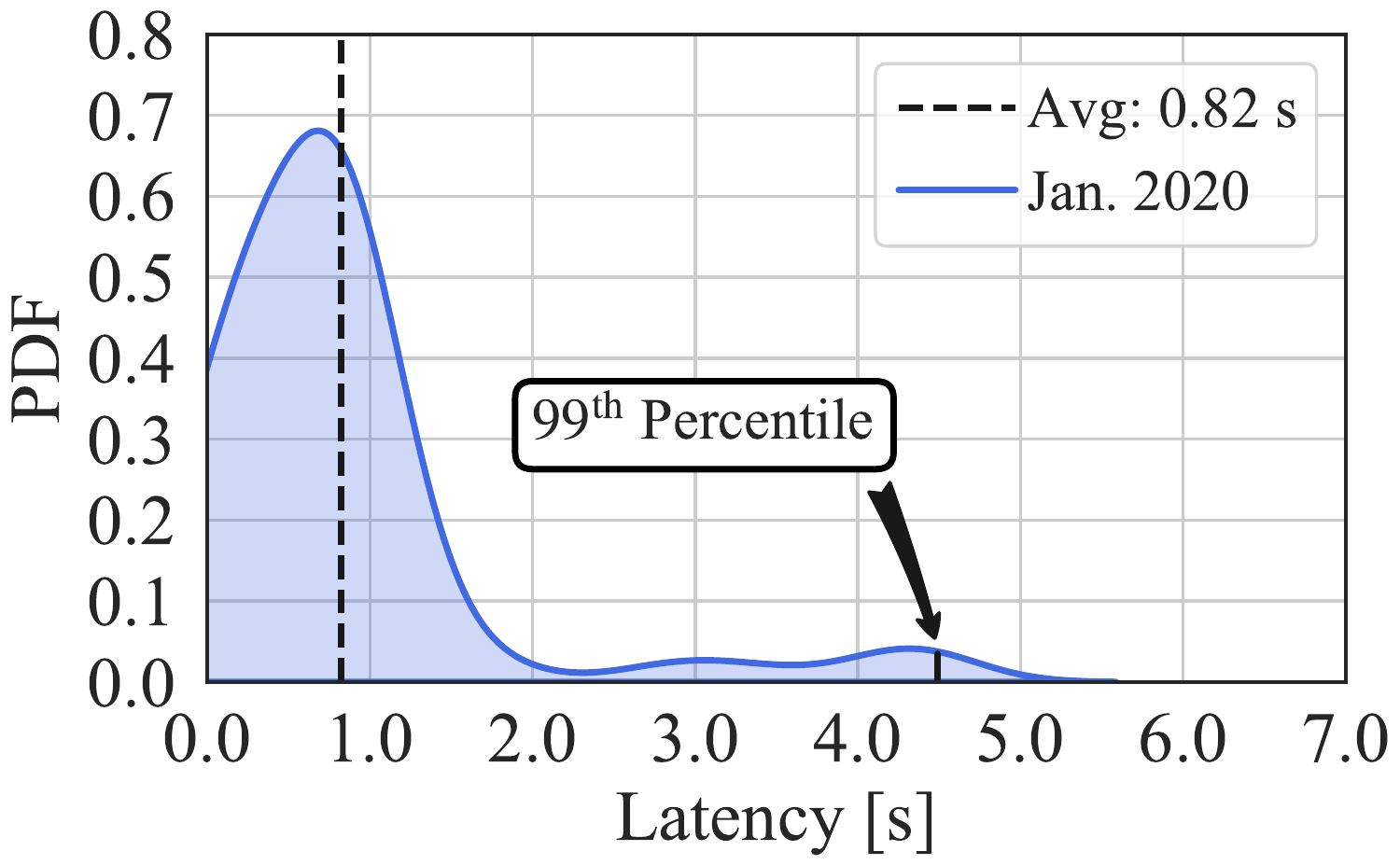}
	}
	\hspace{-2mm}
	\subfigure [Detection latency composition.]{
        \label{graph:latency:detect-comp}
		\includegraphics[width=0.22\textwidth]{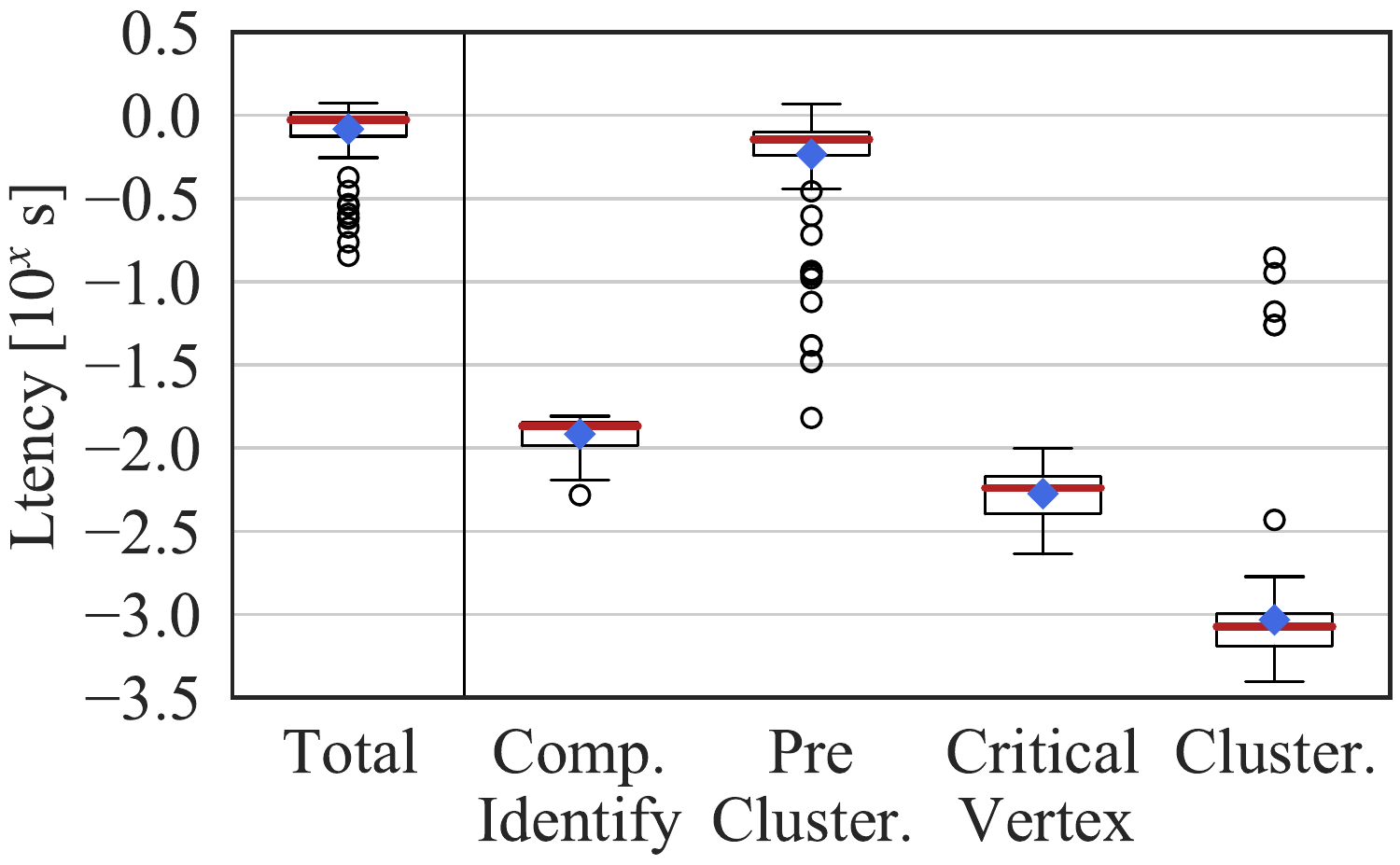}
	}
	\vspace{-2mm}
    \caption{Latency of graph construction and detection.} 
    \label{graph:latency}
    \end{center}
    \vspace{-2mm}
\end{figure}

\begin{figure}[!t]
    \subfigcapskip=-2mm
    \vspace{-1mm}
    \begin{center}
	\subfigure[Runtime memory usages.]{
        \label{graph:usage:memory}
		\includegraphics[width=0.22\textwidth]{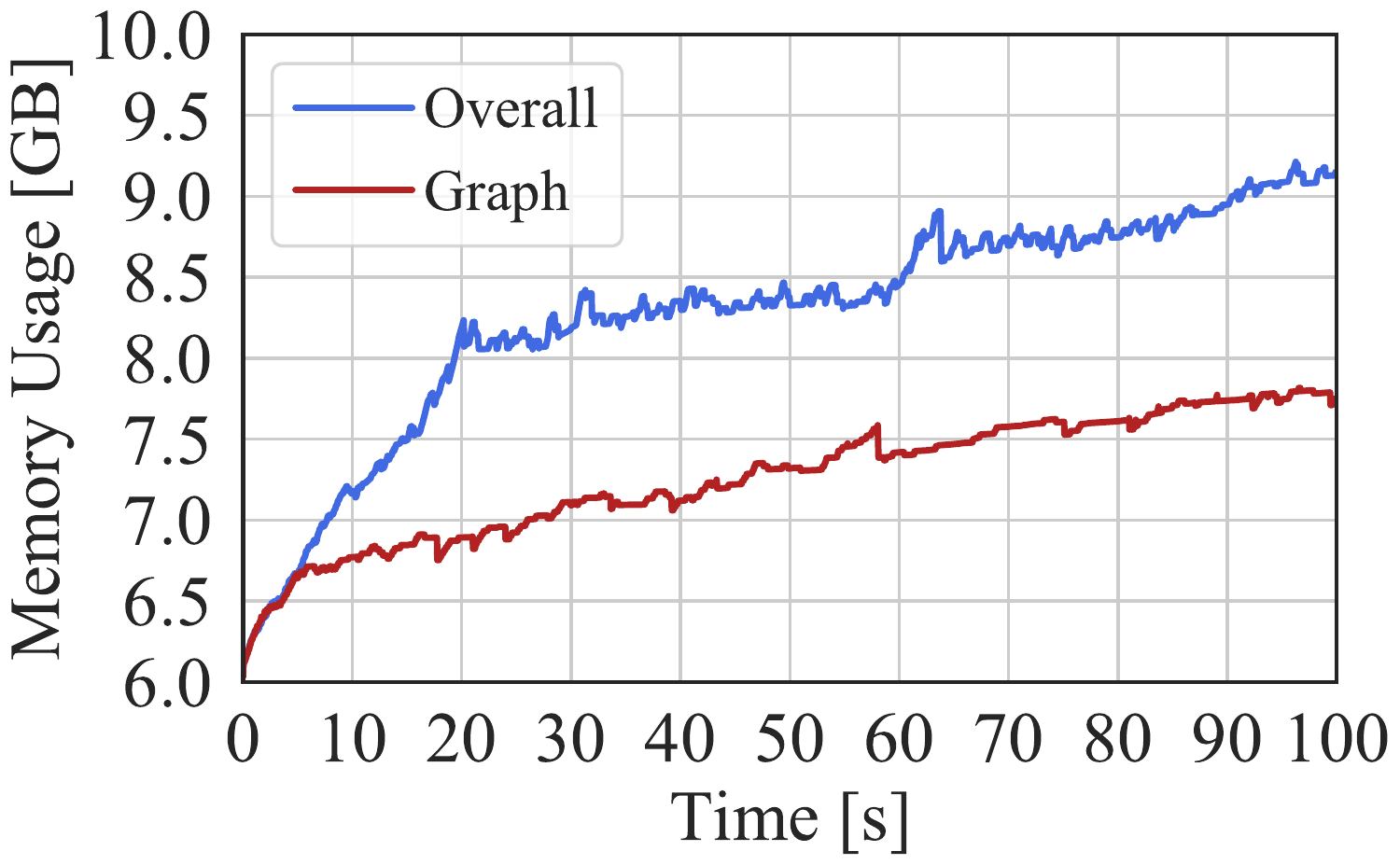}
	}
	\hspace{-2mm}
	\subfigure [Graph storage usages.]{
        \label{graph:usage:disk}
		\includegraphics[width=0.22\textwidth]{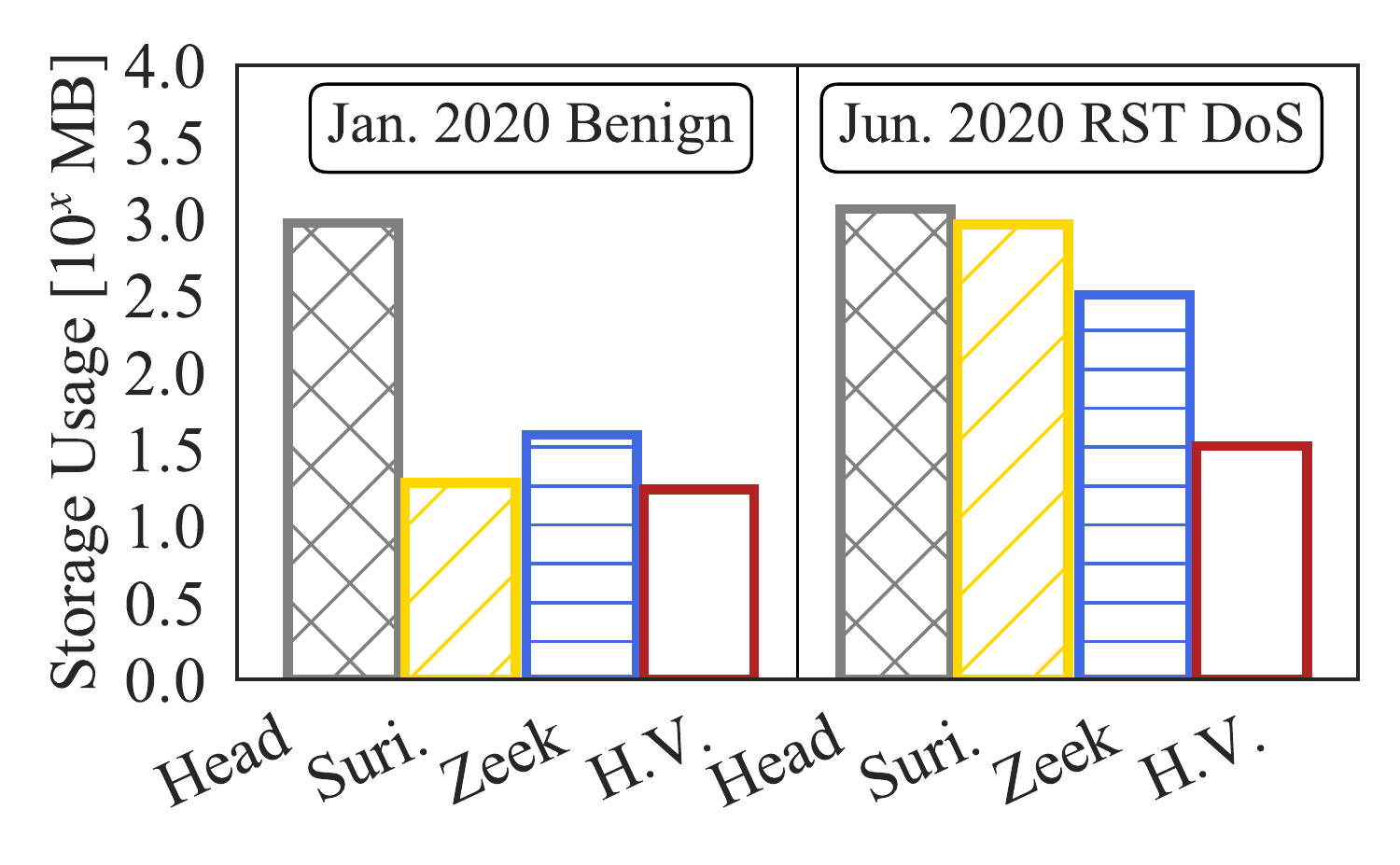}
	}
	\vspace{-2mm}
    \caption{Hardware resource usages of \name.} 
    \label{graph:usage}
    \end{center}
    \vspace{-6mm}
\end{figure}

\noindent\textbf{Encrypted Malware Traffic.} We show the detection accuracy of encrypted malware traffic in Figure~\ref{graph:malware}. Note that, the encrypted malware traffic is hard to detect for the baselines because it is slow and persistent. However, \name accurately detects the malware campaigns with at least 0.964 AUC and 0.891 F1. Specifically, it captures the C\&C servers of spyware for exfiltration as abnormal critical vertices that are connected by massive infected hosts in the graph. As a result, it detects the encrypted malicious traffic of the malware with at least 0.942 F1. For example, to detect Sality P2P botnet shown in Figure~\ref{graph:visual:sality}, \name collects the interactions among similar P2P bots, aggregates the encrypted short flows as edges, and finally clusters the edges with higher loss than benign interaction patterns. Similarly, it can capture the static servers of adware, malware component delivery servers, the infected miner pools as abnormal vertices. Note that, the low-rate malicious flows (at least 0.814 pps) are represented as the edges associated with short flows connected to critical vertices. Meanwhile, the massive long flows with almost 100\% encrypted packet proportion are represented as the edges associated with long flows to the vertices. Therefore, a critical vertex connected with the edges indicates the malware campaign that is significantly different from benign vertices with large degrees, e.g., benign websites.

\subsection{Performance Results} \label{section:evaluation:efficiency}
\noindent\textbf{Throughput.} We truncate the packets to the first 200 bytes on the physical testbed and increase the sending rates until the graph construction module reaches maximum throughput. Figure~\ref{graph:throughput} shows the throughput of the graph construction and the detection. Figure~\ref{graph:throughput:audit-avg} presents the distribution of average throughput within a 1.0s time window. We observe that \name constructs the graph for 28.21 Gb traffic per second. Figure~\ref{graph:throughput:audit-max} presents the maximum throughput in each time window with all the backbone traffic datasets used in the experiments. \name achieves 32.43 $\sim$ 39.71 peak throughput on average. Moreover, we measure the throughput of the graph learning module, which inspects flow interactions. According to Figure~\ref{graph:throughput:detect-avg}, we observe that it can analyze 121.14 Gb traffic per second on average. Note that, the detection throughput is 4.2 times higher than the construction so that the detection can analyze the recorded traffic iteratively to consider the past interaction information. We observe that the average throughput exhibits a bimodal distribution. The peak of low throughput (around 75 Gb/s) is caused by lacking the information on the graph for analyzing during cold start stages. Figure~\ref{graph:throughput:detect-max} illustrates the throughput when the performance of the system is stable. We observe that it achieves 80.6 $\sim$ 148.9 Gb/s throughput. Note that, the throughput on Apr. and Jun. 2020 datasets is lower because of their low original traffic volume.

\noindent\textbf{Latency.} We measure the latency caused by graph construction and detection. Figure~\ref{graph:latency:audit-avg} presents the PDF of the maximum latency for constructing each edge within a 1.0s window. We observe that \name has 1.09s $\sim$ 1.04s average construction latency with an upper bound of 1.93s. The distribution is a significant bimodal one because the receive side scaling (RSS) on the Intel NIC is unbalanced on the threads. The light-load threads have only 0.75s latency. We analyze the composition of the latency in Figure~\ref{graph:latency:audit-comp} (where the error bar is $10^{\mathrm{th}}$ and $90^{\mathrm{th}}$ percentile) and find that the flow classification, short flow aggregation, and long flow distribution fitting share 50.95\%, 35.03\%, and 14.0\% latency, respectively. We measure the average detection latency. Figure~\ref{graph:latency:detect-avg} shows that the learning module has a 0.83s latency on average with a $99^{\mathrm{th}}$ percentile of 4.48s. We also analyze the latency in each step (see Figure~\ref{graph:latency:detect-comp}). We see that 75.8\% of the latency comes from pre-clustering (i.e., 0.66s on average). However, the pre-clustering step reduces the processing overhead of the subsequent processing, i.e., selecting critical vertex and clustering, for $5.5 \times 10^{-3}$s (0.64\%) and $3.4 \times 10^{-3}$s (0.40\%).

\noindent\textbf{Resource Consumption.} Figure~\ref{graph:usage:memory} presents the memory usage of \name. Note that, the DPDK huge pages require 6GB memory and thus we measure the consumption when the usage reaches 6GB. We observe that the increasing rate of memory for maintaining the graph is only 13.1 MB/s. Finally, \name utilizes 1.78 GB memory to maintain the flow interaction patterns extracted from 2.82 TB ongoing traffic. \name incurs low memory consumption because the feature distribution fitting for long flow and short flow aggregation make the in-memory graph compact which ensures low-latency detection and long-term recording. Moreover, the memory consumption of the learning algorithm is 1.452 $\sim$ 1.619 GB. \name can export the graph to disk for forensic analysis. Figure~\ref{graph:usage:disk} shows the storage used for recording the first 45s traffic of the MAWI dataset by different methods, i.e., \name, event based network monitors (i.e., Suricata~\cite{Suricata} and Zeek~\cite{Zeek}), and raw packet headers. We observe that \name achieves 8.99\%, 55.7\%, 98.1\% storage reduction over the baselines, respectively. Meanwhile, our analysis shows that \name retains more traffic information than the existing tools (see Section~\ref{section:analysis}). Thus, the graph based analysis is more efficient than these existing tools.

\section{Related Work} \label{section:related}
\noindent\textbf{Graph Based Anomaly Detection.} Graph based structures have been used for task-specific traffic detection. These methods heavily rely on DPI and thus cannot be applied to detect encrypted traffic~\cite{CONEXT12-BotFinder}. Kwon~\etal analyzed the download relationship graph to identify malware downloading~\cite{CCS15-dropper}, which is similar to WebWitness~\cite{USEC15-WebWitness}. Eshete~\etal constructed HTTP interaction graphs to detect malware static resources~\cite{DSN17-DynaMiner}, and Invernizzi~\etal used a graph constructed from plain-text traffic to identify malware infrastructures~\cite{NDSS14-Nazca}. Different from these works, \name constructs the interaction graph without parsing specific application layer headers and thus achieves task-agnostic encrypted traffic detection. Note that, the provenance graph based attack forensic analysis~\cite{NDSS21-UIScope,NDSS21-WATSON} is orthogonal to our traffic detection.

\noindent\textbf{DTMC Based Anomaly Detection.} 
Discrete-Time Markov Chain (DTMC) has been used to model the behaviors of users/devices~\cite{WiSec20-Peek,ACSAC19-Aegis,USEC17-6thSense}. These methods aim to predict behaviors of users and devices by utilizing DTMC. For instance, Peek-a-Boo predicted user activities~\cite{WiSec20-Peek}, Aegis predicted user behaviors for abnormal event detection~\cite{ACSAC19-Aegis}, and 6thSense predicted sensor behaviors for detecting sensor-based attacks~\cite{USEC17-6thSense}. Different to these methods, our work utilizes DTMC to quantify the benefits of building the compact graph for detecting various unknown attacks.

\noindent\textbf{ML Based Malicious Traffic Detection.} ML based detection can detect zero-day attacks~\cite{ACM-CS-16-survey} and achieve higher accuracy than the traditional signature based methods~\cite{OSDI20-FPGADPI}. For example, Fu~\etal leveraged frequency domain features to realize realtime detection~\cite{CCS21-Whisper}. Barradas~\etal developed Flowlens to extract flow distribution features on data-plane and detect attacks by applying random forest~\cite{NDSS21-Flowlens}. Stealthwatch detected attacks by analyzing flow features extracted from NetFlow~\cite{CiscoETA}. Mirsky~\etal developed Kitsune to learn the per-packet features by adopting auto-encoders~\cite{NDSS18-Kitsune}. For task-specific methods, Nelms~\etal\cite{USEC15-WebWitness}, Invernizzi~\etal\cite{NDSS14-Nazca}, and Bilge~\etal\cite{ACSAC12-Disclose} detected traffic in the different stages of malware campaigns by using statistical ML. Bartos~\etal\cite{USEC16-Variants} and Tang~\etal\cite{INFO20-ZeroWall} detect malformed HTTP request traffic. Holland~\etal\cite{CCS21-nPrint} developed an automatic pipeline for traffic detection. All these methods cannot effectively detect attacks based on encrypted traffic. 

\noindent\textbf{Task-Specific Encrypted Traffic Detection.} The existing encrypted traffic detection relies on domain knowledge for short-term flow-level features~\cite{AISEC16-Encrypted,CiscoETA,C&S21-EncryptedTraffic}. For example, Zheng~\etal leveraged SDN to achieve crossfire attack detection~\cite{TIFS18-Qi}, and Xing~\etal designed the primitives for the programmable switch to detect link flooding attacks~\cite{USEC21-Ripple}. For encrypted malware traffic, Bilge~\etal~\cite{ACSAC12-Disclose} leveraged the traffic history to detect C\&C server, and Tegeler~\etal developed supervised learning using time-scale flow features extracted from malware binaries~\cite{CONEXT12-BotFinder}. Anderson~\etal studies the feasibility of detecting malware encrypted communication via malformed TLS headers~\cite{KDD17-EncryptedTraffic}. To the best of our knowledge, our \name\ is the first system that enables unsupervised detection for the encrypted traffic with unknown patterns.

\noindent\textbf{Encrypted Traffic Classification.} \name aims to identify the malicious behaviors according to encrypted traffic. It is different from encrypted traffic classifications that decide if the traffic is generated by certain applications or users~\cite{TIFS21-DAPP}. For instance, Rimmer~\etal leveraged DL for web fingerprint, which de-anonymizes Tor traffic by classifying encrypted web traffic~\cite{NDSS18-Auto-FP}. Siby~\etal showed that classifying encrypted DNS traffic can jeopardize the user privacy~\cite{NDSS20-DNS-FP}. Similarly, Bahramali~\etal classified the encrypted traffic of instant messaging applications~\cite{NDSS20-instantaneous-FP}. Ede~\etal designed semi-supervised learning for mobile applications fingerprinting~\cite{NDSS20-mobile-app-FP}. All these classifications are orthogonal to \name.

\section{Conclusion} \label{section:conclusion}
In this paper, we present \name, an ML based realtime detection system for encrypted malicious traffic with unknown patterns. \name utilizes a compact in-memory graph to retain flow interaction patterns, while not requiring prior knowledge on the traffic. Specifically, \name uses two different strategies to represent the interaction patterns generated by short and long flows and aggregates the information of these flows. We develop an unsupervised graph learning method to detect the traffic by utilizing the connectivity, sparsity, and statistical features in the graph. Moreover, we establish an information theory based analysis framework to demonstrate that \name preserves near-optimal information of flows for effective detection. The experiments with 92 real-world attack traffic datasets demonstrate that \name achieves at least 0.86 F1 and 0.92 AUC with over 80.6 Gb/s detection throughput and average detection latency of 0.83s. In particular, 44 out of the attacks can evade all five state-of-the-art methods, which demonstrate the effectiveness of \name.


\bibliographystyle{IEEEtranS}
\bibliography{input}

\appendix
\subsection{Details of Implementations}\label{section:appendix:algorithm}\label{section:appendix:tables}
We present the details of the flow classification and short flow aggregation algorithm  in Algorithm 1 and 2, respectively. The features used for edge pre-clustering and clustering  are shown in Table~\ref{table:feature}. And Table~\ref{table:configure} shows the hyper-parameters used in \name and the recommended values. 

\renewcommand{\arraystretch}{1.0}
\begin{table}[h]
    \scriptsize
    \centering
    \setlength\tabcolsep{2.3pt}
    \vspace{-2mm}
    \caption{The features of edges used in \name.}
    \vspace{-4mm}
    \begin{center}
    \begin{threeparttable}
    \begin{tabular}{c|c|c|c}
    \toprule
    \textbf{\tabincell{c}{Edge}} & \textbf{\tabincell{c}{Group}} & \textbf{\tabincell{c}{Data}} & \textbf{Description} \\
    \midrule
    \multirow{13}{*}{\rotatebox{90}{\tabincell{c}{Edge Denoting Short Flows}}} & \multirow{8}{*}{\rotatebox{90}{structural}} & bool & Denoting short flows with the same source address. \\
    &  & bool & Denoting short flows with the same source port. \\
    &  & bool & Denoting short flows with the same destination address. \\
    &  & bool & Denoting show flows with the same destination port. \\
    &  & int & The in-degree of the connected source vertex. \\
    &  & int & The out-degree of the connected source vertex. \\
    &  & int & The in-degree of the connected destination vertex. \\
    &  & int & The out-degree of the connected destination vertex. \\
    \cline{2-4}
    & \multirow{5}{*}{\rotatebox{90}{statistical}} & int & The number of flows denoted by the edge. \\
    &  & int & The length of the feature sequence associated with the edge. \\
    &  & int & The sum of packet lengths in the feature sequence. \\
    &  & int & The mask of protocols in the feature sequence. \\
    &  & float & The mean of arrival intervals in the feature sequence. \\
    \midrule
    \multirow{11}{*}{\rotatebox{90}{\tabincell{c}{Edge Denoting Long Flows}}} & \multirow{4}{*}{\rotatebox{90}{structural}} & int & The in-degree of the connected source vertex. \\
    &  & int & The out-degree of the connected source vertex. \\
    &  & int & The in-degree of the connected destination vertex. \\
    &  & int & The out-degree of the connected destination vertex. \\
    \cline{2-4}
    & \multirow{8}{*}{\rotatebox{90}{statistical}} & float & The flow completion time  of the denoted long flow. \\
    &  & float & The packet rate of the denoted long flow. \\
    &  & int & The number of packets in the denoted long flow. \\
    &  & int & The maximum bin size for fitting packet length distribution.\\
    &  & int & The length associated with the maximum bin size.\\
    &  & int & The maximum bin size for fitting protocol distribution. \\
    &  & int & The protocol associated with the maximum bin size. \\
    \bottomrule
    \end{tabular}
    \end{threeparttable}
    \end{center}
    \label{table:feature}
    \vspace{-3mm}
\end{table}
\newpage

\renewcommand{\arraystretch}{0.96}
\begin{table}[H]
     \vspace{-2mm}
     \scriptsize
     \centering
     \setlength\tabcolsep{1.2pt}
     \caption{Details of malicious traffic datasets.}
     \vspace{-4mm}
     \begin{center}
     \begin{threeparttable}
     \begin{tabular}{c|c|c|c|cc|cc}
     \toprule
     \multicolumn{2}{c|}{\tabincell{c}{\textbf{Class}}} & \textbf{\tabincell{c}{Dataset\\Label}} & \textbf{\tabincell{c}{Description}} & \textbf{\tabincell{c}{Att.}}\tnote{1}  & \textbf{\tabincell{c}{Vic.}} & \textbf{\tabincell{c}{B.W.}}\tnote{2} & \textbf{\tabincell{c}{Enc.\\Ratio}}\\
     \midrule
     \multirow{24}{*}{\rotatebox{90}{Malware Related Encrypted Traffic}} & \multirow{6}{*}{\rotatebox{90}{Spyware}} & Magic. & Magic Hound spyware. & 2 & 479 & 0.34 & 0.13\% \\
     & & Trickster & Encrypted C\&C connections. & 2 & 793 & 0.63 & 10.0\% \\
     & & Plankton & Pulling components from CDN. & 3 & 579 & 59.2 & 23.8\% \\
     & & Penetho & Wifi cracking APK spyware. & 1 & 516 & 3.57 & 100\% \\
     & & Zsone & Multi-round encrypted uploads. & 1 & 479 & 5.98 & 93.0\% \\
     & & CCleaner & Unwanted software downloads. & 4 & 466 & 28.1 & 4.09\% \\
     \cline{2-8}

     & \multirow{4}{*}{\rotatebox{90}{Adware}} & Feiwo & Encrypted ad API calls. & 3 & 1.00K & 19.8 & 100\% \\
     & & Mobidash & Periodical statistic ad updates. & 3 & 624 & 6.08 & 100\% \\
     & & WebComp. & WebCompanion click tricker. & 3 & 281 & 8.38 & 55.2\% \\
     & & Adload & Static resources for PPI adware. & 1 & 280 & 1.04 & 1.09\% \\
     \cline{2-8}

     & \multirow{5}{*}{\rotatebox{90}{\tabincell{c}{Ransom-\\ware}}} & Svpeng & Periodical C\&C interactions (10s). & 2 & 403 & 1.21 & 1.26\% \\
     & & Koler & Invalid TLS connections. & 3 & 333 & 2.22 & 100\% \\
     & & Ransombo & Executable malware downloads. & 5 & 369 & 58.6 & 42.7\% \\
     & & WannaL. & Wannalocker delivers components. & 2 & 275 & 7.49 & 30.3\% \\
     & & Dridex & Victim locations uploading. & 1 & 429 & 4.10 & 100\% \\
     \cline{2-8}
     
     & \multirow{3}{*}{\rotatebox{90}{Miner}} & BitCoinM. & Abnormal encrypted channels. & 1 & 1.54K & 0.79 & 100\% \\
     & & TrojanM. & Long SSL connections to C\&C. & 3 & 1.37K & 2.39 & 89.4\% \\
     & & CoinM. & Periodical connections to pool. & 1 & 1.40K & 0.21 & 100\% \\
     \cline{2-8}
     
     & \multirow{6}{*}{\rotatebox{90}{Botware}} & THBot & Getting C\&C server addresses. & 4 & 103 & 1.72 & 2.71\% \\
     & & Emotet & Communication to C\&C servers. & 6 & 1.17K & 1.43 & 68.6\% \\
     & & Snojan & PPI malware downloading. & 3 & 326 & 8.94 & 100\% \\
     & & Trickbot & Connecting to alternative C\&C. & 4 & 347 & 0.57 & 100\% \\
     & & Mazarbot & Long C\&C connections to cloud. & 3 & 409 & 6.13 & 30.9\% \\
     & & Sality & A P2P botware. & 20 & 247 & 2.19 & 100\% \\
     
     \midrule
     \multirow{15}{*}{\rotatebox{90}{Encrypted Flooding Traffic}} & \multirow{6}{*}{\rotatebox{90}{Link Flooding}} & CrossfireS. & \multirow{3}{*}{\tabincell{c}{We use the botnet cluster sizes\\and the ratio of decony servers\\(HTTPS) in~\cite{SP13-Crossfire}.}} & 100 & 313 & 197 & 100\% \\
     & & CrossfireM. &  & 200 & 313 & 278 & 100\% \\
     & & CrossfireL. &  & 500 & 313 & 503 & 100\% \\
     \cline{3-8}
     & & LrDoS 0.2 & \multirow{3}{*}{\tabincell{c}{We use the traffic of an encrypted\\video application and the settings\\in WAN experiments~\cite{SIGCOMM03-LRTCPDOS}}} & 1 & 1 & 5.57 & 100\% \\
     & & LrDoS 0.5 &  & 1 & 1 & 3.25 & 100\%  \\
     & & LrDoS 1.0 &  & 1 & 1 & 1.90 & 100\% \\
     
     \cline{2-8}
     & \multirow{3}{*}{\rotatebox{90}{\tabincell{c}{SSH\\Inject}}} & ACK Inj. & SSH injection via ACK rate-limits. & 1 & 2 & 1.78 & - \\
     & & IPID Inj. & SSH injection via IPID counters. & 1 & 2 & 0.28 & - \\
     & & IPID Port & Requires of the SSH injection. & 1 & 1 & 1.83 & - \\
     
     \cline{2-8}
     & \multirow{6}{*}{\rotatebox{90}{\tabincell{c}{Password\\Cracking}}} & Telnet S. & Telnet servers in AS38635. & 1 & 19 & 0.63 & 100\% \\
     & & Telnet M. & Telnet servers in AS2501. & 1 & 43 & 1.70 & 100\% \\
     & & Telnet L. & Telnet servers in AS2500. & 1 & 83 & 2.76 & 100\% \\
     & & SSH S. & SSH servers in AS9376. & 1 & 35 & 1.39 & 100\% \\
     & & SSH M. & SSH servers in AS2500. & 1 & 257 & 2.49 & 100\% \\
     & & SSH L. & SSH servers in AS2501. & 1 & 486 & 5.53 & 100\% \\
     
     \midrule
     \multirow{13}{*}{\rotatebox{90}{Encrypted Web Traffic}} & \multirow{10}{*}{\rotatebox{90}{Web Attacks}} & Oracle & TLS padding Oracle. & 1 & 1 & 3.99 & 100\% \\
     & & XSS & Xsssniper XSS detection. & 1 & 1 & 31.8 & 100\% \\
     & & SSLScan & SSL vulnerabilities detection. & 1 & 1 & 15.0 & 100\% \\
     & & Param.Inj. & Commix parameter injection. & 1 & 1 & 17.1 & 100\% \\
     & & Cookie.Inj. & Commix cookie injection. & 1 & 1 & 39.6 & 100\% \\
     & & Agent.Inj. & Commix agent-based injection. & 1 & 1 & 19.7 & 100\% \\
     & & WebCVE & Exploiting CVE-2013-2028. & 1 & 1 & 2.30 & 100\% \\
     & & WebShell & Exploiting CVE-2014-6271. & 1 & 1 & 11.2 & 100\% \\
     & & CSRF & Bolt CSRF detection. & 1 & 1 & 7.73 & 100\% \\
     & & Crawl & A crawler using scrapy. & 1 & 1 & 29.7 & 100\% \\
     \cline{2-8}
     & \multirow{3}{*}{\rotatebox{90}{\tabincell{c}{SMTP\\SSL}}} & Spam1 & Spam using SMTP-over-SSL. & 1 & 1 & 36.2 & 100\% \\
     & & Spam50 & Encrypted spam with 50 bots. & 50 & 1 & 61.7 & 100\% \\
     & & Spam100 & Brute spam using 100 bots. & 100 & 1 & 88.9 & 100\% \\
     
     \midrule
     \multirow{28}{*}{\rotatebox{90}{Traditional Brute Force Attack}} & \multirow{7}{*}{\rotatebox{90}{Brute Scanning}} & ICMP & \multirow{7}{*}{\tabincell{c}{We use the brute force scanning\\rates identified by darknet\\in~\cite{USEC14-ScanScan}. We reproduce the \\scan using Zmap which targets\\the peers and customers\\of AS 2500.}} & 1 & 211K & 5.61 & - \\
     & & NTP &  & 1 & 99.3K & 3.87 & - \\
     & & SSH &  & 1 & 205K & 5.79 & - \\
     & & SQL &  & 1 & 112K & 3.04 & - \\
     & & DNS &  & 1 & 198K & 6.61 & - \\
     & & HTTP &  & 1 & 93.7K & 2.68 & - \\
     & & HTTPS &  & 1 & 209K & 4.89 & - \\
     
     \cline{2-8}
     & \multirow{4}{*}{\rotatebox{90}{\tabincell{c}{Source\\Spoof}}} & SYN & \multirow{4}{*}{\tabincell{c}{We use the protocol types and\\ the packet rates in~\cite{IMC17-DoS}.}} & 6.50K & 1 & 11.41 & - \\
     & & RST &  & 32.5K & 1 & 5.79 & - \\
     & & UDP &  & 6.50K & 1 & 54.3 & - \\
     & & ICMP &  & 3.20K & 1 & 0.13 & - \\
     
     \cline{2-8}
     & \multirow{7}{*}{\rotatebox{90}{\tabincell{c}{Amplification\\Attack}}} & NTP & \multirow{7}{*}{\tabincell{c}{We use the packet rates and\\the vulnerable protocols\\observed in~\cite{IMC17-DoS}.\\ And we use the number of\\the reflectors in~\cite{IMC19-ScanDoS}.}} & 650 & 1 & 95.8 & - \\
     & & DNS &  & 200 & 1 & 82.7 & - \\
     & & CharGen &  & 200 & 1 & 175 & - \\
     & & SSDP &  & 1.30K & 1 & 7.23 & - \\
     & & RIPv1 &  & 500 & 1 & 7.04 & - \\
     & & Memcache &  & 1.60K & 1 & 63.5 & - \\
     & & CLDAP &  & 1.30K & 1 & 36.8 & - \\
     
     \cline{2-8}
     & \multirow{10}{*}{\rotatebox{90}{\tabincell{c}{Probing Vulnerable\\Application}}} & Lr. SMTP & \multirow{10}{*}{\tabincell{c}{We use the sending rates of\\vulnerable application discovery\\disclosed by a darknet~\cite{USEC14-ScanScan}. We\\estimate the number of scanners\\by the number of visible active\\addresses from the vantage\\(i.e., realword measurements)\\and the size of the darknet.}} & 11 & 158K & 7.97 & - \\
     & & Lr.NetBios &  & 28 & 444K & 17.3 & - \\
     & & Lr.Telnet &  & 156 & 1.23M & 49.0 & - \\
     & & Lr.VLC &  & 22 & 352K & 20.5 & - \\
     & & Lr.SNMP &  & 6 & 110K & 6.51 & - \\
     & & Lr.RDP &  & 172 & 1.30M & 53.0 & - \\
     & & Lr.HTTP &  & 94 & 640K & 38.0 & - \\
     & & Lr.DNS &  & 28 & 428K & 25.0 & - \\
     & & Lr.ICMP &  & 268 & 1.82M & 63.3 & - \\
     & & Lr.SSH &  & 72 & 994K & 5.63 & - \\
     
    \bottomrule
    \end{tabular}
    \begin{tablenotes}
          \scriptsize
          \item[1] Att. and Vic. indicate the number of attackers and victims. 
          \item[2] B.W. is short for total bandwidth in the unit of Mb/s.
    \end{tablenotes}
    \end{threeparttable}
    \end{center}
    \label{table:datasetdetail}
    \vspace{-4mm}
\end{table}
\newpage

\renewcommand{\arraystretch}{1.05}
\begin{table}[!t]
    \vspace{-1.9mm}
    \scriptsize
    \centering
    \caption{Recommended hyper-parameter configuration.}
    \vspace{-4mm}
    \begin{center}
        \begin{tabular}{c|c|c|c}
        \toprule
        \textbf{Group} & \textbf{Hyper-Parameter} & \textbf{Description} & \textbf{Value} \\
        \midrule
        \multirow{3}{*}{\tabincell{c}{Graph\\Construction}} & \textsc{pkt\_timeout} & Flow completion time threshold. & 10.0s \\
        & \textsc{flow\_line} & Flow classification threshold. & 15 \\
        & \textsc{agg\_line} & Flow aggregation threshold. & 20 \\
        \midrule
        \multirow{2}{*}{\tabincell{c}{Graph Pre-\\Processing}} & $\epsilon$ & \multirow{2}{*}{\tabincell{c}{DBSCAN hyper-parameters for\\ clustering components and edges.}} & $4\times 10^{-3}$ \\
          & $\mathsf{minPoint}$ & & 40 \\
        \midrule
        \multirow{5}{*}{\tabincell{c}{Traffic\\Detection}} & $K$ & K-means hyper-parameter. & 10 \\
        & $T$ & Loss threshold for malicious traffic. & 10.0 \\
        \cline{2-4}
        & $\alpha$ & \multirow{3}{*}{\tabincell{c}{Balancing the terms in\\the loss function.}} & 0.1 \\
        & $\beta$ & & 0.5 \\
        & $\gamma$ & & 1.7 \\
        \bottomrule
        \end{tabular}
    \label{table:configure}
    \end{center}
    \vspace{-1.2mm}
\end{table}

\newcommand\mycommfont[1]{\footnotesize\rmfamily\textcolor{black}{#1}}
\SetCommentSty{mycommfont}
\SetKwInput{KwInput}{Input}
\SetKwInput{KwOutput}{Output}
\SetAlCapNameFnt{\footnotesize}
\SetAlCapFnt{\footnotesize}

\begin{algorithm}[!t]
\scriptsize
\DontPrintSemicolon
\KwInput{Per-packet features: $\mathsf{PktInfo}$, the hash table for flow collecting: $\mathsf{FlowHashTable}$.}
\KwOutput{Classified flows: $\mathsf{ShortFlow}$ and $\mathsf{LongFlow}$.}
$\mathsf{time\_now} := \mathsf{PktInfo[0].time}$, $\mathsf{last\_check} := \mathsf{time\_now}$. \\
\For(){$\mathsf{pkt}$ in $\mathsf{PktInfo}$} {
     \tcp{Aggregate packets into flows.}
     \If{$\mathsf{Hash(pkt)}$ not in $\mathsf{FlowHashTable}$} {
        $\mathsf{FlowHashTable}$ adds an entry for $\mathsf{pkt}$. \\
     }
     $\mathsf{FlowHashTable[Hash(pkt)]}$ appends $\mathsf{pkt}$. \\
     \If{$\mathsf{time\_now} - \mathsf{last\_check} > $ \textsc{judge\_interval}} {
          \For(){$\mathsf{flow}$ in $\mathsf{FlowHashTable}$} {
               \tcp{Judge the completion of flows.}
               \If {$\mathsf{time\_now} - \mathsf{flow}[-1].\mathsf{time} > $ \textsc{pkt\_timeout}} {
                    \tcp{Classify the flow via the number of packets.}
                    \If{$\mathsf{flow}.\mathsf{size} > $ \textsc{flow\_line}}{
                         $\mathsf{ShortFlow}$ adds $\mathsf{flow}$.} \Else {
                         $\mathsf{LongFlow}$ adds $\mathsf{flow}$.
                    }
                    $\mathsf{FlowHashTable}$ clears the states of $\mathsf{flow}$.
               }
          }
          $\mathsf{last\_check} \gets \mathsf{time\_now}$. \tcp{Record the time of checking.}
    }
    $\mathsf{time\_now} \gets \mathsf{pkt.time}$. \tcp{Update the timer.}
}
\caption{Secure flow classification.}
\label{algorithm:classification}
\end{algorithm}

\begin{algorithm}[!t]
\scriptsize
\DontPrintSemicolon
     \KwInput{Short flows: $\mathsf{ShortFlow}$.}
     \KwOutput{Constructed edges: $\mathsf{ShortEdge}$.}
     Initialize $\mathsf{ProtoHashTable}$ as an empty table.  \\
     \tcp{Select candidate protocols for the aggregation.}
     \For(){$\mathsf{flow}$ in $\mathsf{ShortFlow}$} {
          \tcp{Calculate the protocol mask of a short flow.}
           $\mathsf{flow\_proto} := (\mathsf{flow}[0].\mathsf{proto}| ... | ... | \mathsf{flow}[-1].\mathsf{proto})$. \\
          \If{$\mathsf{Hash(flow\_proto)}$ not in $\mathsf{ProtoHashTable}$} {
               $\mathsf{ProtoHashTable}$ adds an entry for $\mathsf{flow\_proto}$. \\
          }
          Append $\mathsf{flow}$ to $\mathsf{ProtoHashTable[Hash(flow\_proto)]}$. \\
     }
     \tcp{Perform the source aggregation.}
     \For(){$\mathsf{flows}$ in $\mathsf{ProtoHashTable}$ with same protocols} {
          $\mathsf{SrcAddrTable}$ collects the flows with same sources in $\mathsf{flows}$. \\
          \For(){$\mathsf{sflow}$ in $\mathsf{SrcAddrTable}$} {
               \tcp{The flows can be aggregated and denoted by one edge.}
               \If {$\mathsf{sflow.size} > $ \textsc{agg\_line}} {
                    $\mathsf{edge.features}$ := $\mathsf{sflow[0].features}$. \\
                    $\mathsf{edge.source}$ := $\mathsf{sflow[0].source}$. \\
                    \If {an unique destination in $\mathsf{sflow}$}{
                         \tcp{Source and destination aggregation.}
                         $\mathsf{edge.destination}$ saves the unique destination.
                    } \Else (){
                         \tcp{Source aggregation only.}
                         Record each destination in $\mathsf{sflow}$. \\
                    }
                    Add the constructed $\mathsf{edge}$ to $\mathsf{ShortEdge}$. \\
                    $\mathsf{SrcAddrTable}$ evicts $\mathsf{sflow}$.
               }
          }
          $\mathsf{DstAddrTable}$ collects flows with same destinations.\\
          Inspect the flows with the same destinations similarly. \\
          \tcp{Process short flows which cannot be aggregated.}
          $\mathsf{ShortEdge}$ adds flows in $\mathsf{SrcAddrTable}$ and $\mathsf{DstAddrTable}$.
    }
    \caption{Short flow aggregation.}
    \label{algorithm:aggregation}
\end{algorithm}

\subsection{Details of Experiments}\label{section:appendix:experiment}
\subsubsection{Details of Datasets}
We present the detailed properties of the 80 newly collected datasets in Table~\ref{table:datasetdetail}, including the number of attackers and victims, the packet rates of attack flows, and the ratios of encrypted traffic. All the datasets are collected and labeled using the same method as MAWI datasets~\cite{WIDE} and CIC datasets~\cite{CIC17,CIC19}.

\subsubsection{Detection Accuracy of Other Datasets}\label{section:appendix:experiment:dataset}
We use 12 existing datasets to eliminate the impact of dataset bias. Overall, \name achieves 7.8\%, 11.0\%, 5.1\% F1 improvements over the best accuracy of the baselines on Kitsune datasets~\cite{NDSS18-Kitsune}, CIC-IDS2017 datasets~\cite{CIC17}, and CIC-DDoS2019 datasets~\cite{CIC19}, respectively. From the Kitsune datasets, we validate the correctness of the deployed baselines.



\subsubsection{Long-run Performances}\label{section:appendix:experiment:longrun}
By using the CIC datasets~\cite{CIC17,CIC19}, we validate the long-run performances of \name. Specifically, the experiments show that \name achieves over 0.95 F1 and 0.99 AUC in long-run detection (6$\sim$8 hours). The results also verify that the accumulation of detection errors cannot interfere with \name, and \name can detect multiple attacks simultaneously even in the presence of attacks with changed addresses. Moreover, the memory consumption of the compact graph is bounded by 15.6 GB.

\subsubsection{Robustness Against Obfuscation Techniques}\label{section:appendix:experiment:robust}
We validate our method under evasion attacks with different obfuscation techniques according to a recent study~\cite{CCS21-Whisper}, i.e., injecting three kinds of benign traffic. The results demonstrate that the accuracy decrease incurred by the obfuscation is bounded by 4.3\% F1. Specifically, when benign TLS traffic, UDP video traffic, and normal ICMP traffic is injected into brute force attack traffic, the average F1 decreases by 1.7\%, 0.9\%, and 2.4\%, respectively.

The reason why the obfuscation techniques incur negligible accuracy decrease is that they only manipulate patterns of a single flow. \name can still detect the evasion attacks by learning the interaction patterns among various flows.

\subsection{Details of Theoretical Analysis}\label{section:appendix:theory}
\subsubsection{Analysis of Event based Mode}\label{section:appendix:prove-eve}
Let random variable $\mathrm{I_{Eve.}}$ indicate if the event based mode records an event for a flow denoted by a random variable sequence, $\langle s_1, s_2, \dots, s_L \rangle$, $L \sim G(q)$. And we assume that the mode can merge repetitive events. First, we obtain the probability distribution of the random variable $\mathrm{I_{Eve.}}$:

\newcommand{\pr}{\mathbb{P}}
\begin{small}
\begin{equation}
\begin{aligned}
     \pr[\mathrm{I_{Eve.}} = 1] &= 1- \pr[\mathrm{I_{Eve.}} = 0], \\
     \pr[\mathrm{I_{Eve.}} = 0] &= \sum_{l=1}^{\infty}\pr[L = l] \cdot \pr[\mathrm{I_{Eve.}} = 0|L = l] \\
     &= \sum_{l=1}^{\infty} (1-q)^{l-1} \cdot q \cdot (1-p^s)^l \\
     &= \frac{q(1-p^s)}{1- (1-q)(1-p^s)}.
\end{aligned}
\end{equation}
\end{small}

Then, we obtain the entropy of the random variable $\mathrm{I_{Eve.}}$:

\begin{small}
\begin{equation}
\begin{aligned}
     \mathcal{H}_{\mathrm{Eve.}} &= \mathcal{H}[\mathrm{I_{Eve.}}] = \\
     -\pr[\mathrm{I_{Eve.}} = 0] \ln\pr[\mathrm{I_{Eve.}} = 0] &- \pr[\mathrm{I_{Eve.}} = 1] \ln\pr[\mathrm{I_{Eve.}} = 1].
\end{aligned}
\end{equation}
\end{small}

We observe that $\frac{\partial \mathcal{H}[\mathrm{I_{Eve.}}]}{\partial q} \approx 0$ when $q > 0.5$. Thus, we use the second-order taylor series of $q$ to approach $\mathcal{H}_{\mathrm{Eve.}}$:

\begin{small}
\begin{equation}
\begin{aligned}
     \mathcal{H}_{\mathrm{Eve.}} &= \frac{2q(1-p^s)\ln[\frac{(p^s-1)q}{p^s(q-1)-q}]}{p^s(q-1)-q} = -2\theta \ln \theta,
\end{aligned}
\end{equation}
\end{small}

\noindent where $\theta=\frac{\zeta}{\eta}$, $\zeta=q-qp^s$, and $\eta=q-p^s(q-1)$. Similarly, we obtain the expected data scale $\mathcal{L}_{\mathrm{Eve.}}$ and the information density $\mathcal{D}_{\mathrm{Eve.}}$:

\begin{small}
\begin{equation}
\begin{aligned}
     \mathcal{L}_{\mathrm{Eve.}} &= \pr[\mathrm{I_{Eve.}} = 1] = \frac{p^s}{p^s(1-q)+q} = -\frac{p^s}{\eta}, \\
     \mathcal{D}_{\mathrm{Eve.}} &= \frac{\mathcal{H}_{\mathrm{Eve.}}}{\mathcal{L}_{\mathrm{Eve.}}} = \frac{2\zeta}{p^s} \cdot \ln \theta.
\end{aligned}
\end{equation}
\end{small}

\noindent Here, we complete the analysis for the event based mode.

\subsubsection{Analysis of Sampling based Mode}\label{section:appendix:prove-samp}
We use $X_{\mathrm{Samp.}}$ to denote the random variable to be recorded as the flow information in the sampling based mode which is the sum of the observed per-packet features denoted by the random variable sequence. We can obtain the distribution of $X_{\mathrm{Samp.}}$ as follows:

\begin{small}
\vspace{-1mm}
\begin{equation}
\begin{aligned}
     X_{\mathrm{Samp.}}=\sum_{i=1}^{L}s_i,\quad s_i \sim B(s,p) \Rightarrow X_{\mathrm{Samp.}} \sim B(Ls,p).
\end{aligned}
\end{equation}
\end{small}

The amount of the information recorded by the sampling based mode is the Shannon entropy of $X_{\mathrm{Samp.}}$. We decompose the entropy as conditional entropy and mutual information:

\begin{small}
\vspace{-1mm}
\begin{equation}
\begin{aligned}
     \mathcal{H}_{\mathrm{Samp.}} &=\mathcal{H}[X_{\mathrm{Samp.}}] \\ 
     &= \mathcal{H}[X_{\mathrm{Samp.}}|L] + \mathcal{I}(X_{\mathrm{Samp.}};L).
\end{aligned}
\end{equation}
\end{small}

We assume that the mutual information between the sequence length $L$ and the accumulative statistic $X_{\mathrm{Samp.}}$ is close to zero. It implies the impossibility of inferring the statistic from the length of the packet sequence. Then we obtain a lower bound of the entropy as an estimation which is verified to be a tight bound via numerical analysis:

\begin{small}
\begin{equation*}
\vspace{-2mm}
\begin{cases}
     \mathcal{H}_{\mathrm{Samp.}} = \mathcal{H}[X_{\mathrm{Samp.}}|L] &= \sum\limits_{l=1}^{\infty}\pr[L=l] \cdot \mathcal{H}[X_{\mathrm{Samp.}}|L=l] \\
     \qquad \mathcal{H}[X_{\mathrm{Samp.}}|L=l] &= \frac{1}{2} \ln 2\pi elsp(1-p), \\
\end{cases}
\end{equation*}
\begin{equation}
     \Rightarrow\mathcal{H}_{\mathrm{Samp.}} = \frac{1}{2}\ln 2\pi esp(1-p) + \frac{q}{2}\sum\limits_{l=1}^{\infty}(1-q)^{l-1}\ln l.
\end{equation}
\end{small}

We observed that the second-order taylor series can accurately approach the second term of the entropy:

\begin{small}
\begin{equation}
\begin{aligned}
     \mathcal{H}_{\mathrm{Samp.}} = \frac{1}{2}\ln 2\pi esp(1-p) + \frac{\ln 2}{2} q (1 - q).
\end{aligned}
\end{equation}
\end{small}

Finally, we obtain the expected data scale and the information density similar to the analysis for the event based mode and complete the analysis for the sampling based mode.

\subsubsection{Analysis of Graph based Mode in \name}\label{section:appendix:prove-hv}
\name applies different recording strategies for short and long flows, i.e., when $L > K$ it extracts the histogram for long flow feature distribution fitting, and when $L \leq K$ it records detailed per-packet features and aggregates short flows. Let $\mathcal{X}_{\mathrm{\Name}}$ denote the random set of the recorded information. For short flows, all the random variables are collected in $\mathcal{X}_{\mathrm{\Name}}$. For long flows, $\mathcal{X}_{\mathrm{\Name}}$ collects $s$ counters of the histogram for each state on the state diagram of the DTMC. First, we decompose the entropy of the graph based recording mode as the terms for short and long flows:

\begin{small}
\vspace{-2mm}
\begin{equation}
\begin{aligned}
     \mathcal{H}_{\mathrm{\Name}} &= \mathcal{H}[\mathcal{X}_{\mathrm{\Name}}|L] = \sum\limits_{l=1}^{\infty} \pr[L=l] \cdot \mathcal{H}[\mathcal{X}_{\mathrm{\Name}}|L=l] \\
     &= \mathcal{H}[\mathcal{X}_{\mathrm{\Name}}^{\mathrm{S}}|L] + \mathcal{H}[\mathcal{X}_{\mathrm{\Name}}^{\mathrm{L}}|L]
\end{aligned}
\label{equation:all-result}
\end{equation}
\begin{equation*}
\begin{cases}
     \mathcal{H}[\mathcal{X}_{\mathrm{\Name}}^{\mathrm{S}}|L] &= \sum\limits_{l=1}^{K} \pr[L=l] \cdot \mathcal{H}[\mathcal{X}_{\mathrm{\Name}}|L=l]  \\
     \mathcal{H}[\mathcal{X}_{\mathrm{\Name}}^{\mathrm{L}}|L] &= \sum\limits_{l=K+1}^{\infty} \pr[L=l] \cdot \mathcal{H}[\mathcal{X}_{\mathrm{\Name}}|L=l]. \\
\end{cases}
\end{equation*}
\end{small}

\noindent \textbf{Short Flow Information.} \name records detailed per-packet feature sequences for short flows which is the same as the brute recording in the idealized mode. Thus, the increasing rate of information equals the entropy rate of the DTMC:

\begin{small}
\begin{equation}
     \mathcal{H}[\mathcal{X}_{\mathrm{\Name}}|L=l] = l \cdot \mathcal{H}[\mathcal{G}],
\end{equation}
\begin{equation}
\begin{aligned}
     \mathcal{H}[\mathcal{X}_{\mathrm{\Name}}^{\mathrm{S}}|L] &= \sum\limits_{l=1}^{K} \pr[L=l]\cdot l \cdot \mathcal{H}[\mathcal{G}] \\
     &= q \cdot \mathcal{H}[\mathcal{G}] \cdot \sum_{l=1}^{K} (1-q)^{l-1} \cdot l \\
     &= \frac{1-(Kq+1)(1-q)^K}{q} \cdot \mathcal{H}[\mathcal{G}].
\end{aligned}
\label{equation:shortflow-result}
\end{equation}
\end{small}

\noindent \textbf{Long Flow Information.} When $L > K$, the random set collects the counters for distribution fitting. When the DTMC has $s$ states, the histogram has $s$ counters $\upsilon_1, \upsilon_2, \dots, \upsilon_s$, i.e., $\mathcal{X}_{\mathrm{\Name}} = \lbrace \upsilon_1, \upsilon_2, \dots, \upsilon_s \rbrace$. We assume that the counters are independent:

\begin{small}
\begin{equation}
\begin{aligned}
     \upsilon_i = \sum\limits_{j=1}^{L} \delta_j, \qquad
     \delta_j = \left\{
          \begin{aligned}
          1 & ,\quad \textrm{if $s_j$ is the $i^{\mathrm{th}}$ state} \\
          0 & ,\quad \textrm{else.}
          \end{aligned}
          \right.
\end{aligned}
\end{equation}
\end{small}

We observe that $\langle \upsilon_1, \upsilon_2, \dots, \upsilon_s \rangle$ is a binomial process:

\begin{small}
\begin{equation}
\begin{aligned}
     \upsilon_i &\sim B(L, \pr[s_i = i]) \\
          &\sim B(L, C_s^i p^i (1-p)^{s-i}).
\end{aligned}
\end{equation}
\end{small}

To obtain the closed-form solution, we use $\frac{(sp)^i e^{-sp}}{i!}$ as an estimation of $C_s^i p^i (1-p)^{s-i}$. Moreover, the length of the per-packet feature sequence of a long flow is relatively large which implies $\upsilon_i$ approaches a Poisson distribution:

\begin{small}
\begin{equation}
\begin{aligned}
     \upsilon_i \sim& \pi(L \cdot \pr[s_i = i]) \\
     \sim& \pi(\lambda_i), \quad \lambda_i =  \frac{(sp)^i e^{-sp}}{i!}.
\end{aligned}
\end{equation}
\end{small}

Basing on the distribution of the collected counters, we obtain the entropy of the random set:

\begin{small}
\begin{equation}
\begin{aligned}
     \begin{cases}
          \mathcal{H}[\upsilon_i|L=l] &= \quad \frac{1}{2} \ln 2\pi el \frac{(sp)^i e^{-sp}}{i!} \\
          \mathcal{H}[\mathcal{X}_{\mathrm{\Name}}^{\mathrm{L}}|L=l] &= \quad \sum\limits_{i=1}^s \mathcal{H}[\upsilon_i|L=l],
     \end{cases} 
\end{aligned}
\label{equation:longflow}
\end{equation}
\begin{equation*}
\begin{aligned}
          \mathcal{H}[\mathcal{X}_{\mathrm{\Name}}^{\mathrm{L}}|L] &= \sum\limits_{l=K+1}^{\infty} \pr[L=l] \cdot \mathcal{H}[\mathcal{X}_{\mathrm{\Name}}^{\mathrm{L}}|L=l] \\
          &= \sum\limits_{l=K+1}^{\infty} q(1-q)^{l-1} \cdot \sum_{i=1}^{s} \frac{1}{2} \ln 2\pi el \frac{(sp)^i e^{-sp}}{i!} \\
          &= \frac{(1-q)^K}{2} [s\ln 2\pi e + \frac{s(s+1)}{2}\ln sp\\
          &\quad - sp^2 - \sum\limits_{i=1}^s \ln i!] + \frac{qs}{2}[\sum_{l=K+1}^{\infty} (1-q)^{l-1} \ln l].
\end{aligned}
\end{equation*}
\end{small}

The assumption of $q > 0.5$ implies $K^{\mathrm{th}}$ order taylor series can accurately approach the last term in \eqref{equation:longflow}. Moreover, we utilize the quadric term of $s$ in the taylor series of $\sum_{i=1}^{s}\ln i!$ to approach the entropy of long flows ($\gamma$ is Euler–Mascheroni constant):

\begin{small}
\begin{equation}
\begin{aligned}
     \mathcal{H}[\mathcal{X}_{\mathrm{\Name}}^{\mathrm{L}}|L] &= \frac{1}{4}s(1-q)^K[(1+s)\ln ps + \\ & 2\ln2\pi e + 2q\ln K - 2s(1 + p + \gamma)].
\end{aligned}
\label{equation:longflow-result}
\end{equation}
\end{small}

Finally, we take \eqref{equation:shortflow-result} and \eqref{equation:longflow-result} in \eqref{equation:all-result} and complete the analysis for the entropy of the graph based recording mode. Similarly, we obtain the expected data scale by analyzing the conditions of short and long flows separately:

\begin{small}
\begin{equation}
\begin{aligned}
         \mathcal{L}_{\mathrm{\Name}} &= \mathrm{E}[\mathcal{L}_{\mathrm{\Name}}^{\mathrm{S}}|L] + \mathrm{E}[\mathcal{L}_{\mathrm{\Name}}^{\mathrm{L}}|L] \\
         &= \sum_{l=1}^K \pr[L=l] \cdot \frac{L}{C} + \sum_{l=K+1}^{\infty} s \cdot \pr[L=l] \\
         &= s(1-q)^K + \frac{1-(Kq+1)(1-q)^K}{Cq}, \\
\end{aligned}
\end{equation}
\end{small}

\noindent where $C$ is the average number of flows denoted by an edge associated with short flows. Also, we obtain the expected information density by its definition: $\mathcal{D}_{\mathrm{\Name}} = \mathcal{H}_{\mathrm{\Name}} / \mathcal{L}_{\mathrm{\Name}}$ and complete the analysis for the graph based recording mode used by \name.

\end{document}